\documentclass[twocolumn,times,tighten]{aastex631}

\usepackage{bm}  
\usepackage{enumitem}  
\usepackage{color}
\usepackage{amsmath}
\usepackage{graphicx}
\usepackage{grffile}  

\begin{document}

\title{Secondary Whistler and Ion-cyclotron Instabilities driven by Mirror Modes in Galaxy Clusters}

\author[0000-0002-8820-8177]{Francisco Ley}
\affiliation{Department of Astronomy, University of Wisconsin-Madison, Madison, Wisconsin 53706, USA}

\author[0000-0003-4821-713X]{Ellen G. Zweibel}
\affiliation{Department of Astronomy, University of Wisconsin-Madison, Madison, Wisconsin 53706, USA}
\affiliation{Department of Physics, University of Wisconsin-Madison, 1150 University Avenue, Madison,WI, USA 53706}

\author{Drake Miller}
\affiliation{Department of Astronomy, University of Wisconsin-Madison, Madison, Wisconsin 53706, USA}

\author[0000-0003-2928-6412]{Mario Riquelme}
\affiliation{Departamento de F\'{i}sica, Facultad de Ciencias F\'{i}sicas y Matem\'{a}ticas, Universidad de Chile, Chile}


\newcommand{\EZ}[1]{\color{blue} #1}

\newcommand{\Ley}[1]{\color{purple} #1}

\shorttitle{Secondary IC and Whistlers in the ICM}

\begin{abstract}

Electron cyclotron waves (whistlers), are commonly observed in plasmas near Earth and the solar wind. In the presence of nonlinear mirror modes, bursts of whistlers, usually called lion roars, have been observed within low magnetic field regions associated to these modes. In the intracluster medium (ICM) of galaxy clusters, the excitation of the mirror instability is expected, but it is not yet clear whether electron and ion cyclotron waves can also be present under conditions where gas pressure dominates over magnetic pressure (high $\beta$). In this work, we perform fully kinetic particle-in-cell (PIC) simulations of a plasma subject to a continuous amplification of the mean magnetic field $\textbf{B}(t)$ to study the nonlinear stages of the mirror instability and the ensuing excitation of whistler and ion cyclotron (IC) waves under ICM conditions. Once mirror modes reach nonlinear amplitudes, both whistler and IC waves start to emerge simultaneously, with sub-dominant amplitudes, propagating in low-$\textbf{B}$ regions, and quasi-parallel to $\textbf{B}(t)$. We show that the underlying source of excitation is the pressure anisotropy of electrons and ions trapped in mirror modes with loss-cone type distributions. We also observe that IC waves play an essential role in regulating the ion pressure anisotropy at nonlinear stages. We argue that whistler and IC waves are a concomitant feature at late stages of the mirror instability even at high-$\beta$, and therefore expected to be present in astrophysical environments like the ICM. We discuss the implications of our results for collisionless heating and dissipation of turbulence in the ICM.
\end{abstract}



\section{Introduction} \label{sec:intro}

Several classes of astrophysical plasmas display fully developed turbulent states and a weak collisionality, in the sense that the particles' mean free path is several orders of magnitude larger than the typical radius at which they gyrate around the ambient magnetic field. These two characteristics alone can make the transport properties and global evolution of the astrophysical environment in question challenging and dependent on the local evolution at particles' scales. Therefore a detailed study of the behavior of these plasmas at the kinetic level becomes a necessity. 

That is the case of the intracluster medium of galaxy clusters (ICM). The ICM is a hot, magnetized (\cite{Bonafede2010}), weakly collisional and turbulent (\cite{Schuecker2004,Zhuravleva2014,Hitomi2016}) gas in the plasma state where the thermal pressure greatly exceeds the magnetic pressure ($\beta \equiv 8\pi P/B^2 \sim 10-100$, $P$ is the isotropic thermal pressure and $B$ the magnetic field strength). In these conditions, departures from thermodynamic equilibrium, such as pressure anisotropies, are easy to achieve. For example, slow compression of the magnetic field increases particle kinetic energy perpendicular to the magnetic field such that the magnetic moment (or, the magnetic flux through the particle gyro-orbit) remains constant, leading to an excess of perpendicular pressure $P_{\perp}$ over parallel pressure $P_{\parallel}$. However, pressure anisotropy cannot grow unchecked. Pressure anisotropies can easily destabilize microinstabilities such as mirror, firehose, ion-cyclotron and whistler (\cite{Schekochihin2005,Schekochihin2006}). The back reaction of these instabilities on the particles can maintain pressure anisotropy near its marginally unstable value, and are thought to play an important role in several aspects of ICM transport and heating (\cite{Kunz2011,Berlok2021,Drake_2021,Perrone2022a,Perrone2022b,Ley_2023,Tran_2023}). 

In a similar vein, the solar wind and some regions of the Earth's magnetosheath and magnetosphere host plasmas that are also collisionless and turbulent. Even when the plasma $\beta$ is lower than in the ICM ($\beta_i \sim 1 - 10$, $\beta_e \sim 1$), we can encounter some similarities. In particular, the plasma is also pressure anisotropic, and the same microinstabilities above mentioned are found to be present, usually in their fully developed, nonlinear stage (\cite{Bale2009}). Particularly important to this work is the presence of the mirror instability (\cite{Chandrasekhar1958,RudakovSagdeev1961,Hasegawa1969,Southwood&Kivelson1993,KivelsonSouthwood1996,Pokhotelov2002,Pokhotelov2004}) and its interplay with the whistler and (potentially) ion-cyclotron instabilities (\cite{Gary1992},\cite{GaryWang1996}). An example of this has been observed in these space plasmas, and termed whistler lion roars. 

Whistler lion roars are short bursts of right-hand polarized waves, with frequencies below the electron cyclotron frequency ($\omega_{c,e}$) commonly observed in the Earth's magnetosheath and magnetosphere (\cite{Smith1969,Tsurutani1982,Baumjohann1999,Breuillard2018,Giagkiozis2018,Kitamura2020,Zhang2021}), therefore identified as whistler waves. They have also been observed in Saturn's magnetosheath (\cite{Pisa2018}) and the solar wind. They are observed in regions of locally low magnetic field strength (magnetic troughs, or magnetic holes) of magnetic fluctuations. These magnetic troughs are usually identified as structures produced by mirror instability modes, which are able to trap electrons with low parallel velocity within these regions due to the aforementioned invariance of magnetic moment (\cite{Southwood&Kivelson1993}). 

Several mechanisms have been proposed to explain the excitation of whistler lion roars. They usually invoke the pressure anisotropy $P_{\perp,e}> P_{\parallel,e}$ that electrons generate while trapped inside the magnetic troughs ($P_{\perp,e}$ and $P_{\parallel,e}$ are, respectively, the electron pressure perpendicular and parallel with respect to the local magnetic field $\textbf{B}$). Other mechanisms have also been proposed involving counter-propagating electron beams inside these regions, and butterfly distributions in pitch-angle (\cite{Zhang2021,Jiang2022}). As the waves propagate out from the magnetic troughs, they are thought to interact with electrons, regulating the number of trapped electron inside magnetic troughs and also the global anisotropy of electrons in the magnetosheath. This way, there would be a causal connection between an ion-scale mirror instability with an electron scale whistler instability at nonlinear stages, providing valuable insight into the interaction of mirror modes with electrons.

The question arises as to whether a similar interplay can be expected in the ICM. Such  behavior would imply a more complex scenario in which several microinstabilities would be causally connected and coexisting with each other, and several channels of turbulent energy dissipation would open, leading to a much richer dynamics. 

Mirror instability and its consequences have been extensively studied using particle-in-cell (PIC) simulations of moderately and high-$\beta$ plasmas, both hybrid (\cite{Kunz2014,Melville2016,Arzamasskiy2023}) and fully kinetic (\cite{SironiNarayan2015,Riquelme2015,Riquelme2016,Ley_2023}), up to nonlinear stages. Consistent with early theoretical works (\cite{Southwood&Kivelson1993,KivelsonSouthwood1996}), it has been demonstrated that mirror modes are efficient in trapping ions inside regions of low magnetic field strength during their secular growth (\cite{Kunz2014}). When mirror modes reach amplitudes of order $\delta B/B \sim 1$, they reach a saturated stage and the ions eventually undergo scattering, allowing them to escape. This trapping process is similar for electrons, and it has been shown to have important consequences in the electron viscosity and thermal conduction of the plasma (\cite{Riquelme2016,Roberg-Clark2016,Roberg-Clark2018}). Interestingly, \cite{Riquelme2016} reported the observation of whistler waves in the nonlinear, saturated stages of mirror modes in their simulations, along with ion-cyclotron (IC) waves, although they did not pinpoint the cause of the excitation.

In this work, we use PIC simulations to investigate the nonlinear stages of the mirror instability at moderate and high-$\beta$, focusing on the abovementioned excitation of whistler and IC waves. We observe that, indeed, both right hand and left hand polarized, quasi parallel-propagating waves are excited at the end of mirror's secular growth and during its saturated stage, and provide evidence for their excitation mechanism associated to the pressure anisotropy electrons and ions within magnetic troughs of mirror modes. The right- and left-handed circular polarization of these waves lead to their identification as electron-cyclotron (i.e. whistlers) and ion-cyclotron (IC) waves. We also provide some additional discussion about their nature. We describe the interaction of these waves with electrons and ions, and their effect on the regulation of the pressure anisotropy at late stages. 

This paper is organized as follows. Section \S\ref{sec:simulation setup} describes our simulation setup and the runs we perform. Section \S\ref{sec:results} shows our simulation results starting from the excitation of the mirror instability, an early whistler burst and then the late excitation of the electron and ion cyclotron waves at nonlinear stages of the mirror instability. We also detail the mechanism by which these cyclotron waves are excited during the saturated stage of mirror modes, by tracking ions and electrons throughout the simulations. We also describe the subsequent interaction of these waves with the ions and electrons at late stages. In section \S\ref{sec:MassRatioDependence} we discuss the dependence of our results on the mass ratio used in our simulations and show that they are fairly insensitive to it. In section \S\ref{sec:betadependence} we present results of simulations at different initial ion plasma beta, and show these cyclotron waves are also present at lower and higher betas as well. Finally, we discuss the implication of our work in the context of galaxy clusters and present our conclusions in section \S\ref{sec:discussion}.

\section{Simulation Setup}\label{sec:simulation setup}

\begin{figure}[t]
    \centering
    \includegraphics[width=\linewidth]{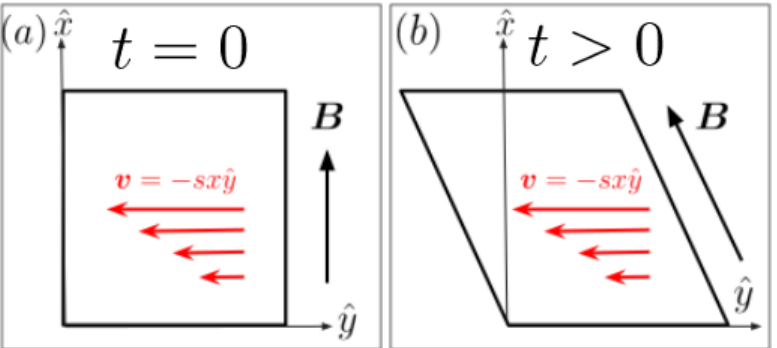}
    \caption{The evolution of the simulation domain. Panel $a$: Initially, the box is straight, the magnetic field is initialized pointing in the $\hat{x}$ direction and a shear velocity field $\textbf{v} = -sx\hat{y}$ is imposed in the y--direction (red arrows). Panel $b$: The velocity field shears the box continuously throughout the simulation, amplifying the magnetic field and changing its direction in the process due to magnetic flux conservation.}
    \label{fig:shearingbox}
\end{figure}

\begin{table}[ht]
\centering
\caption{Simulation List: The physical parameters of the simulations are: the initial ion plasma beta $\beta\equiv 8\pi P_i^{\text{init}}/B^2$, where $P_i^{\text{init}}$ is the initial ion pressure, the mass ratio between ions and electrons $m_i/m_e$, and the magnetization $\omega_{c,i}/s$. The numerical parameters are the number of particles per cell $N_{\text{ppc}}$ and the domain size in terms of the initial ion Larmor radius $L/R_{L,i}^{\text{init}}$. Our fiducial simulation is highlighted in bold.}
\begin{tabular}{lcccccc}
\hline
Runs               & \multicolumn{1}{l}{$\beta_{i}^{\text{init}}$} & \multicolumn{1}{l}{$m_i/m_e$} & \multicolumn{1}{l}{$\omega_{c,i}^{\text init}/s$} & \multicolumn{1}{l}{$\frac{k_BT}{m_ic^2}$} & \multicolumn{1}{l}{$N_{\text{ppc}}$} & \multicolumn{1}{l}{$L/R_{L,i}^{\text{init}}$} \\ \hline\hline
\textbf{b20m8w800} & \textbf{20}                                   & \textbf{8}                    & \textbf{800}                         & \textbf{0.02}                             & \textbf{600}                         & \textbf{54}                                   \\
b20m32w800         & 20                                            & 32                            & 800                                  & 0.01                                      & 300                                  & 50                                            \\
b20m64w800         & 20                                            & 64                            & 800                                  & 0.01                                      & 200                                  & 40                                            \\
b40m8w800           & 40                                             & 8                             & 800                                  & 0.02                                      & 300                                  & 49                                            \\ 
b2m8w800           & 2                                             & 8                             & 800                                  & 0.02                                      & 300                                  & 68                                            \\ \hline
\end{tabular}
\label{table:SimulationList}
\end{table}

We perform fully kinetic, 2.5D particle-in-cell (PIC) simulations using TRISTAN-MP (\cite{Buneman1993,Spitkovsky2005}), in which we continuously shear a collisionless, magnetized plasma composed of ions and electrons (\cite{Riquelme2012}). The magnetic field is initially spatially uniform and starts pointing along the $x$--axis. A shear velocity field is imposed with $\textbf{v} = -sx\hat{y}$ (red arrows in fig. \ref{fig:shearingbox}), where $x$ is the distance along the $x$--axis and $s$ is a constant shear rate. We solve the PIC system of equations using shearing coordinates, as implemented in \cite{Riquelme2012} (The suitability of this approach to studying ion Larmor scale phenomena is also discussed in \cite{Riquelme2015}). The conservation of magnetic flux implies that the $y$--component of the magnetic field $\textbf{B}$ evolves as $dB_y/dt = -sB_0$, whereas $dB_x/dt=0$ and $dB_z/dt=0$. The action of the shear then continuously amplifies the magnetic field strength such that its magnitude evolves as $B(t) = B_0\sqrt{1 + s^2t^2}$.

In our simulations, ions and electrons are initialized with Maxwell-J\"{u}ttner distributions (the relativistic generalization of the Maxwell-Boltzmann distribution, \cite{Juttner1911}) with equal initial temperatures $T_i^{\text{init}}=T_e^{\text{init}}$, and $k_BT_{i}^{\text{init}}/m_ic^2$ between $0.01$ and $0.02$. The physical parameters of our simulations are the initial temperature of ions and electrons ($T_i^{\text{init}}=T_e^{\text{init}}$), the initial ion plasma beta, $\beta_i^{\text{init}}$ , the mass ratio between ions and electrons $m_i/m_e$, and the ratio between the initial ion cyclotron frequency and the shear frequency, $\omega_{c,i}^{\text{init}}/s$, that we call the ``scale-separation ratio''. The numerical parameters in our simulations are the number of macroparticles per cell, $N_{\text{ppc}}$, the plasma skin depth in terms of grid point spacing, $c/\sqrt{\omega_{p,e}^2+\omega_{p,i}^2}/\Delta x$, and the domain size in terms of the initial ion Larmor radius, $L/R_{L,i}^{\text{init}}$, where $R_{L,i}^{\text{init}}=v_{\text{th},i}/\omega_{c,i}^{\text{init}}$ and $v_{th,i}^2 = k_B T_i/m_i$.  These physical and numerical parameters are listed in Table \ref{table:SimulationList}. We fix $c/\sqrt{\omega_{p,e}^2+\omega_{p,i}^2}/\Delta x = 3.5$ in the simulations presented in Table \ref{table:SimulationList}. 

In the bulk of the paper we discuss a representative, fiducial simulation with $m_i/m_e=8$, $\beta_{i}^{\text{init}}=20$ (thus $\beta^{\text{init}}=\beta_i^{\text{init}}+\beta_e^{\text{init}}=40$) and $\omega_{c,i}^{\text{init}}=800$ (simulation b20m8w200 in Table \ref{table:SimulationList}, highlighted in boldface). We vary the above parameters in a series of simulations, all listed in Table \ref{table:SimulationList}. Importantly, given the available computational capabilities, performing a simulation with realistic mass ratio $m_i/m_e=1836$ becomes prohibitively expensive. Therefore, a range of values of ion-to-electron mass ratio are presented in order to ensure that our results do not strongly depend on this parameter. The effects of varying these parameters are discussed in \S\S \ref{sec:MassRatioDependence} \& \ref{sec:betadependence}.

In the absence of a scattering mechanism and/or collisions, the ion and electron magnetic moments $\mu_j\equiv p^2_{\perp,j}/(2m_jB)$ and longitudinal action $\mathcal{J}_j \equiv \oint p_{j,\parallel} d\ell$ are adiabatic invariants ($p_{\perp,j}$ and $p_{\parallel,j}$ are the components of the momentum of a particle of species $j$ perpendicular and parallel to the local magnetic field, respectively, and $j=i,e$), and therefore are conserved as the system evolves, provided that the variation of $\textbf{B}$ is sufficiently slow compared to the particle cyclotron frequencies; in our case, 
$s \ll \omega_{c,j}$, where $\omega_{c,j}=eB/m_jc$ is the cyclotron frequency of particles of species $j$, $c$ is the speed of light, and $e$ is the magnitude of the electric charge.

The continuous amplification of the magnetic field $\textbf{B}$ implies that the particles' adiabatic invariance drives a pressure anisotropy in the plasma such that $P_{\perp,j}>P_{\parallel,j}$. In the very early stages of the simulation, we expect the evolution of $P_{\perp,j}$ and $P_{\parallel,j}$ to be dictated by the double-adiabatic scalings (\cite{CGL1956}). Soon after this stage, however, the pressure anisotropy acts as a free energy source in the plasma and is able to excite several kinetic microinstabilities after surpassing their excitation thresholds, which are proportional to $\beta^{-\alpha}$, $(\alpha \sim 0.5-1)$ (\cite{Hasegawa1969,GaryLee1993,GaryWang1996}). These microinstabilities break the adiabatic invariants and act upon the pressure anisotropy to regulate the anisotropy growth in the nonlinear stages.

In our simulations, and given our initial physical parameters (namely, $\beta_{i}^{\text{init}}\equiv 8\pi P^{\text{init}}_{i}/B^{2\text{init}} = 20$), we expect the dominant instability to be the mirror instability. Mirror modes are purely growing (i.e. zero real frequency), with the fastest growing modes propagating highly obliquely with respect to the mean magnetic field. Their most unstable wavenumbers satisfy $k_{\perp}R_{L,i} \sim 1$, where $R_{L,i}$ is the ion Larmor radius. This instability presents Landau resonances with particles of very small parallel momentum, $p_{\parallel} \approx 0$, that become trapped in between mirror modes, and contribute to regulating the pressure anisotropy. 

In addition to the mirror instability, we also observe wave activity that we associate with the ion-cyclotron (\cite{Gary1992}) and whistler (\cite{GaryWang1996}) instabilities at ion and electron scales, respectively, during the late stages of our simulations. Ion cyclotron (IC) modes are left circularly polarized and have real frequency below the ion-cyclotron frequency $\omega_{c,i}$, with modes of maximum growth rate propagating parallel to the mean magnetic field \textbf{B}. Similarly, whistler modes are right circularly polarized and have real frequency below the electron cyclotron frequency $\omega_{c,e}$, with modes of maximum growth rate also propagating parallel to \textbf{B}. As we will see, this wave activity is associated with the ion and electron trapping processes that mirror modes generate.

\section{Results}\label{sec:results}
\begin{figure*}[ht]
    \centering
    \begin{tabular}{c}
         \includegraphics[width=0.98\linewidth]{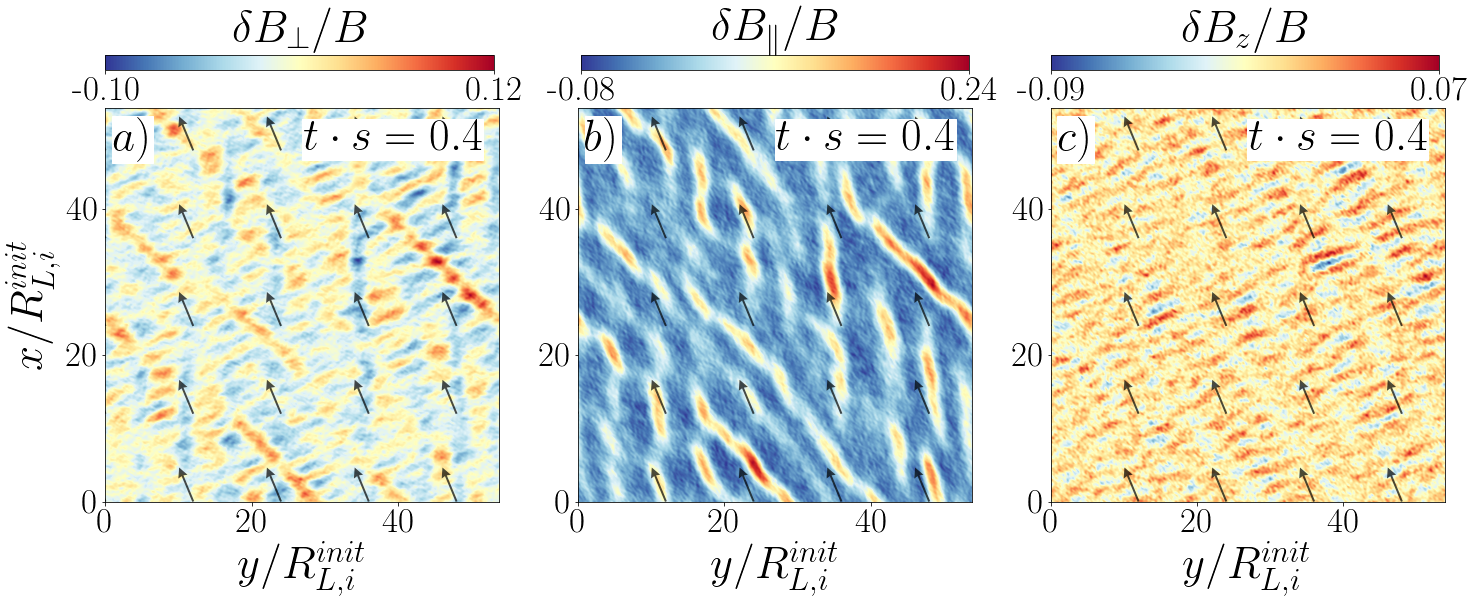}  \\
         \includegraphics[width=0.98\linewidth]{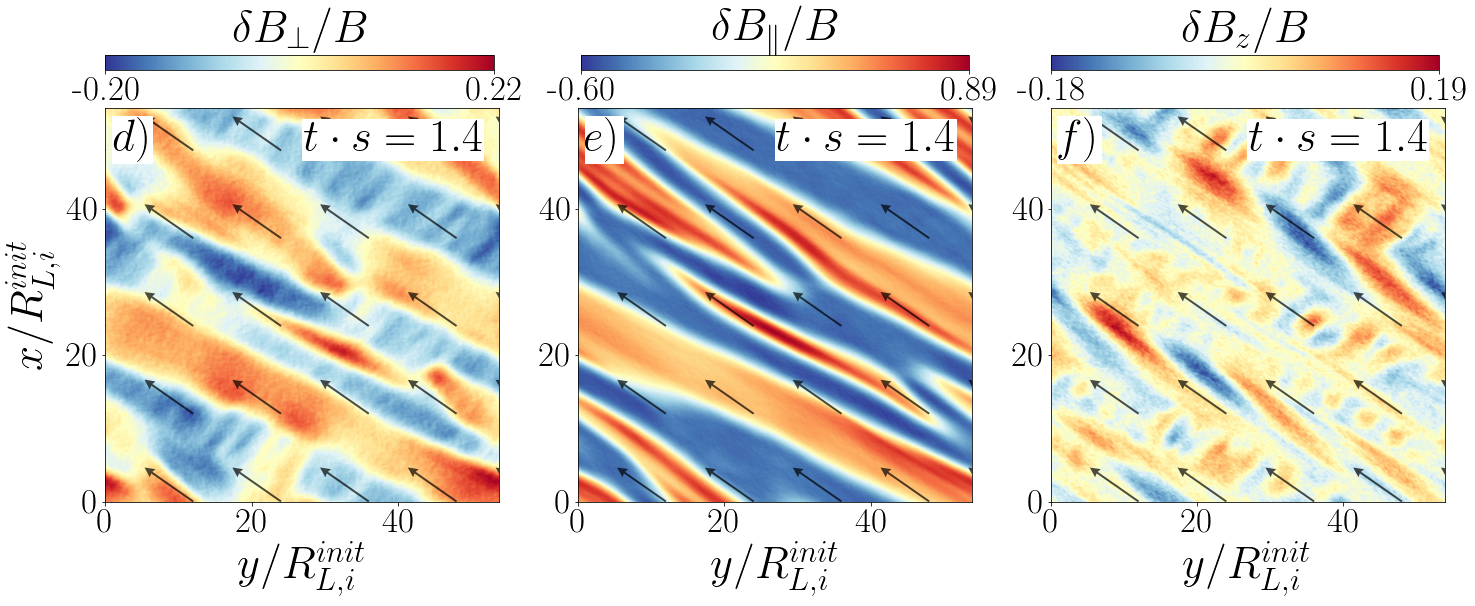}  \\
         \includegraphics[width=0.98\linewidth]{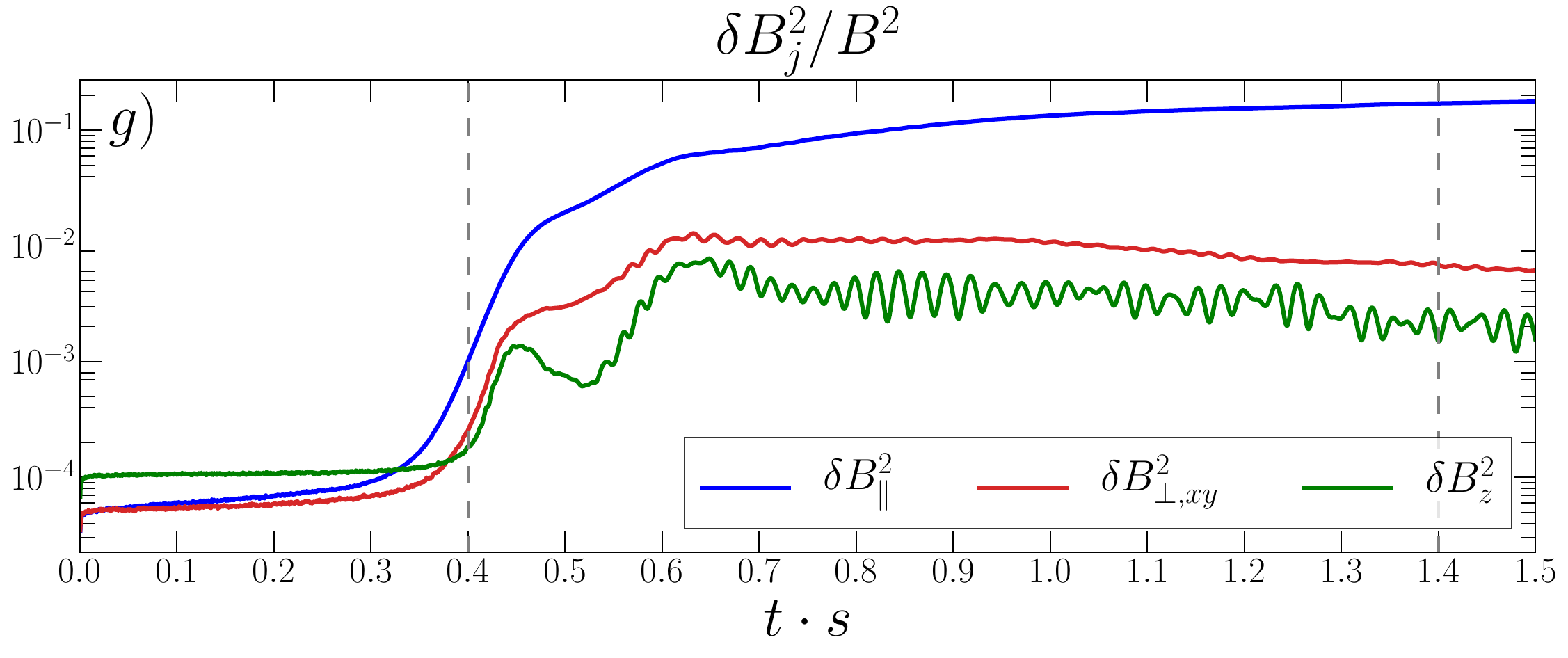}  \\
    \end{tabular}
    \caption{\textbf{First row:} The different component of magnetic fluctuations $\delta \textbf{B} = \textbf{B} - \langle\textbf{B}\rangle$ for run b20m8w800 in the simulation domain at $t\cdot s=0.4$: $\delta B_{\perp}$ (Panel $a$) is the component perpendicular to the main field $\langle \textbf{B} \rangle$ in the $x$--$y$ plane of the simulation, $\delta B_{\parallel}$ (panel $b$) is the component parallel to $\langle \textbf{B} \rangle $ and $\delta B_z$ (panel $c$) is the component perpendicular to $\langle \textbf{B} \rangle$ in the direction out of the plane of the simulation. \textbf{Second row:} Panels $d$, $e$ and $f$ show the same as panels $a$, $b$ and $c$, but  but at $t\cdot s=1.4$. \textbf{Third row:} The evolution of the energy in the three component of the magnetic field fluctuations $\delta \textbf{B}$ normalized to $B(t)^2$, $\delta B_{\parallel}^2$ (blue line), $\delta B_{\perp,xy}^2$ (red line) and $\delta B_z^2$ (green line). The dashed gray lines indicate the time at which the fluctuations in the first and second row are shown. An animation is available in the online version.}
    \label{fig:MagneticFluctuations}
\end{figure*}

\begin{figure}[hbtp]
    \centering
    \begin{tabular}{c}
         \includegraphics[width=0.84\linewidth]{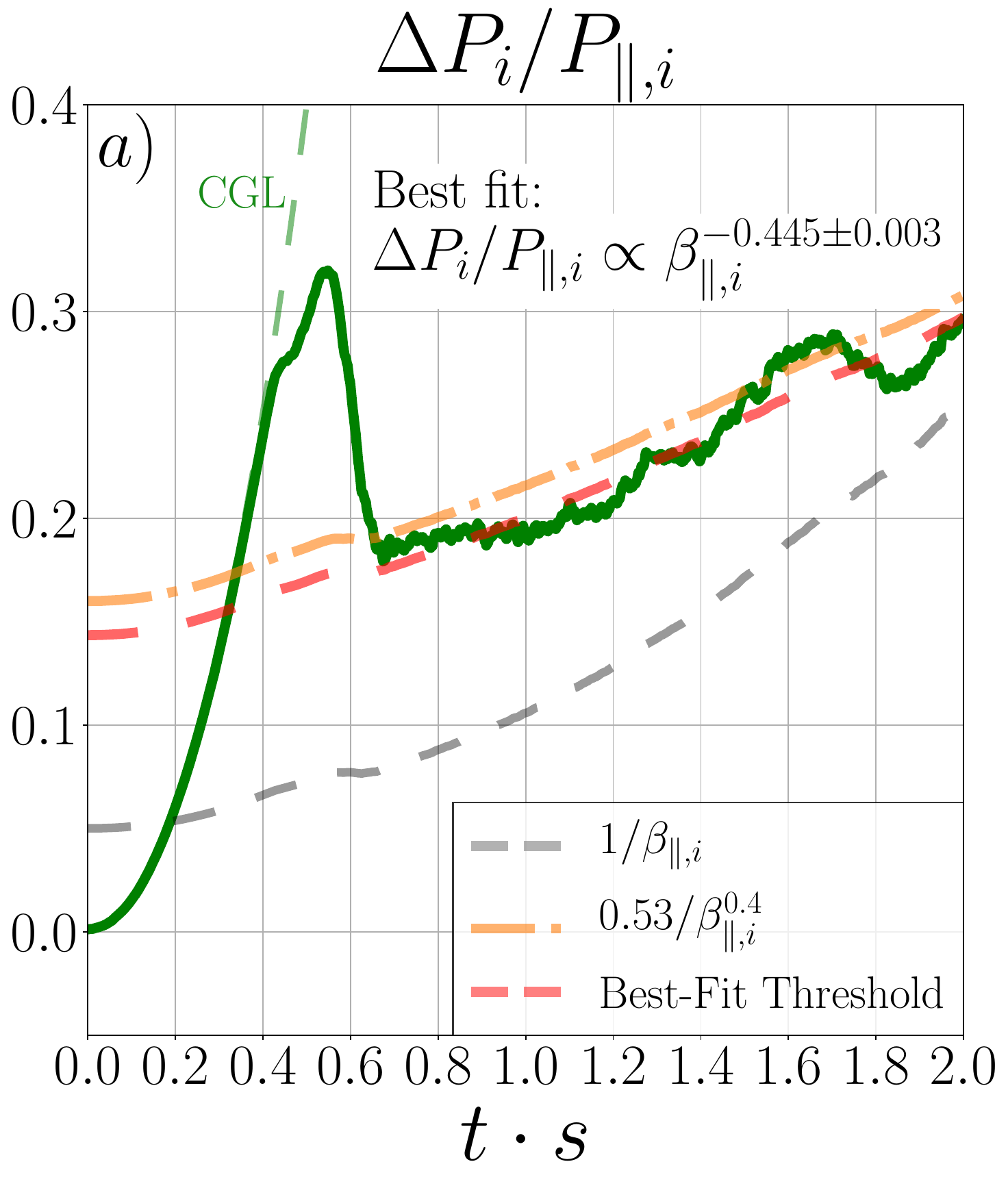}  \\
          \includegraphics[width=0.84\linewidth]{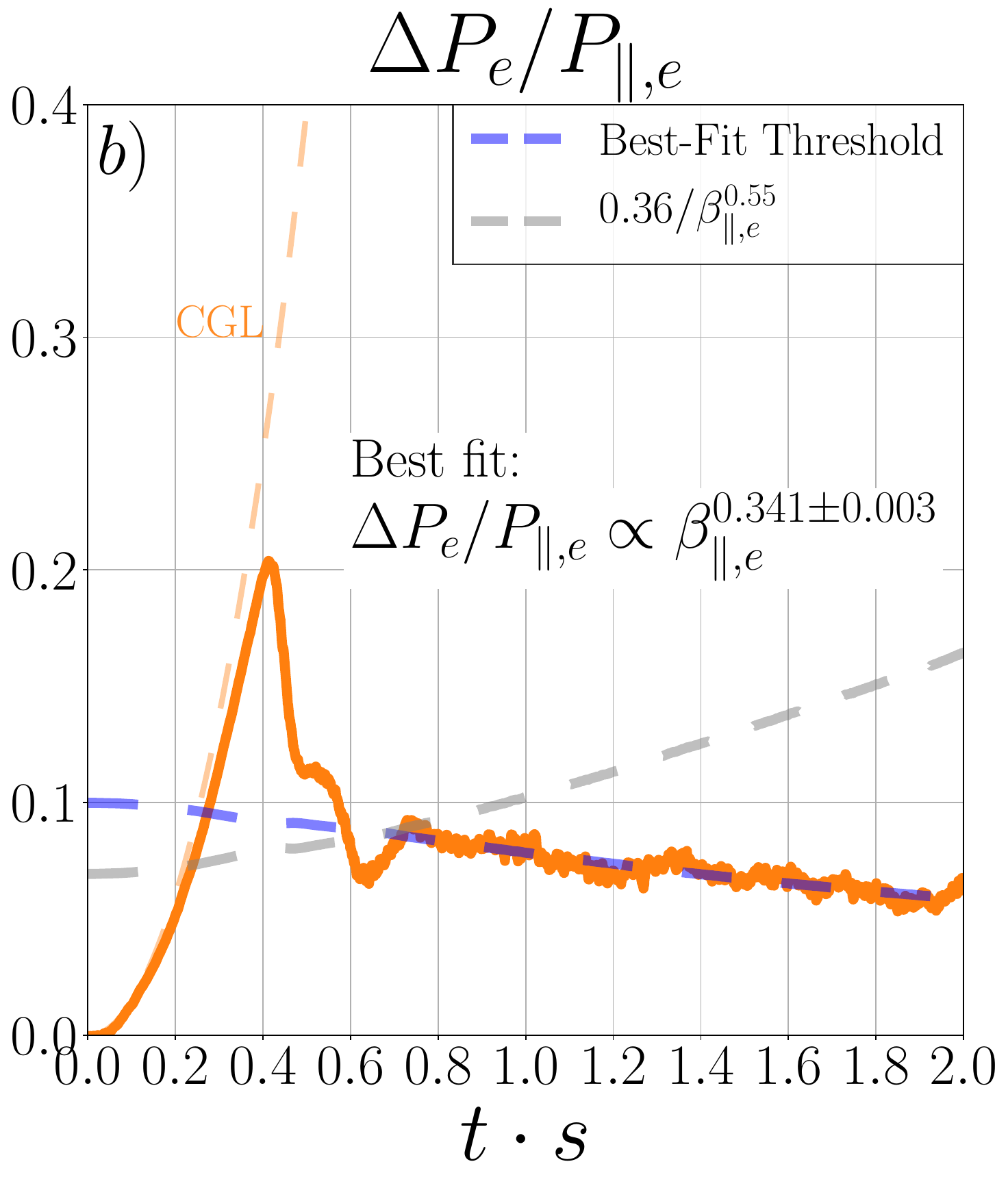}
    \end{tabular}
    \caption{Panel $a$: The evolution of the ion pressure anisotropy $\Delta P_i/P_{\parallel,i}$ for run b20m8w800 is shown as a solid green line. The dashed green line shows the double-adiabatic evolution of $\Delta P_i/P_{\parallel,i}$ (\cite{CGL1956}). The dashed gray line shows the approximate threshold for the mirror instability: $1/\beta_{\parallel,i}$ (\cite{Hasegawa1969}). The dotted-dashed orange line shows the threshold for the IC instability from \cite{GaryLee1993} for $\gamma_{IC}/\omega_{c,i} = 10^{-2}$ ($\gamma_{IC}$ is the IC growth rate). The red dashed line shows the best-fit to $\Delta P_{i}/P_{\parallel,i} = A_i \beta_{\parallel,i}^{\alpha_i}$ from $t \cdot s = 0.7 $ to $t\cdot s = 2.0$, with $A_i=0.544 \pm 0.003$ and $\alpha_i=0.445 \pm 0.003$. Panel $b$: The evolution of the electron pressure anisotropy $\Delta P_e/P_{\parallel,e}$ is shown as solid orange line. The dashed orange line shows the double-adiabatic evolution of $\Delta P_e/P_{\parallel,e}$. The dashed blue line shows the best-fit to $\Delta P_{e}/P_{\parallel,e} = A_e \beta_{\parallel,e}^{\alpha_e}$ from $t \cdot s = 0.7 $ to $t\cdot s = 2.0$, with $A_e = 0.036 \pm 0.0002$ and $\alpha_e = 0.341\pm 0.003$. The dashed gray line shows the linear threshold for the anisotropic whistler instability from (\cite{GaryWang1996}) for growth rate $\gamma_W/\omega_{c,e}=0.01$. ($\gamma_W$ is the whistler growth rate).}
    \label{fig:IonLecAnisotropy}
\end{figure}

Figures \ref{fig:MagneticFluctuations} and \ref{fig:IonLecAnisotropy} summarize the evolution of magnetic field fluctuations and particle pressure anisotropy over time.

Figure \ref{fig:MagneticFluctuations} shows the fluctuations in the magnetic field $\delta \textbf{B} \equiv \textbf{B} - \langle \textbf{B} \rangle$ (where $\langle \cdot \rangle$ denotes a volume average over the entire simulation domain) in its three different components at two different times: $t\cdot s = 0.4$ (first row, panels $a$,$b$ and $c$) and at $t\cdot s = 1.4$ (second row, panels $d$, $e$ and $f$). The black arrows in panels $a$--$f$ denote the direction of the mean magnetic field $\langle \textbf{B} \rangle$ at those particular times. The components of $\delta \textbf{B}$ are defined as parallel with respect to the main field $\langle \textbf{B}\rangle$ ($\delta B_{\parallel}$, panels $b$ and $e$), perpendicular to $\langle \textbf{B}\rangle$ in the plane of the simulation ($\delta B_{\perp,xy}$, panels $a$ and $d$) and perpendicular to $\langle \textbf{B}\rangle$ in the direction out of the simulation plane ($\delta B_z$, panels $c$ and $f$). Additionally, figure \ref{fig:MagneticFluctuations}$g$ shows the evolution of the energy in each of the three components of $\delta \textbf{B}$, normalized by $B(t)^2$; $\delta B_{\parallel}^2$ (blue line), $\delta B_{\perp,xy}^2$ (red line), and $\delta B_z^2$ (green line).

Figure \ref{fig:IonLecAnisotropy}$a$ shows the evolution of the ion pressure anisotropy $\Delta P_i \equiv P_{\perp,i}-P_{\parallel,i}$ for run b20m8w800, and the dashed gray line shows the approximate instability threshold for the mirror instability (\cite{Hasegawa1969,Hellinger2007}). We can see that the ion anisotropy surpasses the mirror threshold very early in the simulation, and reaches its maximum value at $t\cdot s \approx 0.5$ (we will call this stage the anisotropy overshoot hereafter). We will show that this is consistent with the beginning of the secular growth of mirror modes (\cite{Kunz2014}, \cite{Riquelme2016}). Figure \ref{fig:IonLecAnisotropy}$b$ shows the same for the electron
pressure anisotropy, which we will show relaxes by efficient scattering.

\subsection{Mirror Instability Evolution}
\label{sec:MirrorEvolution}

Since mirror modes are highly oblique, their evolution is well represented by the time trace of $\delta B_{\parallel}^2$ shown in fig. \ref{fig:MagneticFluctuations}$g$. We identify both a linear, exponentially growing phase until $t\cdot s \approx 0.45$, and a subsequent nonlinear, slower growing secular phase, consistent with the different evolutionary phases of the ion and electron pressure anisotropies described above. Besides the break in the mirror mode's evolution at $t\cdot s \approx 0.45$, a second break in the secular growth occurs around $t \cdot s=0.6$ followed by a shallower slope of growth. We will show that this break coincides with the excitation of both whistler and IC waves in $\delta B_{\perp,xy}^2$ and $\delta B_z^2$, implying that whistler and IC waves, albeit smaller in amplitude, modulate the evolution of mirror modes during nonlinear stages.

\subsubsection{Linear, exponentially growing mirror phase}

After an early CGL phase of the pressure anisotropy $\Delta P_j$ ($j=i,e$, see fig. \ref{fig:IonLecAnisotropy}), fig. \ref{fig:MagneticFluctuations}$g$ shows the excitation of the mirror instability starting at $t\cdot s \approx 0.35$, mainly in the parallel component of the magnetic fluctuations, $\delta B_{\parallel}$ (blue line), consistent with theoretical expectations (\cite{Southwood&Kivelson1993,Pokhotelov2004}). Figure \ref{fig:MagneticFluctuations}$g$ also shows that $\delta B_{\parallel}$ grows first and it has the largest amplitude throughout the entire simulation, meaning that the mirror instability is indeed the dominant instability.

Figure \ref{fig:MagneticFluctuations}$b$ (i.e. $\delta B_{\parallel}^2$) shows the linear, exponentially growing phase of mirror modes at $t\cdot s = 0.4$, where small filamentary structures of high local magnetic field amplitude start to emerge and slowly grow, in between wider regions of low local magnetic field amplitude. The obliqueness of the modes is readily apparent, as well as the fact that the mirror generated magnetic fluctuations lie mainly in the ($\textbf{k}$,\textbf{B}) plane (they can be seen in $\delta B_{\perp,xy}^2$ too, but not in $\delta B_z^2$, as expected from linear theory (\cite{Pokhotelov2004})). The oblique nature of mirror modes can also be seen in fig. \ref{fig:PowerSpectrum_dBpar_Space}$a$, where we show the power spectrum in space of $\delta B_{\parallel}$ at $t\cdot s=0.4$. The solid and dashed lines represent the directions parallel and perpendicular to the mean magnetic field $\langle \textbf{B}\rangle$, respectively. Therefore, we can see that at $t\cdot s = 0.4$, the power is mostly concentrated between wavevectors $0.44 \lesssim kR_{L,i}^{\text{init}} \lesssim 1.35$ and angles of $52^{\circ} \lesssim \theta_{k} \lesssim 77^{\circ}$, where $\theta_{k} \equiv \cos^{-1}(\textbf{k}\cdot \langle \textbf{B}\rangle /kB)$ is the angle between mirror modes' wavevector and the mean magnetic field $\langle \textbf{B}\rangle$.

It should be emphasized that the ion-cyclotron wave activity only starts at $t\cdot s = 0.6$, and not before. There is no sign of an early excitation of the ion-cyclotron instability competing with the mirror instability for the available free energy in $\Delta P_i$. Instead, at earlier stages, only the mirror instability is excited, consistent with our initial conditions of high-beta ($\beta_i^{\text{init}}=20$), where the mirror instability is expected to dominate (e.g. \cite{Riquelme2015}).

The absence of ion-cyclotron waves early in the simulation ($0< t\cdot s < 0.6$) is clearly seen in fig. \ref{fig:PowerSpectrum_whistler1}$a$, where we show the power spectrum in time and space of $\delta B_z(\omega,k_{\parallel})+i\delta B_{\perp,xy}(\omega,k_{\parallel})$ at early stages: $0.3 < t\cdot s < 0.5$. This particular combination of the two perpendicular components of $\delta \textbf{B}$ allows us to disentangle the parallel-propagating waves (with respect to the main magnetic field $\langle \textbf{B}\rangle$, e.g. ion-cyclotron and whistlers), and also their left-handed and right-handed circular polarizations (\cite{Ley2019,Tran_2023}). In this case, the left-hand circularly polarized wave activity is shown for $\omega > 0$, whereas right-hand circularly polarized wave activity is shown for $\omega < 0$. We readily see that, apart from the $\omega \approx 0$ power consistent with mirror modes appearing in $\delta B_{\perp,xy}$, there is no left-handed polarized wave activity throughout $0.3 < t\cdot s < 0.5$, only right-handed polarized waves, which corresponds to an early excitation of the whistler instability, as we will see in section \ref{sec:EarlyWhistlers}.

\subsubsection{Nonlinear, secular mirror phase}

At $t \cdot s \approx 0.45$, we can clearly see the beginning of the secular growth of the mirror instability, where the modes reach nonlinear amplitudes, and keep growing but at a slower rate. This evolution is consistent with previous works (\cite{Kunz2014,Riquelme2016}).

Interestingly, the mirror secular growth is interrupted at $t \cdot s \approx 0.6$, and the slope of $\delta B_{\parallel}^2$ breaks. This is also approximately where the ion pressure anisotropy experiences its fastest decline (fig. \ref{fig:IonLecAnisotropy}). Mirror modes continue to grow, but at a much slower rate. This is consistent with the saturation of energy in the subdominant components $\delta B_{\perp,xy}^2$ and $\delta B_z^2$ (solid red and green line in fig. \ref{fig:MagneticFluctuations}$g$, respectively), which also presents a distinct pattern of oscillations. This activity is a clear evidence of a new burst of waves with components mainly in the direction perpendicular to $\delta \textbf{B}$, and we will see that they are consistent with both electron cyclotron waves (whistlers) and ion cyclotron waves excited by electron and ion populations, respectively, that become trapped within mirror modes (see sec. \ref{sec:LionRoars}). 


\begin{figure}[ht]
\centering
    \begin{tabular}{c}
         \includegraphics[width=0.95\linewidth]{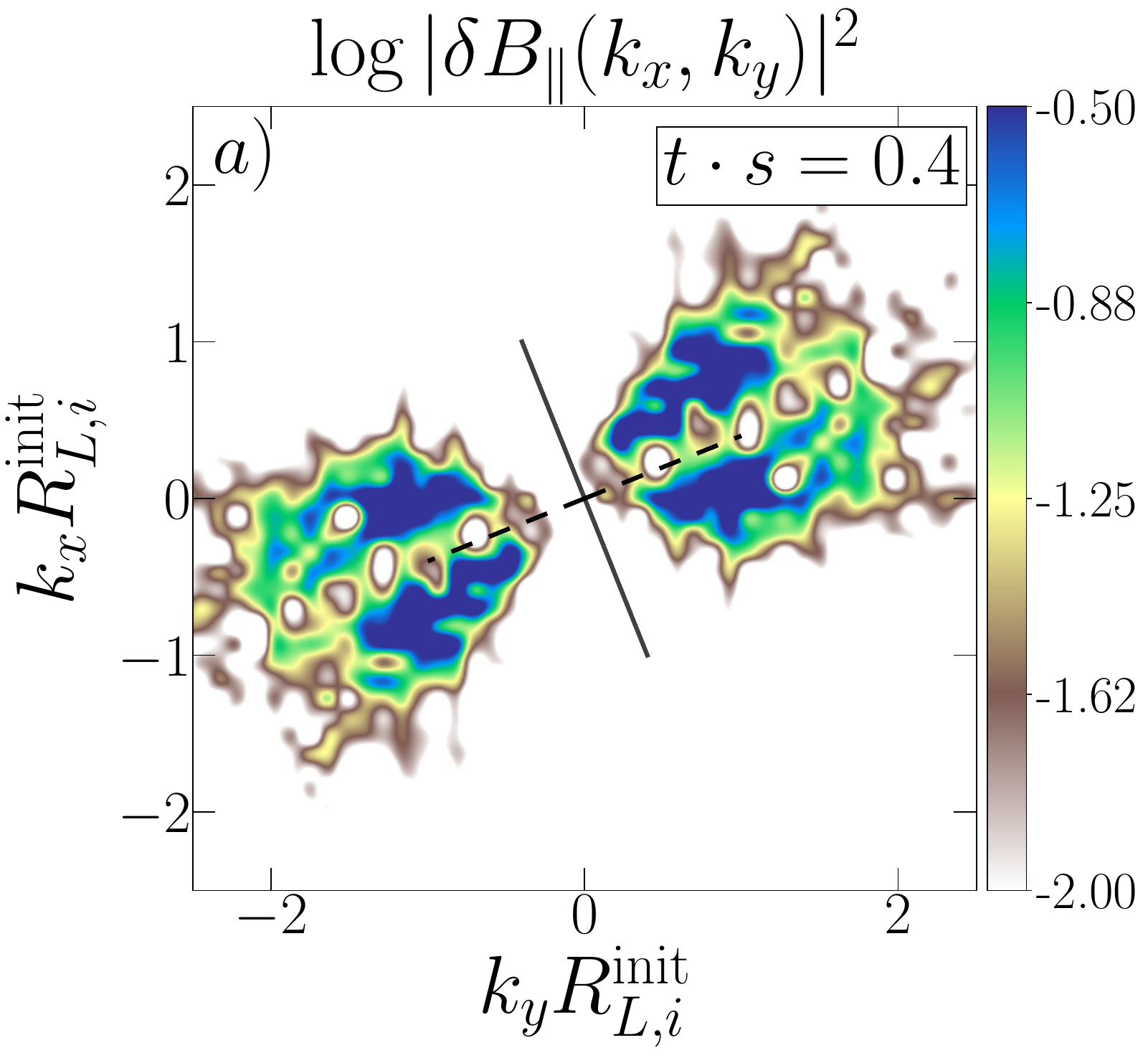}\\
         \includegraphics[width=0.95\linewidth]{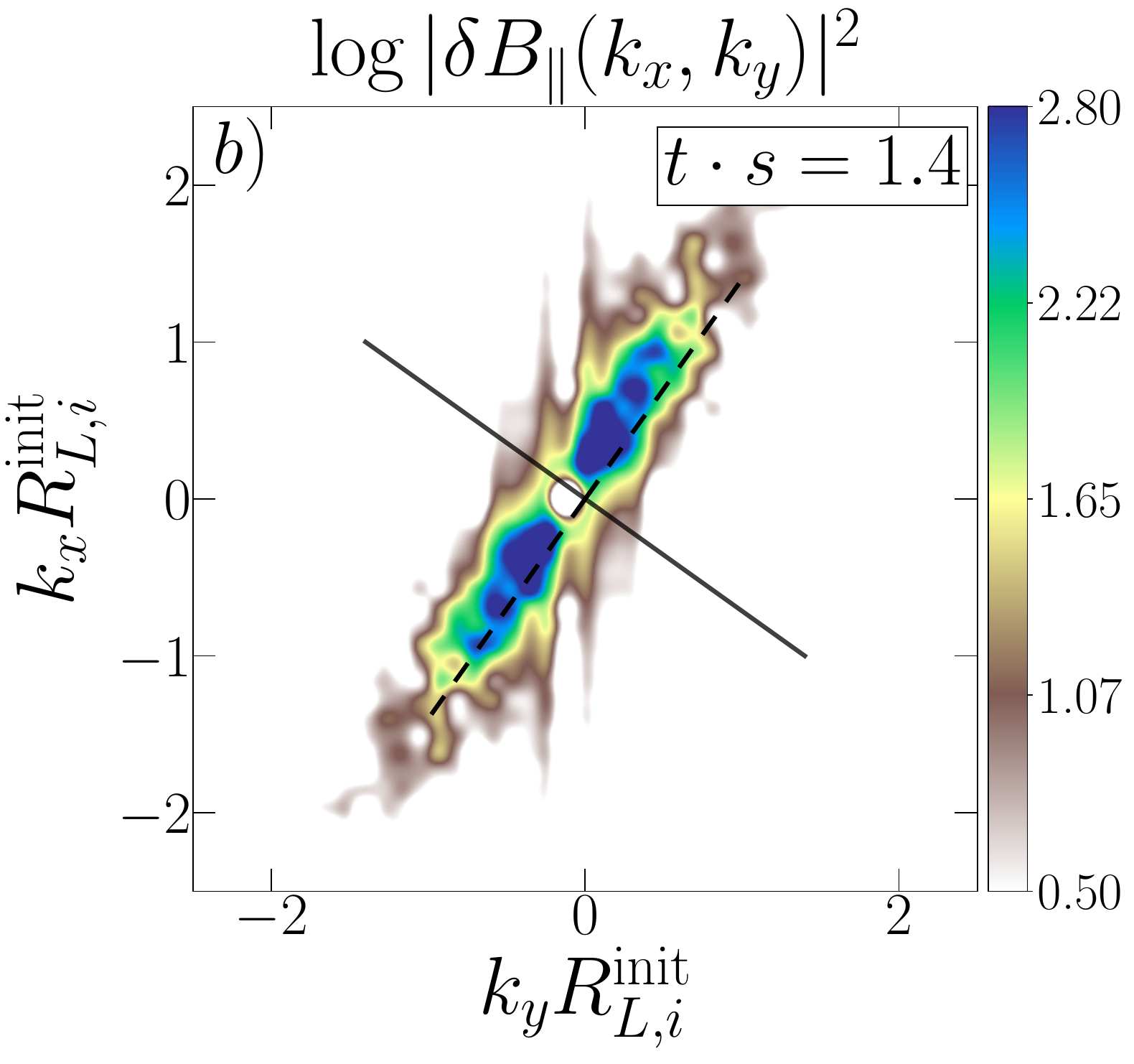}
    \end{tabular}
    \caption{Panel $a$: Power spectrum in space of $\delta B_{\parallel}(k_x,k_y)$ at $t\cdot s = 0.4$. The wavenumbers $k_x,k_y$ are normalized by the initial Larmor radius of the ions, $R_{L,i}^{\text{init}}$. The solid and dashed black lines represent the direction parallel and perpendicular to the main magnetic field at that time, respectively. Panel $b$: Power spectrum in space of $\delta B_{\parallel}(k_x,k_y)$ at $t\cdot s = 1.4$. Note that the scale of colorbars in panel $a$ and $b$ are different.}
    \label{fig:PowerSpectrum_dBpar_Space}
\end{figure}

Figure \ref{fig:MagneticFluctuations}$e$ shows a late, nonlinear stage of the mirror instability, at $t\cdot s = 1.4$. At this time, the regions of high magnetic field of mirror modes (e.g. red filamentary structures seen in fig. \ref{fig:MagneticFluctuations}$b$) have grown significantly and merged with neighboring structures to form wider and sharper regions of high local amplitudes ($\delta B_{\parallel}/B\sim 0.9$), whose sizes are comparable to regions of low magnetic field. At this stage, most of the power is concentrated in wavevectors $0.2 \lesssim kR_{L,i}^{\text{init}} \lesssim 1.1$, and angles $57^{\circ} \lesssim \theta_k \lesssim 85^{\circ}$ (see fig. \ref{fig:PowerSpectrum_dBpar_Space}$b$).

After reaching its overshoot, the ion anisotropy starts to decrease towards marginal stability. However, this decrease stops around $t\cdot s \approx 0.65$ at $\Delta P_i/P_{\parallel,i} \approx 0.18$, well above the approximate mirror threshold (dashed gray line, (\cite{Hasegawa1969,Hellinger2007})). The anisotropy then reaches a marginal stability level that is above the mirror threshold, similar to some previous works using both hybrid and fully kinetic simulations (\cite{SironiNarayan2015,Melville2016,Ley_2023}). 

In order to better characterize the evolution of $\Delta P_i$, we fit a relation $\Delta P_i = A_i\beta_{\parallel,i}^{\alpha_i}$ from $0.7 \leq t\cdot s \leq 2$ (In our simulations, the shear motion continuously amplifies $B$, therefore $\beta_{\parallel,i}$ also evolves.). As shown in fig. \ref{fig:IonLecAnisotropy}$a$, our best-fit parameters are $A_i = 0.544 \pm 0.003$ and $\alpha_i = -0.445 \pm 0.003$. The obtained exponent is consistent with marginal stability threshold given by the ion-cyclotron instability for lower $\beta_i$ (\cite{GaryLee1993}). Indeed, the threshold for the IC instability, $\Delta P_i = 0.53\beta_{\parallel,i}^{-0.4}$, is plotted as dotted-dashed orange line in fig. \ref{fig:IonLecAnisotropy}$a$ for $\gamma_{IC}/\omega_{c,i}=10^{-2}$ (\cite{GaryLee1993}), and we can clearly see the similarity with our best-fit threshold, even at this higher value of initial $\beta_{\parallel,i}^{\text{init}}$. This observation was also reported in \cite{SironiNarayan2015}, and we will see that, indeed, we do observe ion-cyclotron waves as part of the saturated phase of the mirror instability that starts at $t\cdot s = 0.6$. The presence of ion and electron cyclotron waves coexisting with mirror modes at late, nonlinear stages of the mirror instability has been reported in previous works (\cite{Riquelme2016,SironiNarayan2015,Ahmadi2018}). In \S\ref{sec:LionRoars}, we argue that a natural explanation of the source of these cyclotron waves is pressure anisotropy of ions trapped within nonlinear mirror modes. 

\subsection{First Whistler Burst -- $t\cdot s \approx 0.4$}
\label{sec:EarlyWhistlers}

\begin{figure}[hbtp]
    \centering
    \begin{tabular}{c}
          \includegraphics[width=0.95\linewidth]{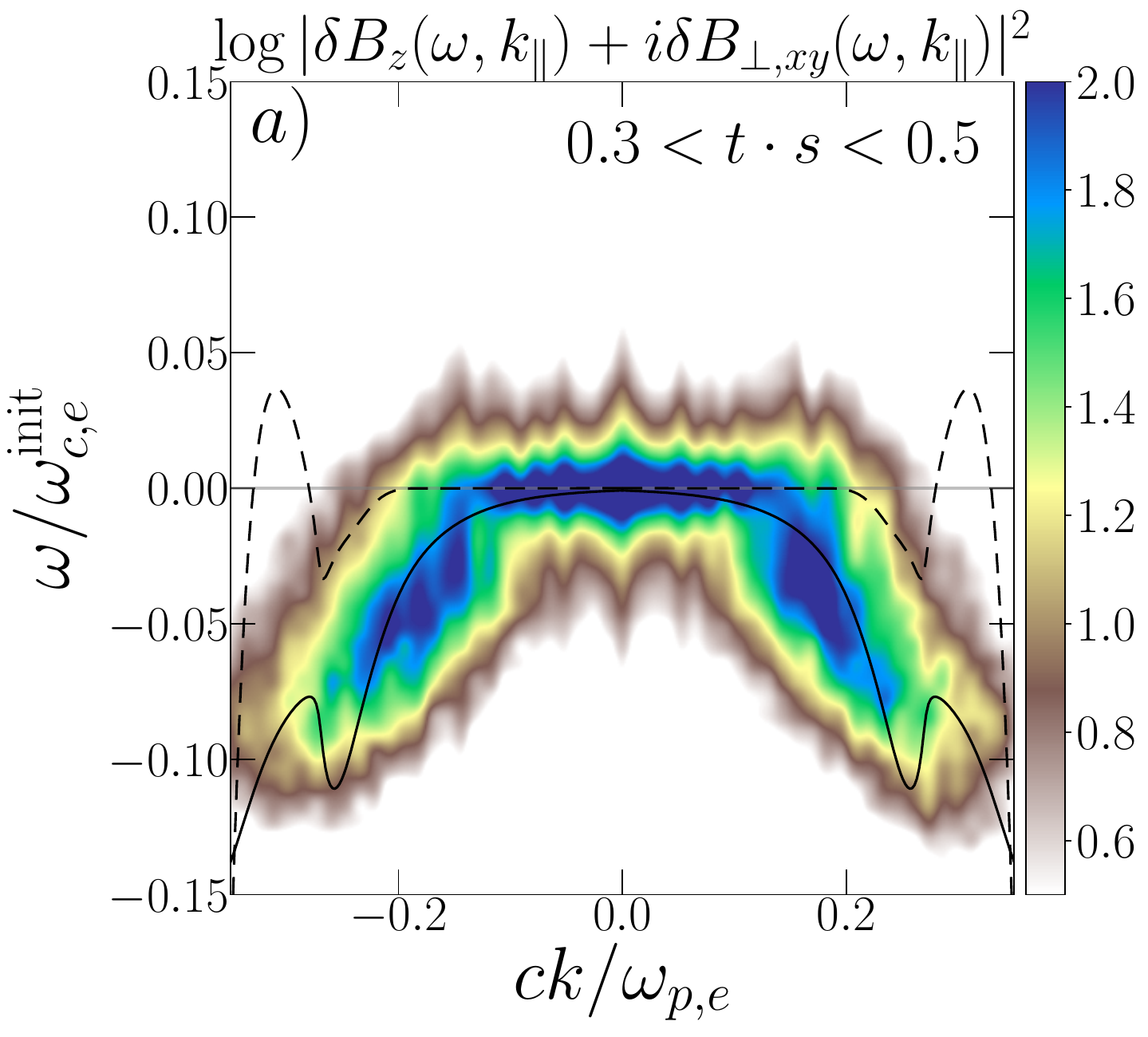} \\
          \includegraphics[width=1\linewidth]{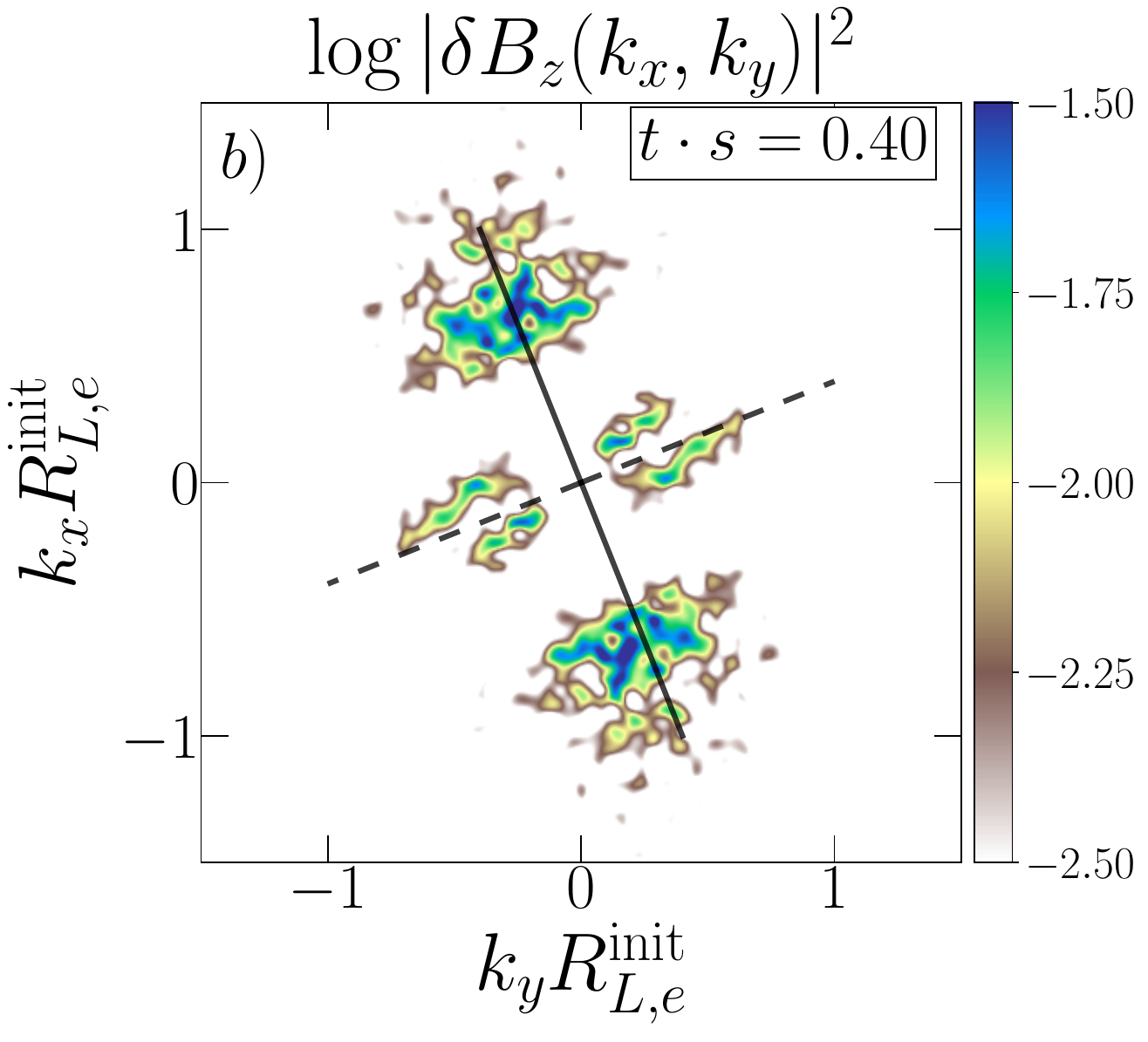}
    \end{tabular}
    \caption{Panel $a$: The power spectrum of $\delta B_z(\omega,k_{\parallel}) + i \delta B_{\parallel,xy}(\omega,k_{\parallel})$ in the entire simulation domain and between $0.3 < t\cdot s < 0.5$. The frequency is normalized by the initial electron cyclotron frequency $\omega_{c,e}$, and the wavevector is normalized by the plasma frequency $\omega_{p,e}$ over the speed of light $c$. The solid black line shows the linear dispersion relation $\omega_r(k)$ for the whistler instability according to our linear dispersion solver, whereas the dashed black line shows its growth rate $\gamma$. Panel $b$: The power spectrum in space of $\delta B_z(k_x,k_y)$ at $t\cdot s = 0.4$. The wavenumbers $k_x,k_y$ are normalized to the initial Larmor radius of the electrons, $R_{L,e}^{\text{init}}$. The solid and dashed black lines represent the direction parallel and perpendicular to the main magnetic field at that time.} 
    \label{fig:PowerSpectrum_whistler1}
\end{figure}

Figure \ref{fig:IonLecAnisotropy}$b$ shows the evolution of the electron pressure anisotropy $\Delta P_e \equiv P_{\perp,e}-P_{\parallel,e}$ for run b20m8w800. Initially, the electrons develop their own pressure anisotropy alongside ions and for the same reasons. The anisotropy follows double-adiabatic (CGL) scaling (dashed orange line) until $t\cdot s \approx 0.4$, when it has already reached a value significantly larger than the theoretical threshold for the growth of whistler modes, marked by grey-dashed lines (\cite{GaryWang1996}). Around this time, the whistler instability starts to grow, as seen by the time trace of $\delta B_z^2$ in fig. \ref{fig:MagneticFluctuations}$g$, which is a rough proxy for whistler waves, and also because there are no left-handed IC waves as shown in fig. \ref{fig:PowerSpectrum_whistler1}$a$. At $t\cdot s \approx 0.45$ the whistler modes saturate and enter a regime of quasi-steady amplitude, which lasts until $t\cdot s \approx 0.53$. During this $t\cdot s \approx 0.4-0.53$ period, $\Delta P_e$ is rapidly drawn down by frequent scattering, reaching a more slowly decreasing regime between $t\cdot s \approx 0.53$ and $0.6$. The draw down of electron anisotropy happens at a time when the ion anisotropy is still growing. This lasts until mirror modes are sufficiently high amplitude to start trapping the electrons ($t\cdot s = 0.6$).


The presence of whistler modes at $t\cdot s = 0.4$ can be seen mainly in the perpendicular components of $\delta \textbf{B}$, namely, $\delta B_{\perp,xy}$ and $\delta B_z$, figures \ref{fig:MagneticFluctuations}$a$ and \ref{fig:MagneticFluctuations}$c$, respectively. They propagate quasi-parallel to the main magnetic field $\textbf{B}$ in a fairly homogeneous way inside the simulation domain. This quasi-parallel propagation can also be seen in fig. \ref{fig:PowerSpectrum_whistler1}$b$, where we show the power spectrum in space of $\delta B_z(k_x,k_y)$ at $t\cdot s=0.4$ for run b20m8w800, and the solid and dashed black lines indicate the directions parallel and perpendicular to the main magnetic field $\langle \textbf{B} \rangle$ at $t\cdot s=0.4$. The power of $\delta B_z(k_x,k_y)$ is concentrated at parallel propagation and wavevectors $0.6 < kR_{L,e}^{\text{init}} < 1$. 

We show the whistler wave frequencies in the power spectrum of $\delta B_z(\omega,k_{\parallel}) + i\delta B_{\perp,xy}(\omega,k_{\parallel})$ in the interval $0.3< t\cdot s < 0.5$ in fig. \ref{fig:PowerSpectrum_whistler1}$a$. We can see that the power is localized in the region $\omega < 0$, i.e. right-handed circularly polarized waves, consistent with the whistler polarization, and within frequencies $0.02 < \omega/\omega_{c,e} < 0.05$. As mentioned above, no IC activity is present during this time period. 

We also calculated the theoretical dispersion relation of the anisotropic whistler instability using a linear dispersion solver assuming an initial bi-maxwellian distribution of electrons (\cite{Tran_2023}), using the initial parameters and values of $T_{\perp,e}, T_{\parallel,e}$ directly from the simulations. The dispersion relation $\omega(k)$ is shown as a solid black line in fig. \ref{fig:PowerSpectrum_whistler1}$a$, whereas the instability growth rate is shown in dashed black lines. We can see that the power in right-hand circularly polarized waves is consistent with the whistler dispersion relation.

This way, the early evolution of the electrons is determined by an early burst of whistler modes associated to the initial electron pressure anisotropy growth. We will see that, once electrons start to become trapped in between mirror modes at $t\cdot s \approx 0.6$, another burst of whistler activity happens, this time associated with the trapping process within mirror modes during their secular and saturated phase.

\subsection{Whistler and Ion-cyclotron Excitations -- $t\cdot s \approx 0.6$}
\label{sec:LionRoars}

At the end of its secular growth, when mirror modes have reached sufficiently high-amplitudes, we simultaneously observe right-hand and left-hand circularly polarized wave activity, which we identify as whistler and ion-cyclotron waves, respectively. We will see below (\S\ref{sec:LionRoars}) that these whistler and ion-cyclotron waves propagate mainly in regions of locally low magnetic field (magnetic troughs). The source of this wave activity is identified to be the pressure anisotropic population of ions and electrons mainly due to trapped particles inside the magnetic troughs. The whistlers and ion cyclotron waves then pitch-angle scatter both the trapped and untrapped particles, contributing to regulation of the global anisotropy.

\begin{figure}[hbtp]
    \centering
    \begin{tabular}{c}
         \includegraphics[width=0.9\linewidth]{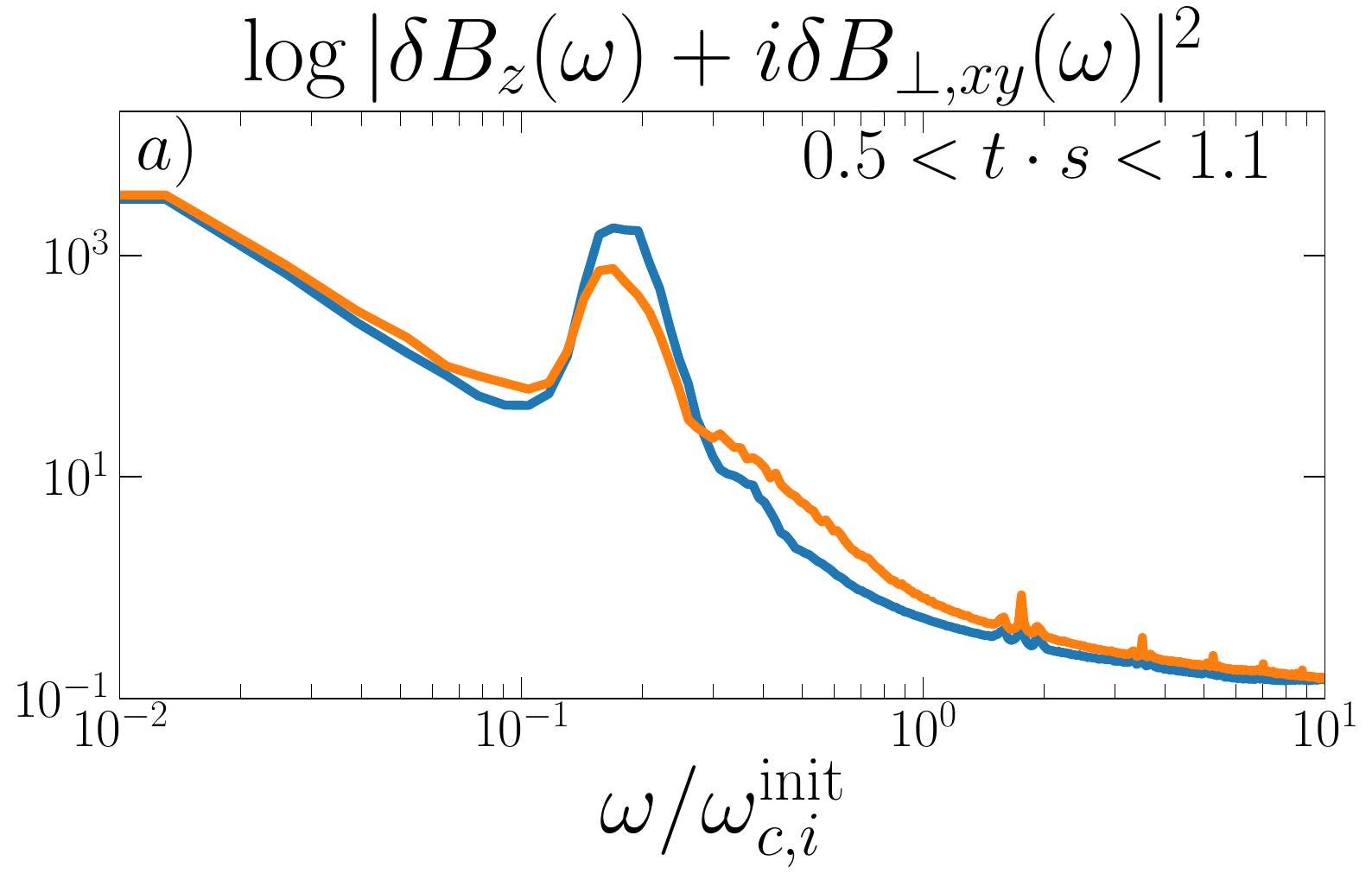} \\
          \includegraphics[width=\linewidth]{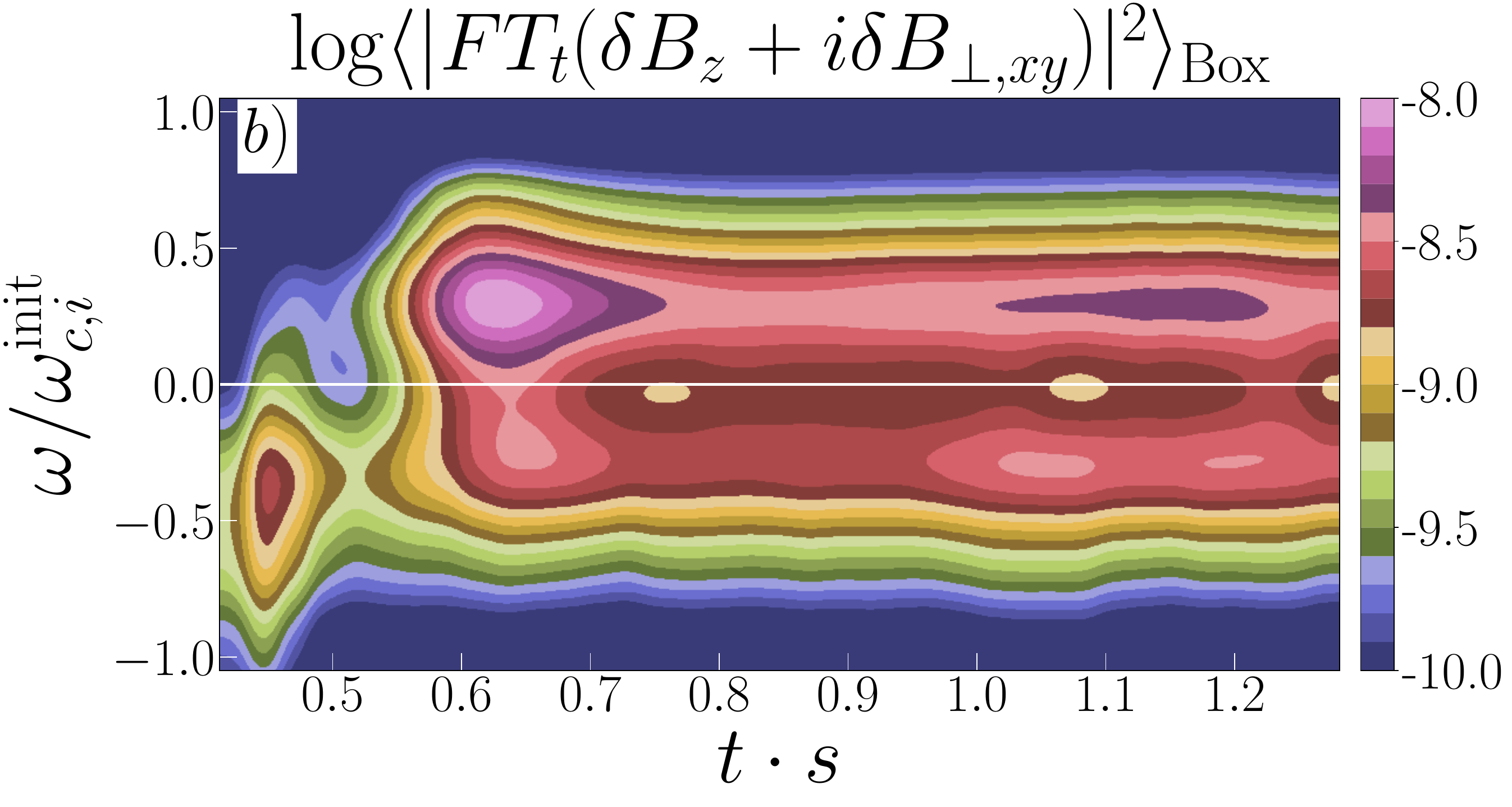} \\
          \includegraphics[width=0.9\linewidth]{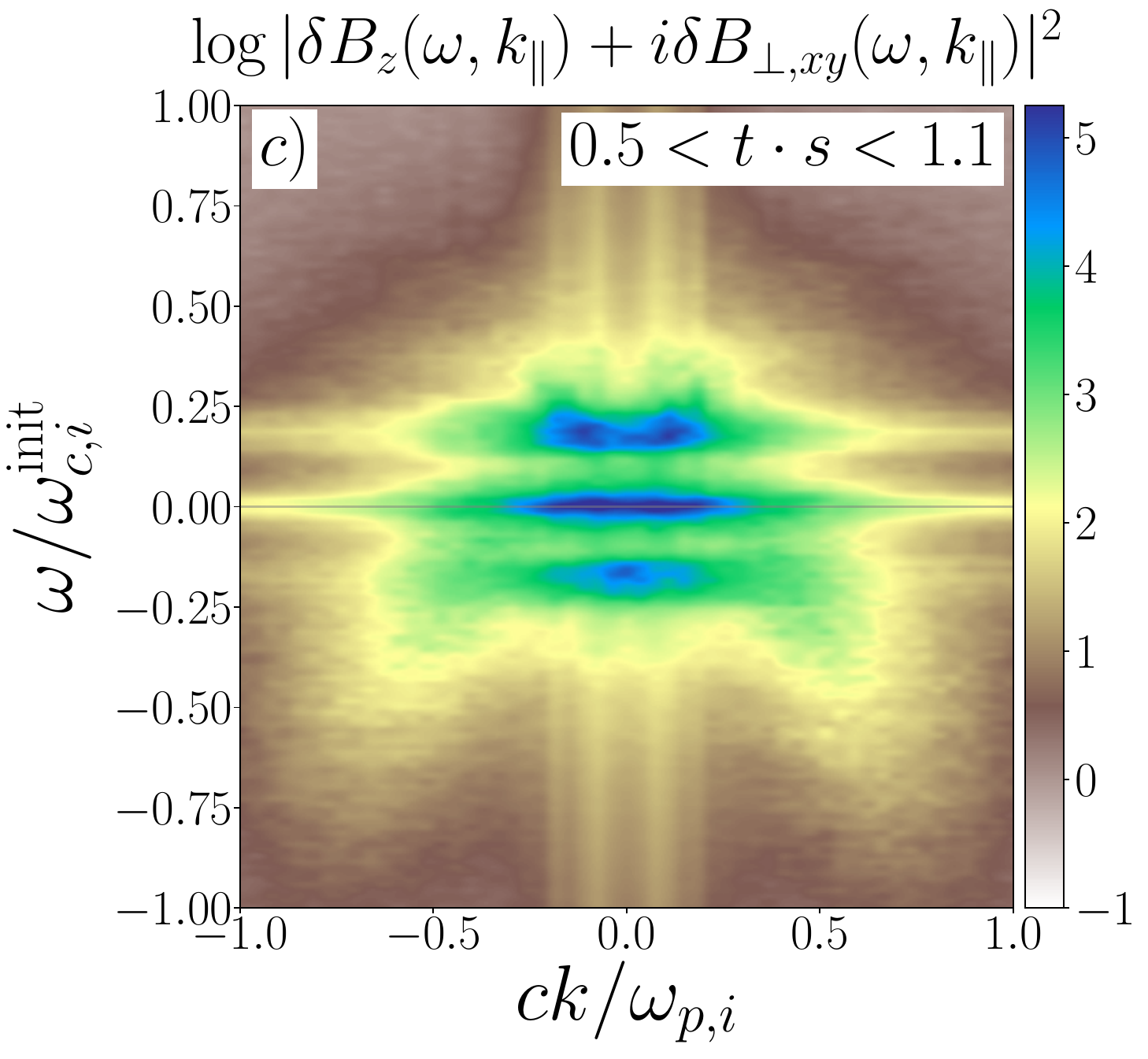}
    \end{tabular}
    \caption{Panel a: The power spectrum of $\delta B_z(\omega) + i\delta B_{\perp,xy}(\omega)$ as a function of frequency. The frequencies are normalized by the initial ion-cyclotron frequency. The power spectrum of left-handed circularly polarized waves ($\omega > 0$) is shown as a solid blue line, whereas the power spectrum corresponding to right-handed circularly polarized waves ($\omega < 0$) is shown as an orange line folded into positive frequencies. Panel b: Spectrogram of $\delta B_z(\omega) + i\delta B_{\perp,xy}(\omega)$ in frequency and time, at $0.4 < t\cdot s < 1.3$. The frequency is normalized by the initial ion-cyclotron frequency. Positive and negatives frequencies corresponds to left-hand and right-hand circularly polarized waves, respectively. Panel $c$: The power spectrum of $\delta B_z(\omega,k_{\parallel}) + i\delta B_{\perp}(\omega,k_{\parallel})$ at $0.5 < t\cdot s < 1.1$. Frequencies are normalized by the initial ion gyrofequency, and wavenumbers are normalized by the initial ion skin depth. Here also, positive and negative frequencies show left-hand and right-hand polarized waves, respectively.}
    \label{fig:PowerSpectrumOmega}
\end{figure}

Figure  \ref{fig:PowerSpectrumOmega} shows different spectral properties of the late burst of waves excited from $t\cdot s \approx 0.6$ onwards. Figure \ref{fig:PowerSpectrumOmega}$a$ shows the power spectrum in time of $\delta B_z(\omega) + i\delta B_{\perp,xy}(\omega)$ between $0.5 < t\cdot s < 1.1$, so we can see both left-hand (solid blue line) and right-hand (solid orange line) circular polarizations. The power spectrum peaks at low-frequencies, consistent with the nature of the dominant mirror modes (mainly appearing in $\delta B_{\perp, xy}$). Additionally, we can clearly see a secondary peak at around $\omega \sim 0.2\omega_{c,i}$, with a spread that goes from $\omega \sim 0.1\omega_{c,i}$ to $\omega \sim 0.3\omega_{c,i}$, in both left and right hand circular polarizations. This constitutes the characteristic feature informing the late burst of wave activity. This peak resembles observations of whistler lion roars in the Earth's Magnetosheath (see e.g. figs. 1 and 2 of \cite{Giagkiozis2018}, fig. 3 of \cite{Zhang2021} for right-hand polarized waves.).

Figure \ref{fig:PowerSpectrumOmega}$b$ shows the spectrogram of $\delta B_z(\omega) + i\delta B_{\perp,xy}(\omega)$ in frequency and time, ranging $0.4 < t\cdot s < 1.3$, with positive frequencies representing left-hand circularly polarized waves, and negative frequencies denoting right-hand circularly polarized waves. Here we can also see the early burst of whistler waves starting at $t\cdot s \approx 0.4$ and peaking at $t\cdot s \approx 0.45$ (see section \S\ref{sec:EarlyWhistlers}), followed by the burst of both left-hand and right-hand circularly polarized waves at $t\cdot s \approx 0.53$ and peaking at $t\cdot s \approx 0.65$. This coincides with the rise in amplitude of $\delta B_z^2$ and $\delta B_{\perp,xy}$ (see fig. \ref{fig:MagneticFluctuations})$g$, and the waves are continuously maintained throughout the simulation at around the same frequencies. 

Finally, figure \ref{fig:PowerSpectrumOmega}$c$ shows the power spectrum of $\delta B_z(\omega, k_{\parallel}) + i\delta B_{\perp,xy}(\omega, k_{\parallel})$ in time and space, at $0.5 < t \cdot s < 1.1$. Frequencies and wavenumbers are normalized by $\omega_{c,i}$ and $\omega_{p,i}/c$, respectively. Here we can also see the power at low frequencies consistent with the dominance of mirror modes appearing in $\delta B_{\perp, xy}$. The burst of left and right hand circularly polarized waves can be seen concentrated around frequencies $\omega \approx 0.2 \omega_{c,i}$ and $\omega \approx -0.15 \omega_{c,i}$, respectively. Their range in wavenumbers is $0.2 \lesssim ck_{\parallel}/\omega_{p,i} \lesssim 0.5$. Overall, the power spectra of both left and right hand polarized waves are very similar to those of ion-cyclotron and electron cyclotron whistlers, and we will identify these waves as such from now on. In the next section, we will confirm that the population of particles that excites these waves have anisotropic distributions that are IC and whistler unstable.

The morphology of IC and whistler waves can also be seen in figures \ref{fig:MagneticFluctuations}$d$ and \ref{fig:MagneticFluctuations}$f$. The short wavelength, wavepacket-like structures are identified with whistler modes, which propagate mainly through regions of low magnetic field strength of mirror modes, as we can see from $\delta B_{\perp,xy}$ ( blue shaded regions in fig. \ref{fig:MagneticFluctuations}$d$). The IC modes, on the other hand, are identified as the longer wavelength, extended modes that can be seen in $\delta B_z$. The IC modes seem to propagate through the entire simulation box, given their ion-scale wavelength, whereas whistler modes clearly propagate within mirrors' magnetic troughs. This also resembles magnetosheath's observations of whistler waves within magnetic troughs (e.g. \cite{Kitamura2020}).

\begin{figure}[t]
    \centering
    \includegraphics[width=\linewidth]{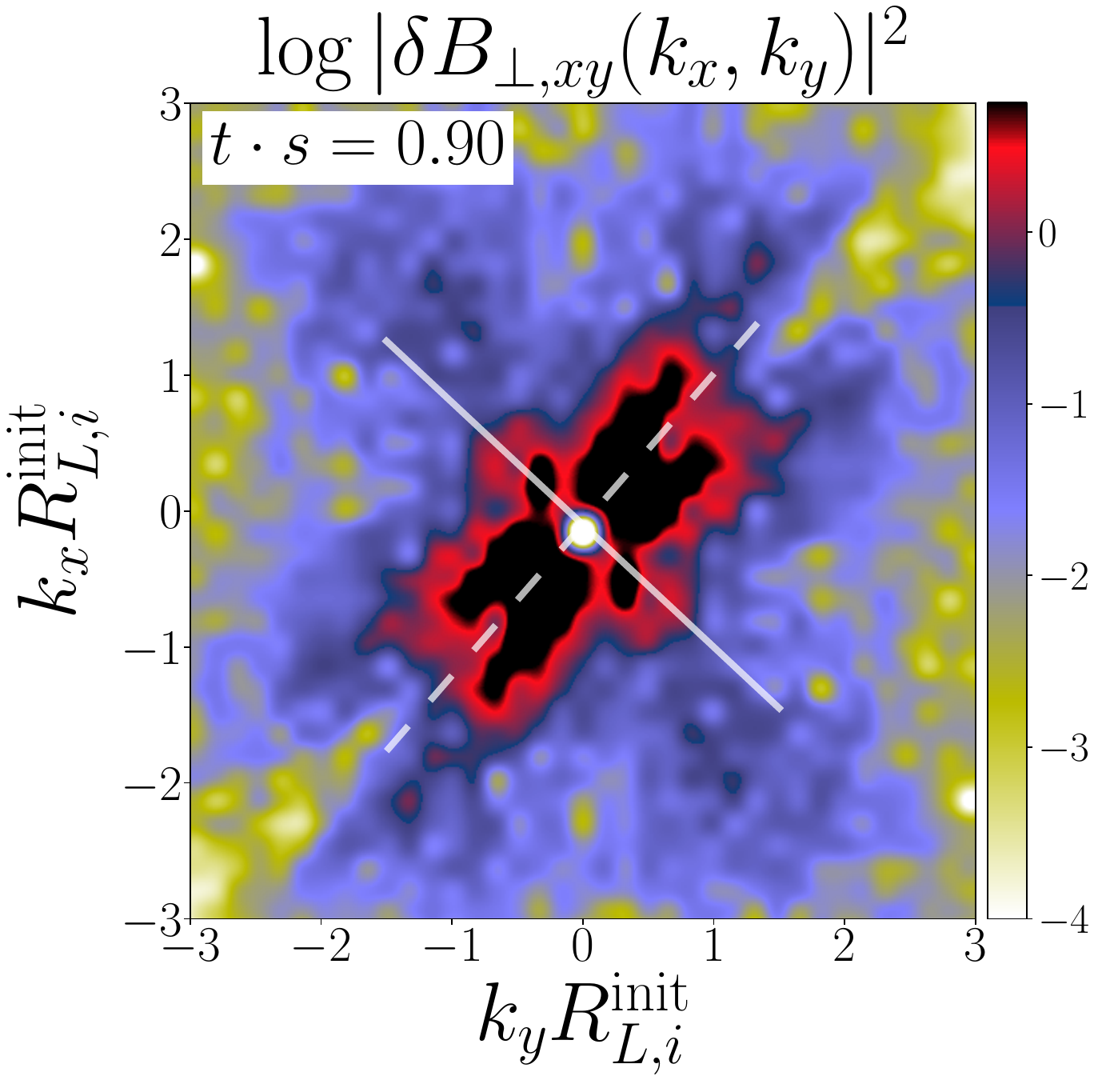}
    \caption{The power spectrum in space of $\delta B_{\perp,xy}(k_x,k_y)$ at $t\cdot s = 0.9$. The wavenumbers $k_x,k_y$ are normalized by the initial ion Larmor radius $R_{L,i}^{\text{init}}$. The solid and dashed white lines represent, respectively, the direction parallel and perpendicular to the main magnetic field at that time.}
    \label{fig:FT_dBperp_k}
\end{figure}

The peak frequencies observed in figure \ref{fig:PowerSpectrumOmega} for both ion-cyclotron and whistler waves can be understood in terms of their dispersion relations. At high-$\beta$ and $k R_{L,e} \sim 1$, and for quasi-parallel propagation, the dispersion relation for whistler waves can be written as (\cite{Stix1992,Drake_2021})

\begin{align}
    \omega_W = \omega_{c,e}k_W^2d_e^2 = \omega_{c,i}k_W^2d_i^2,
\end{align}

where $d_e=c/\omega_{p,e}$ and $d_i=c/\omega_{p,i}$ are the electron and ion skin depths, respectively. Knowing that $d_i^2 = R_{L,i}^2/\beta_i$, we can also write

\begin{align}
    \omega_W = \omega_{c,i}k_W^2R_{L,i}^2/\beta_i.
\end{align}

Similarly, at high-$\beta$ and $kR_{L,i} \sim 1$, and for quasi-parallel propagation, the ion-cyclotron wave dispersion relation is approximately (\cite{Stix1992})

\begin{align}
    \omega_{\text{IC}} = \omega_{c,i}k_{\text{IC}}d_i,
\end{align}

and we can also write

\begin{align}
\omega_{\text{IC}} = \omega_{c,i}k_{\text{IC}}R_{L,i}/\sqrt{\beta_i}.    
\end{align}

We can estimate $k_W$, $k_{\text{IC}}$ by looking at the power spectrum of any of the perpendicular components of the magnetic field fluctuations. Figure \ref{fig:FT_dBperp_k} shows the power spectrum of $\delta B_{\perp,xy}(k_x,k_y)$ at $t\cdot s = 0.9$, where the solid and dashed white lines denote the direction parallel and perpendicular to the mean magnetic field $\textbf{B}$ at that time, respectively. Apart from the power in the perpendicular direction corresponding to the mirror modes, in the power parallel to $\textbf{B}$ (i.e. along the solid black line in fig. \ref{fig:FT_dBperp_k}) we can distinguish large wavenumbers centered at $(k_yR_{L,i}^{\text{init}},k_xR_{L,i}^{\text{init}}) \approx (0.75,-1.5)$ (and also at $(-1.5,0.75)$), corresponding to whistlers, and also smaller wavenumbers centered at $(k_xR_{L,i}^{\text{init}},k_yR_{L,i}^{\text{init}}) \approx (0.5,0.7)$, corresponding to ion-cyclotron waves.

The large wavenumber extent in $k_x,k_y$ observed in fig. \ref{fig:FT_dBperp_k} gives us an approximate range of wavenumbers $1.5 \lesssim k_W R_{L,i}^{\text{init}} \lesssim 3.2$ for whistlers, implying frequencies $0.1 \lesssim \omega_W/\omega_{c,i}^{\text{init}}\lesssim 0.5$ (as $\beta_{i}^{\text{init}}=20$), consistent with the frequencies observed in the negative half of fig. \ref{fig:PowerSpectrumOmega}$c$, corresponding to right-hand polarized waves. Similarly, the small wavenumber extent in $k_x,k_y$ gives us a range of wavenumbers $0.4 \lesssim k_W R_{L,i}^{\text{init}} \lesssim 1.1$, implying frequencies $0.1 \lesssim \omega_{IC}/\omega_{c,i}^{\text{init}} \lesssim 0.25$, also consistent with the frequencies in the positive half of fig. \ref{fig:PowerSpectrumOmega}$c$, corresponding to left-hand polarized waves.

\subsection{2D Particle Distributions}

The specific time at which ion and electron cyclotron wave activity saturates, which coincides with the end of mirror instability's secular growth ($t\cdot s \approx 0.6$), and the propagation of whistler waves within regions of low-magnetic field strength, give a hint towards uncovering the mechanism by which the whistler and IC waves are excited. 

\begin{figure*}[t]
    \centering
    \includegraphics[width=\linewidth]{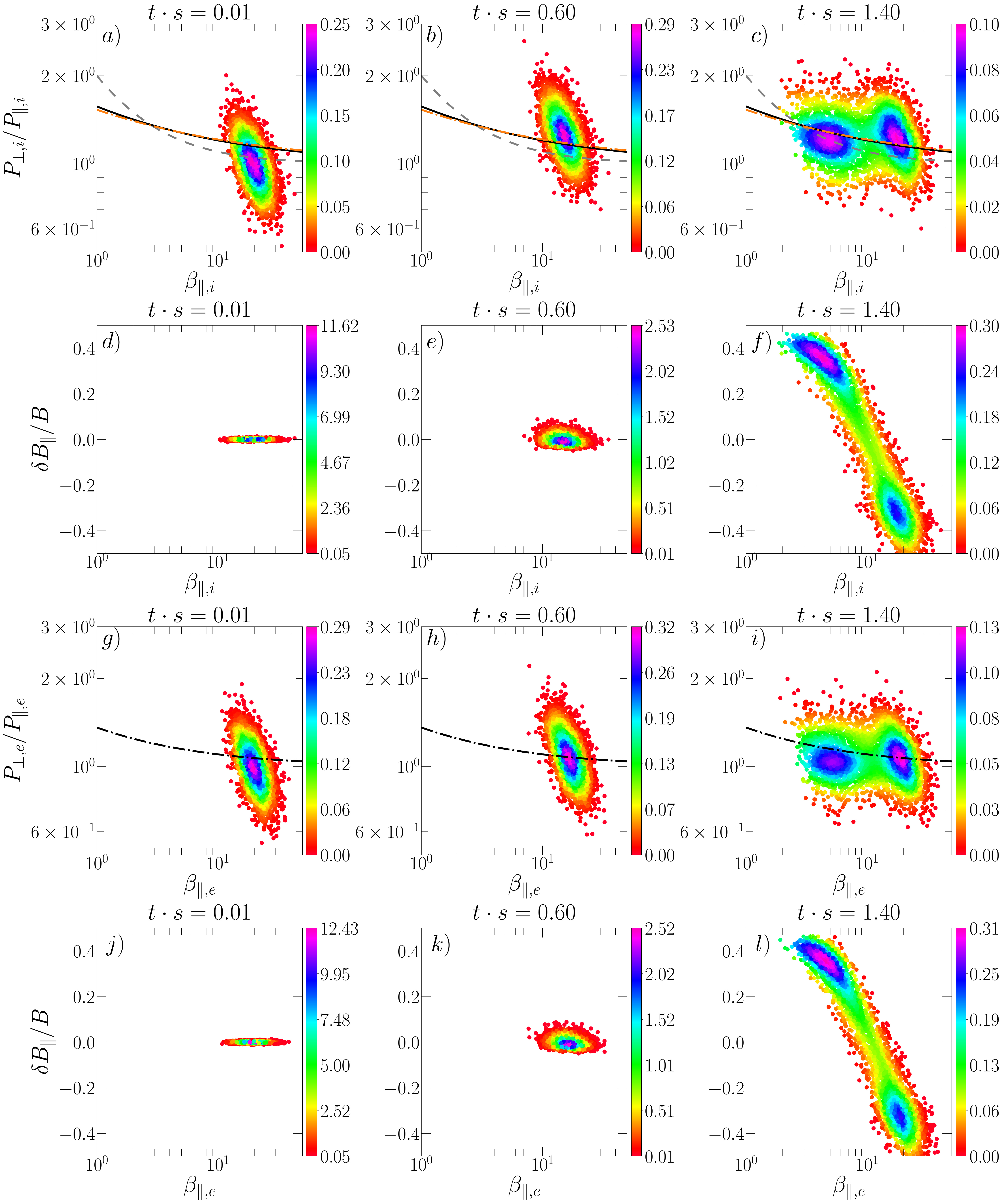}
    \caption{Top row: The distribution of ion $P_{\perp,i}/P_{\parallel,i}$ versus $\beta_{\parallel,i}$ in the simulation domain at different times: $t\cdot s = 0.01$ (left column), $t\cdot s = 0.6$ (middle column), and $t\cdot s = 1.4$ (right column). The dashed gray line represents the approximate mirror instability threshold $1/\beta_{\parallel,i}$ (\cite{Hasegawa1969}), the dotted-dashed orange line represents the IC instability threshold from \cite{GaryLee1993} for $\gamma_{IC}/\omega_{c,i}=10^{-2}$ ($\gamma_{IC}$ is the IC instability growth rate), and the solid black line represents our best-fit threshold from section \ref{sec:MirrorEvolution} (see fig. \ref{fig:IonLecAnisotropy}$a$). Second row: The distribution of $\delta B_{\parallel}/B$ versus ion $\beta_{\parallel,i}$ for the same three times as in the top row. Third row: The distribution of electron $P_{\perp,e}/P_{\parallel,e}$ versus $\beta_{\parallel,e}$ in the simulation domain at the same three times as in the top row. The dotted-dashed black line represents the whistler instability threshold from \cite{GaryWang1996}. Fourth row: The distribution of $\delta B_{\parallel}/B$ versus electron $\beta_{\parallel,e}$ for the same three times as in the top row. An animated version of this plot is available in the online version.}
    \label{fig:AnisotropyBeta}
\end{figure*}

As a first step, we explore the evolution of the pressure anisotropy of ions and electrons at the time at which the IC and whistler waves are excited. At this time, mirror modes have achieved high amplitudes, and created sharp regions of high and low magnetic field strength, making the plasma spatially inhomogeneous. This implies that, in general, the plasma $\beta$ of ions and electrons would not be the same at different locations in the simulation domain, making the anisotropy thresholds for the growth of the modes different in different regions. For this reason, a more appropriate method would be to measure the 2D distribution of pressure anisotropy, $\beta_{\parallel}$ and $\delta B_{\parallel}/B$ in the simulation domain. 

Figure \ref{fig:AnisotropyBeta} shows the distribution of ion and electron pressure anisotropy as a function of ion $\beta_{\parallel,i}$ (panels $a$, $b$, $c$) and electron $\beta_{\parallel,e}$ (panels $g$, $h$, $i$), respectively, and the distribution of $\delta B_{\parallel}/B$ versus $\beta_{\parallel,i}$ (panels $d$, $e$, $f$) and electron $\beta_{\parallel,e}$ (panels $j$, $k$, $l$), respectively. These distributions are shown at three different times: beginning of the simulation ($t\cdot s \approx 0$, left column); end of mirror's secular growth and beginning of ion and electron cyclotron wave activity ($t\cdot s = 0.6$, middle column), and a late stage well into the saturated regime of mirror instability ($t\cdot s = 1.4$, right column). In the top row of fig. \ref{fig:AnisotropyBeta} (i.e. panels $a$, $b$, and $c$), the dashed gray line corresponds to the approximate mirror instability threshold $1/\beta_{\parallel,i}$ (\cite{Hasegawa1969}), the dashed-dotted orange line corresponds to the theoretical IC threshold $0.53/\beta_{\parallel,i}^{0.4}$ from \cite{GaryLee1993} for $\gamma_{IC}/\omega_{c,i}=10^{-2}$, and the solid black line is the best-fit to the global ion anisotropy derived in section \ref{sec:MirrorEvolution} (see fig. \ref{fig:IonLecAnisotropy}$a$). In the third row of fig. \ref{fig:AnisotropyBeta} (panels $g$, $h$, $i$), the dotted-dashed black line shows the whistler instability threshold $0.36/\beta_{\parallel,e}^{0.55}$ from \cite{GaryWang1996}, for $\gamma_W/\omega_{c,e}=10^{-2}$.

Starting with the ions, we can see that, from a stable, isotropic distribution at the very beginning of the simulation (fig. \ref{fig:AnisotropyBeta}$a$), the ions become anisotropic enough to surpass both the mirror and the theoretical IC threshold from \cite{GaryLee1993}, as well as our best-fit instability threshold, as shown in fig. \ref{fig:AnisotropyBeta}$b$. At this point ($t\cdot s = 0.6$), we start to observe the excitation of ion-cyclotron waves that seem to interact with the ions and start driving them towards a marginally stable state. This can be seen in fig. \ref{fig:AnisotropyBeta}$c$, where the distribution becomes bimodal, with one population of ions under both the IC-threshold and our best-fit threshold (centered at $\beta_{\parallel,i}\sim 5$ and $P_{\perp,i}/P_{\parallel,i}\sim 1.2$), meaning that they are driven towards marginal stability with respect to the IC threshold. Interestingly, there exists another ion population that is still unstable (centered at $\beta_{\parallel,i}\sim 18$ and $P_{\perp,i}/P_{\parallel,i}\sim 1.4$), therefore IC waves could then continue being excited even at this late stages. This could explain the sustained amplitude observed in $\delta B_z^2$ and $\delta B_{\perp,xy}^2$ in figure \ref{fig:MagneticFluctuations}$g$. Therefore, we can see that the unstable population has a higher $\beta_{\parallel,i}$, and the marginally stable population moves to lower $\beta_{\parallel,i}$.

For a similar value of $P_{\parallel,i}$, the difference in the values of $\beta_{\parallel,i}$ between the unstable and marginally stable populations should imply a difference in the local magnetic field strength (recall $\beta_{\parallel,i}=8\pi P_{\parallel,i}/B^2$). This gives us a hint on the location of the unstable and marginally stable populations in the domain, as mirror modes generate distinct regions of low and high magnetic field strength. 

As we can see in figs. \ref{fig:AnisotropyBeta}$d$, \ref{fig:AnisotropyBeta}$e$, and \ref{fig:AnisotropyBeta}$f$, the ions also separate into two populations now in $\delta B_{\parallel}/B$. Starting from zero magnetic field fluctuations at the beginning ($t\cdot s \approx 0$, fig. \ref{fig:AnisotropyBeta}$d$), we see how $\delta B_{\parallel}/B$ starts to grow at $t\cdot s = 0.6$ (fig. \ref{fig:AnisotropyBeta}$e$), until we clearly see the bimodal distribution at $t\cdot s = 1.4$, separating the two ion populations: the high-$\beta_{\parallel,i}$ population located in regions of $\delta B_{\parallel}/B<0$ (i.e. low-$B$ strength), and the low-$\beta_{\parallel,i}$ population located in regions of $\delta B_{\parallel}/B > 0$ (i.e. high-$B$ strength). 

We can therefore conclude that, after mirror modes develop and the IC waves are excited ($t\cdot s \gtrsim 0.6$), the ions separate into two populations: one of low-$\beta_{\parallel,i}$, located mainly in high-$B$ strength regions, and marginally stable to IC waves, and the second population with high-$\beta_{\parallel,i}$, low-$B$ strength regions, and still unstable to IC waves. This suggests that the IC wave are excited by the unstable ion populations in regions of low magnetic field strength, and then interact with the ions in such a way that the ions move to regions of high-$B$ strength and low $\beta_{\parallel,i}$. In sections \ref{sec:LionRoarsTrappedPassing} and \ref{sec:LionRoarDistributionFunctions} we will see that the population of ions that contribute most to the anisotropy that destabilize the IC waves are the ones that become trapped within mirror troughs.

In the case of the electrons, we can see a similar evolution. From a stable, isotropic distribution at $t\cdot s \approx 0$ (fig. \ref{fig:AnisotropyBeta}$d$), we can see how part of it becomes now whistler unstable at $t\cdot s = 0.6$ (fig. \ref{fig:AnisotropyBeta}$e$), after which the excited whistler waves interact with the electrons driving again part of the distribution gradually towards marginal stability, also generating a bimodal distribution similar to that of the ions. At $t\cdot s = 1.4$ (fig. \ref{fig:AnisotropyBeta}$f$), we can see that the electron population with low $\beta_{\parallel,e}$ (centered at $\beta_{\parallel,e}\sim 5$ and $P_{\perp,e}/P_{\parallel,e} \sim 1$) is marginally stable with respect to the whistler threshold, whereas the electron population with high $\beta_{\parallel,e}$ (centered at $\beta_{\parallel,e}\sim 18$ and $P_{\perp,e}/P_{\parallel,e} \sim 1.2$) is still unstable with respect to the whistler threshold. This also implies that whistler waves can still be excited at late stages in the simulation.

Analogously, the electrons also separate into two populations with respect to $\delta B_{\parallel}/B$. Similarly to ions, we also see that the population with low-$\beta_{\parallel,e}$ is located in regions of $\delta B_{\parallel}/B < 0$ (low $B$ strength), whereas the high-$\beta_{\parallel,e}$ population is located in regions of $\delta B_{\parallel}/B > 0$ (high $B$ strength). In this sense, we also conclude that in the case of electrons, the unstable population is located mainly in regions of low-$B$ strength and high-$\beta_{\parallel,e}$, where whistler waves are being excited, and the marginally stable population is located mainly in regions of high-$B$ field and low-$\beta_{\parallel,e}$. This also suggests that whistler waves interact with electrons so they move to regions of high-$B$ strength. We will also see in sections \ref{sec:LionRoarsTrappedPassing} and \ref{sec:LionRoarDistributionFunctions} that the electrons that contributes the most to the pressure anisotropy that destabilizes whistler waves are the ones that become trapped within mirror modes.

\subsection{Physical Mechanism of Secondary IC/Whistler Excitation: Trapped and Passing Particles}
\label{sec:LionRoarsTrappedPassing}

In this section, we study the evolution of the ions and electrons that become trapped within mirror modes as part of the mirror instability's interaction with the particles. We characterize the pressure anisotropy and distribution functions of these populations at the moment of trapping, and provide evidence that they are able to destabilize parallel propagating modes that ultimately allow them to escape the mirrors and regulate the overall anisotropy.

\begin{figure}[hbtp]
    \centering
    \begin{tabular}{c}
        \includegraphics[width=\linewidth]{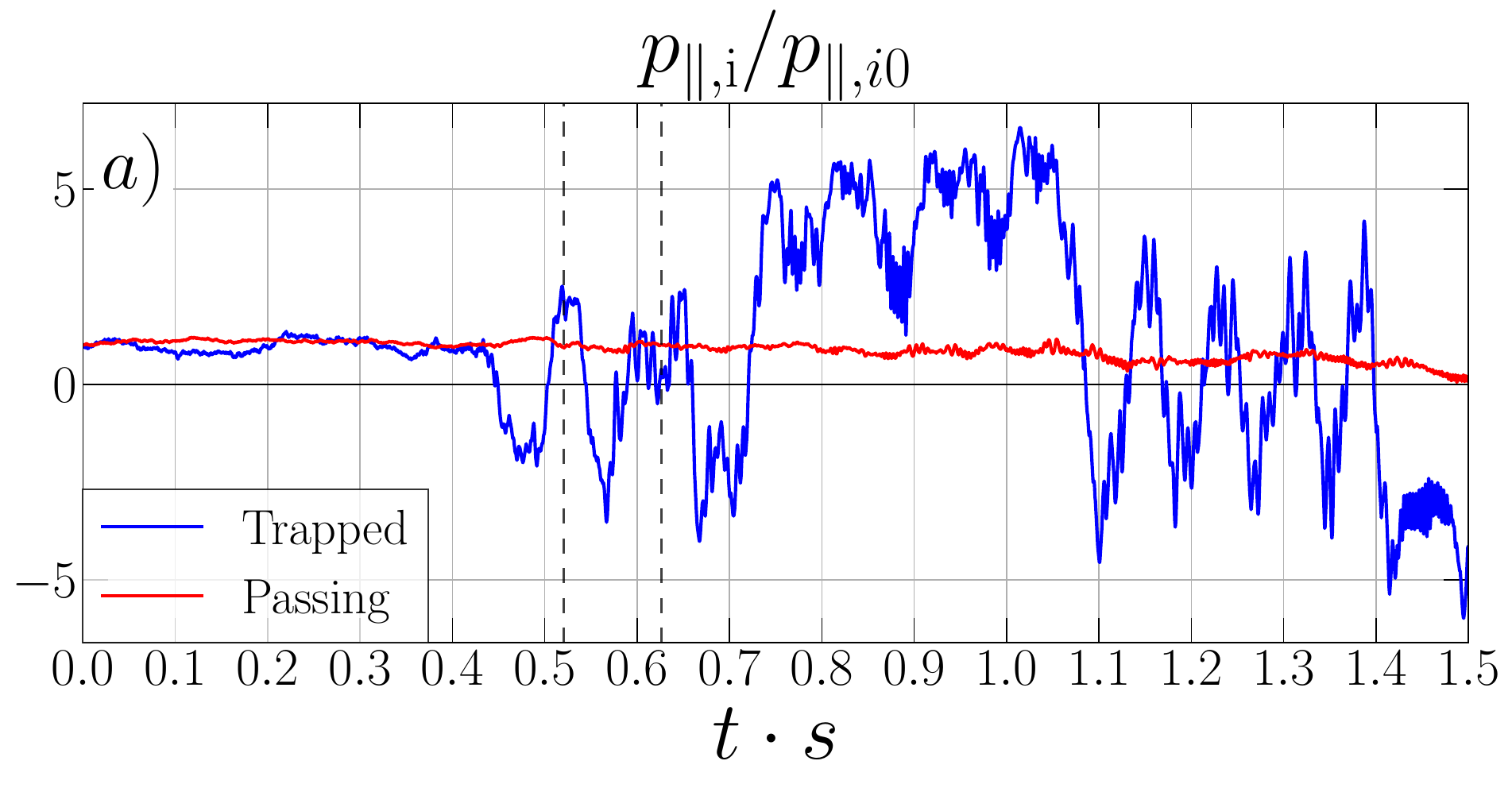}\\
        \includegraphics[width=\linewidth]{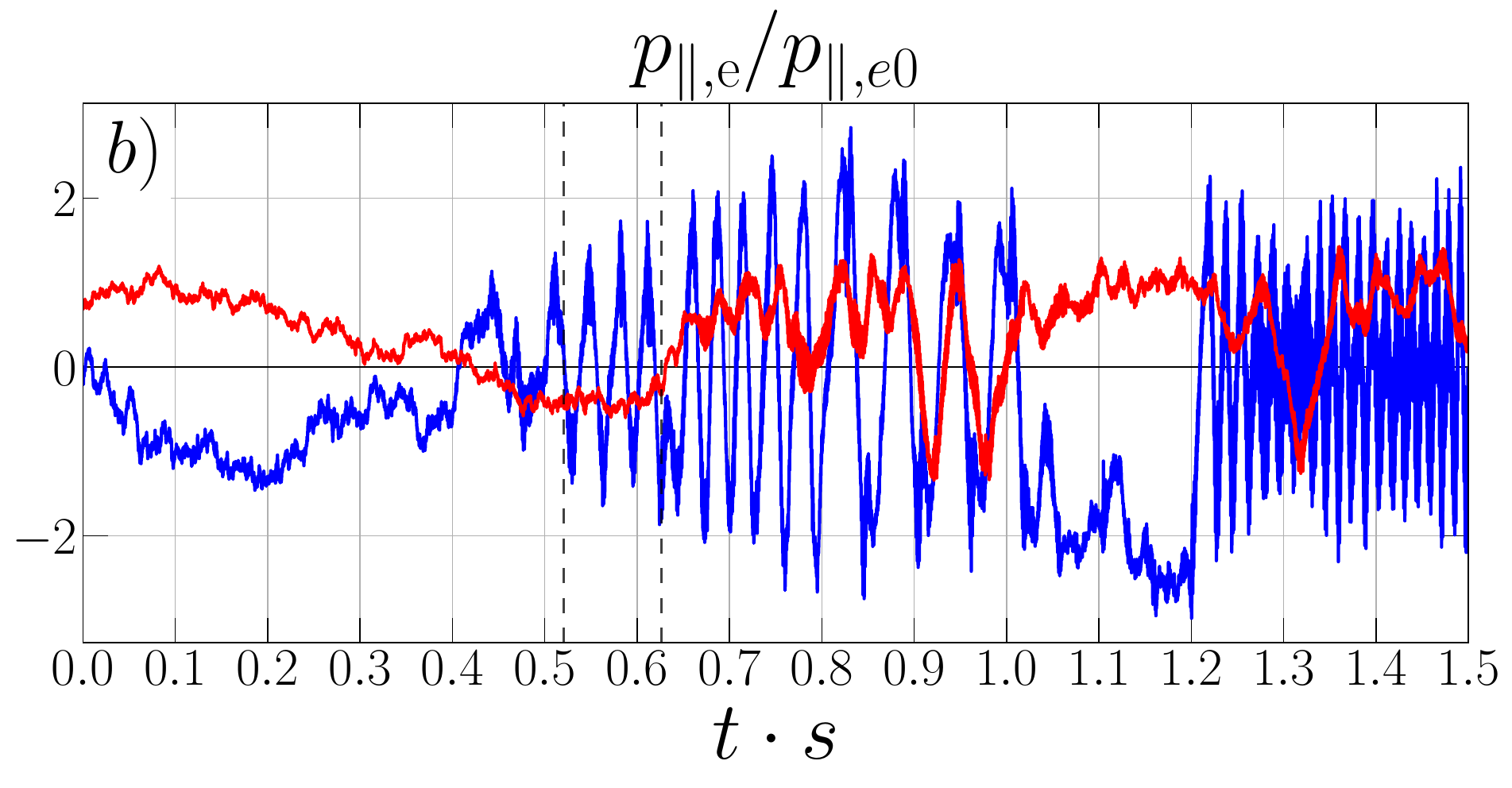}
    \end{tabular}    
    \caption{Panel a: Evolution of the parallel momentum of an individual trapped ion (blue line) and passing ion (red line) for our fiducial simulation b20m8w800. Panel b: Evolution of the parallel momentum of a trapped electron (blue line) and passing electron (red line) for run b20m8w800. The dashed vertical gray lines in each panel indicates the time interval $\Delta \tau_{LR}$.}
    \label{fig:TestParticles_magmom}
\end{figure}

As part of their evolution, and after reaching secular growth, mirror modes start to trap particles of low parallel momentum $p_{\parallel,j}$ $(j=i,e)$ in regions of low local magnetic field strength. The trapped particles bounce between these regions and conserve their magnetic moment in the process (\cite{Southwood&Kivelson1993,Kunz2014}). In order to investigate the relation between this trapping process and the excitation of the these late IC and whistler waves, we select and track a population of ions and electrons throughout the evolution of the simulation, and study the trapped and passing (i.e. untrapped) subpopulations separately. 

We select and track two populations of ions and two populations of electron having relatively small and large parallel momentum at a particular time in the simulation. This way, we make sure that we can capture particles that eventually become trapped and others that remained passing. In our fiducial simulation b20m8w800, the two populations of ions that we track have parallel momentum $-0.12 < p_{\parallel,i}/m_ic < 0.12$ and $0.3395 < p_{\parallel,i}/m_ic < 0.3405$ at $t\cdot s =0.4$. Similarly, the two populations of electrons have $-0.2 < p_{\parallel,e}/m_ec < 0.2$ and $0.4599 < p_{\parallel,i}/m_ic < 0.4601$ at $t\cdot s =0.4$. 


In order to study the behavior of the tracked particles when the IC and whistler activity starts, we ask how many particles become trapped and how many become passing during the interval of time at which this activity happens, which we denote by $\Delta \tau_{LR}$. To answer this, we look at fig. \ref{fig:MagneticFluctuations}$g$ and define $\Delta \tau_{LR}$ as the interval of time $0.52 < t\cdot s < 0.62$, which covers the exponential growth that $\delta B_z^2$ and $\delta B_{\perp,xy}^2$ undergo before saturating. This interval of time also covers the majority of the secular growth of mirror modes (see $\delta B_{\parallel}^2$).

Having this time interval well defined, we now must define the criterion by which we consider a particle to become trapped and passing during $\Delta \tau_{LR}$, and for this we look at the evolution of their parallel momentum. Similarly to \cite{Ley_2023}, we define a particle as trapped during $\Delta \tau_{LR}$ if the median of its parallel momentum over $\Delta \tau_{LR}$ is smaller than one standard deviation over $\Delta \tau_{LR}$. We then define a particle as passing if the median of its parallel momentum over $\Delta \tau_{LR}$ is greater than or equal than one standard deviation over $\Delta \tau_{LR}$. This is a statement of small variation of $p_{\parallel,j}$ over $\Delta \tau_{LR}$, which in turn is a proxy for an oscillatory behavior of $p_{\parallel,j}$, characteristic of a bouncing particle between mirror points. We confirm that this simple criterion gives excellent results separating trapped from passing particles.

Figure \ref{fig:TestParticles_magmom} shows the evolution of the parallel momentum of a trapped and a passing ion (panels $a$) and a trapped and a passing electron (panels $b$), where the dashed vertical gray lines indicate $\Delta \tau_{LR}$. We can see the the oscillation pattern in the evolution of the parallel momentum of the trapped ion during $\Delta \tau_{LR}$ and until $t\cdot s \approx 0.7$, when it escapes. The parallel momentum of the passing ion evolves without major changes as the ion streams through the simulation box. This behavior is consistent with previous works using hybrid and fully kinetic simulations \cite{Kunz2014,Riquelme2016}.  

In figure \ref{fig:TestParticles_magmom}$d$ we can also see the oscillating pattern of the parallel momentum of the trapped electron, indicating bouncing inside mirror modes, which ends at $t\cdot s \approx 1.1$, when it escapes. The parallel momentum of the passing electron does not vary significantly during $\Delta \tau_{LR}$, confirming that it was streaming along field lines at least at that interval.

It is worth noting, however, what happens after $\Delta \tau_{LR}$. Our criterion for identifying particles as trapped and passing was only within $\Delta \tau_{LR}$, and after that period of time particles can continue evolving into the saturated stage of mirror modes, where they can escape, be trapped again or continue streaming unperturbed. Indeed, by looking at its parallel momentum, we can see that after escaping and streaming for a while, the trapped ion shown in figure \ref{fig:TestParticles_magmom}$a$ gets trapped again at $t\cdot s \approx 1.1$, bounces inside a mirror mode and escapes again at $t\cdot s \approx 1.4$. Similarly, we can also see that the trapped electron shown in figure \ref{fig:TestParticles_magmom}$b$ gets trapped again at $t\cdot s \approx 1.2$ and seems to stay trapped until the end of the simulation. Interestingly, the passing electron also gets trapped at around $t\cdot s \approx 0.7$, by looking at its parallel momentum, and then escapes again at $t\cdot s \approx 1.2$. Therefore, in a statistical sense, we can consider the particles as trapped and passing only over the particular period of time $\Delta \tau_{LR}$ that we chose, after which they can continue evolving and turn into passing or trapped again, as long as the mirror saturation persists in the simulation.

\subsection{Physical Mechanism of Secondary IC/Whistler Excitation: Distribution Functions}
\label{sec:LionRoarDistributionFunctions}

\begin{figure}[hbtp]
    \centering
    \begin{tabular}{c}
         \includegraphics[width=0.9\linewidth]{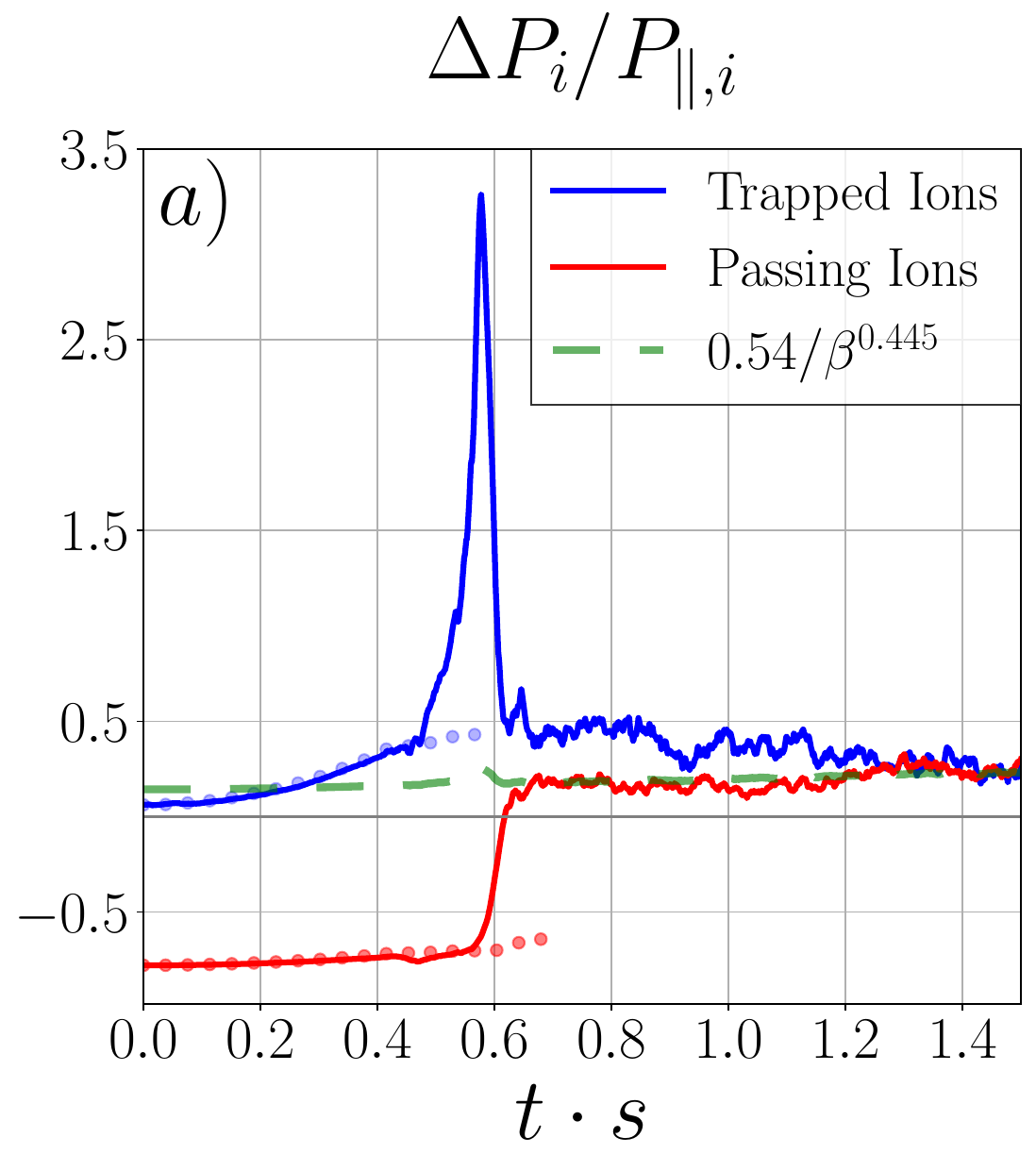}\\
          \includegraphics[width=0.9\linewidth]{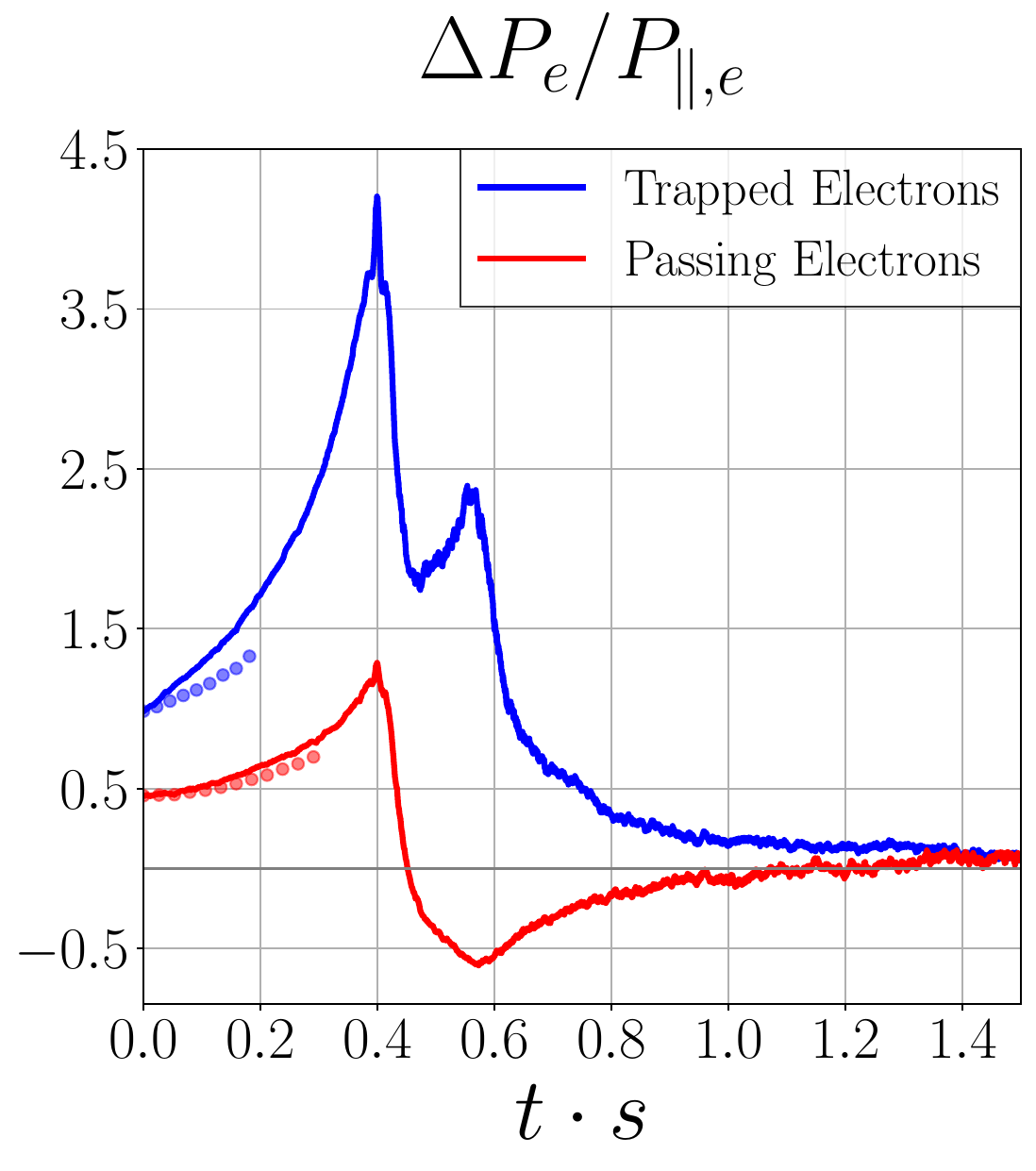}
    \end{tabular}
    \caption{Panel a: Evolution of the pressure anisotropy of ions identified as trapped (blue line) and passing (red line). The dashed green line indicates the best-fit threshold to $\Delta P_{\parallel,i}/P_{\parallel,i}$ shown in fig. \ref{fig:IonLecAnisotropy}$a$, and the dotted blue-gray and red lines show the corresponding double-adiabatic (CGL) evolution of trapped and passing ions, respectively. Panel b: Evolution of the pressure anisotropy of trapped (blue line) and passing (red line) electrons. The dotted blue and red lines show the corresponding CGL evolution of trapped and passing electrons, respectively.}
    \label{fig:AnisotropyTrapped}
\end{figure}

In this section, we look at the evolution of the pressure anisotropy and distribution functions of trapped and passing ions and electrons defined according to the criterion described in section \S\ref{sec:LionRoarsTrappedPassing}. We see that during $\Delta \tau_{LR}$, both trapped ions and trapped electrons contribute most of the pressure anisotropy necessary to destabilize IC and whistler modes. We show that these IC and whistler waves interact in a quasilinear fashion with ions and electrons, respectively, and quickly regulate their pressure anisotropy such that their distributions evolve to a more isotropic state.

Figure \ref{fig:AnisotropyTrapped}$a$ shows the evolution of the pressure anisotropy of trapped and passing ions. We can see that the anisotropy of trapped ions initially follows a double-adiabatic (CGL, dotted blue line) evolution until $t\cdot s \approx 0.5$ (i.e. just starting $\Delta \tau_{LR}$), when the mirror modes start to trap them. We can readily see that during $\Delta \tau_{LR}$, the trapped ions develop a significant anisotropy, peaking at around $t\cdot s \approx 0.55$. The anisotropy is quickly regulated and converges to the best-fit threshold that we derived in section \ref{sec:MirrorEvolution} and show in figure \ref{fig:IonLecAnisotropy}$a$. Similarly, the pressure anisotropy of passing ions evolves in a relatively unperturbed fashion following CGL evolution (dotted red line) through the majority of $\Delta \tau_{LR}$, until $t\cdot s \approx 0.6$, where it passes from negative values (consistent with passing ions having preferentially large parallel momentum) to a positive but, more isotropic state consistent with the best-fit threshold from fig. \ref{fig:IonLecAnisotropy}$a$. 

The behavior of the pressure anisotropy of trapped and passing particles can be understood as follows. Mirror modes interact resonantly with ions and electrons according to the resonance condition $\omega_M - k_{\parallel,M}v_{\parallel} = 0$, where $\omega_M$ and $k_{\parallel,M}$ are the frequency and parallel wavenumber of mirror modes, respectively, and $v_{\parallel}$ is the parallel velocity of the particle. The very low frequency of mirror modes, $\omega_M \sim 0$, implies that the resonant particles are the ones having very low $v_{\parallel}$ ($v_{\parallel}<\gamma_M/k_{\parallel,M}$, where $\gamma_M$ is the mirror growth rate, \cite{Southwood&Kivelson1993,Pokhotelov2002}). These are the particles that become trapped within mirror modes (\cite{KivelsonSouthwood1996}). Consequently, all trapped particles have very low parallel velocity and, as a whole, they should naturally have a pressure anisotropy $P_{\perp,j}>P_{\parallel,j}$ $(j=i,e)$. Similarly, all passing particles have large $v_{\parallel}$, and therefore they have a pressure anisotropy $P_{\parallel,j}>P_{\perp,j}$. In this sense, fig. \ref{fig:AnisotropyTrapped} is consistent with the trapping argument described in \cite{KivelsonSouthwood1996} (see their fig. 1).

The fact that both trapped and passing ions evolve into the average level of ion anisotropy shown in fig \ref{fig:IonLecAnisotropy}$a$ shows that their trapped or passing condition corresponds to a transient state, that passes after a time comparable to $\Delta \tau_{LR}$. Also, notice that the anisotropy of the two populations (and for the whole population for that matter) is significant enough to drive IC waves unstable (see section \ref{sec:LionRoars}), and therefore this can provide evidence for the source of the IC waves that we see. If this is the case, their interaction with ions is the source of the quick regulation of the anisotropy that we see in fig. \ref{fig:AnisotropyTrapped}$a$. Interestingly, under this scenario, the regulation of the pressure anisotropy of passing ions, which happens at the same time as that of the trapped ions, should also be due to the interaction with these IC waves, meaning that the IC waves interact with both populations of trapped and passing ions simultaneously, and therefore regulate the global ion anisotropy. We confirm that this is the case by looking at the evolution of the distribution functions of trapped and passing ions. 

In the case of electrons, we observe a similar evolution in figure \ref{fig:AnisotropyTrapped}$b$. Initially, both trapped and passing electrons detach from their respective CGL evolution (dotted blue and red lines, respectively), and develop a significant anisotropy $\Delta P_e >0$, that peaks at $t\cdot s \approx 0.4$. We also see that trapped electrons detach from their CGL evolution much earlier than passing electrons. This evolution then leads to the early burst of whistler waves, which also quickly regulates and drives anisotropies of both trapped and passing electrons towards a more isotropic state (see section \ref{sec:EarlyWhistlers}). As expected, the anisotropy of trapped electrons is higher than the one of the passing electrons. After this process, and during $\Delta \tau_{LR}$, the anisotropy of trapped electrons increases again, while that of passing electrons continues to decrease. This way, we see that trapped electrons build up a pressure anisotropy $\Delta P_e >0$ that is also quickly regulated after $\Delta \tau_{LR}$, converging to an anisotropy level similar to the one of the general electron populations. The anisotropy $\Delta P_e < 0$ of the passing electrons also gets regulated towards a similar anisotropy level during the same time. This evolution of trapped electrons also suggests that they become anisotropic enough to destabilize whistler waves, and therefore could be the source of the whistler activity observed at $t\cdot s > 0.6$. We provide evidence of this by showing the evolution of the distribution function of electrons.

\begin{figure*}[hbtp]
    \centering
    \begin{tabular}{c}
         \includegraphics[width=0.95\linewidth]{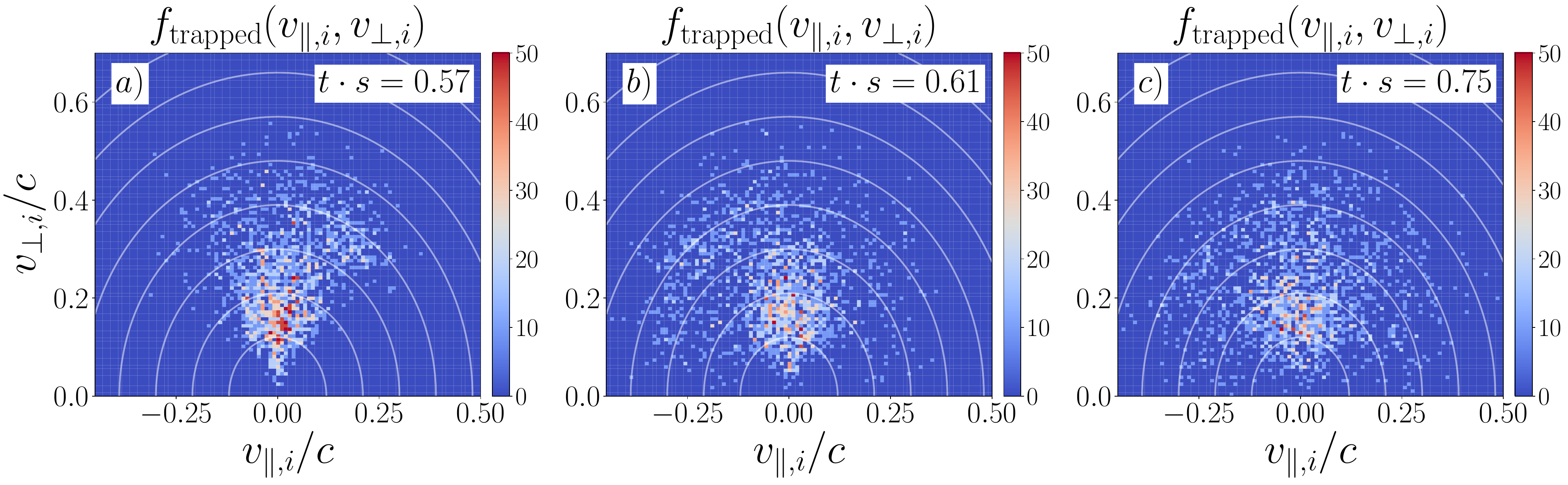}\\
         \includegraphics[width=0.95\linewidth]{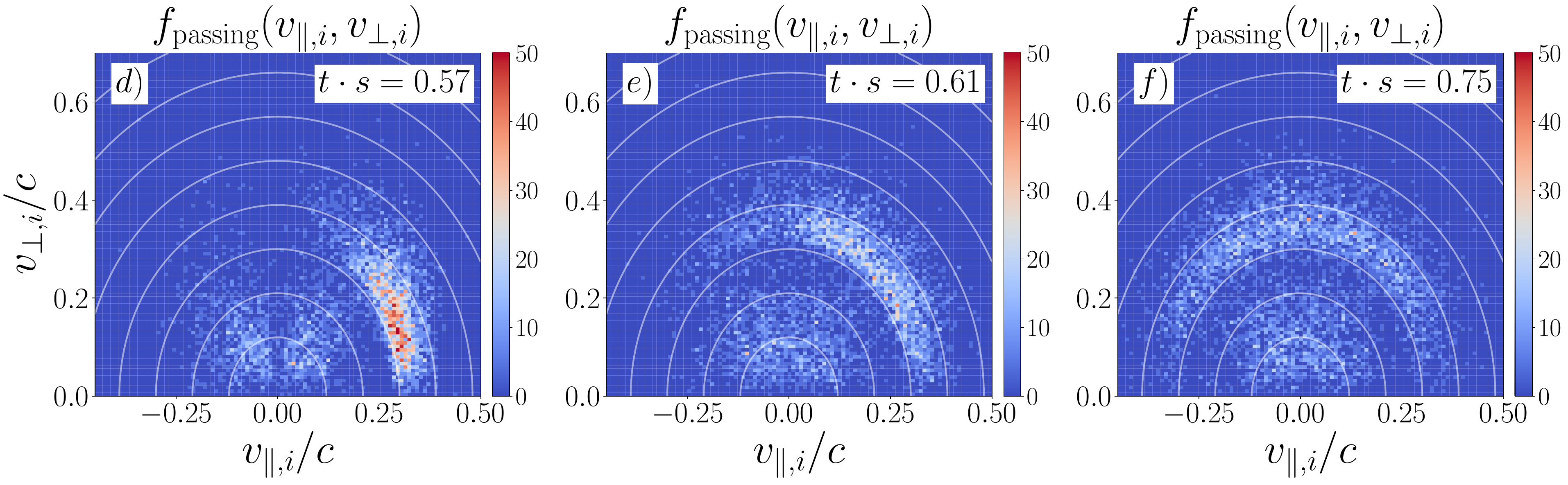}\\
         \includegraphics[width=0.95\linewidth]{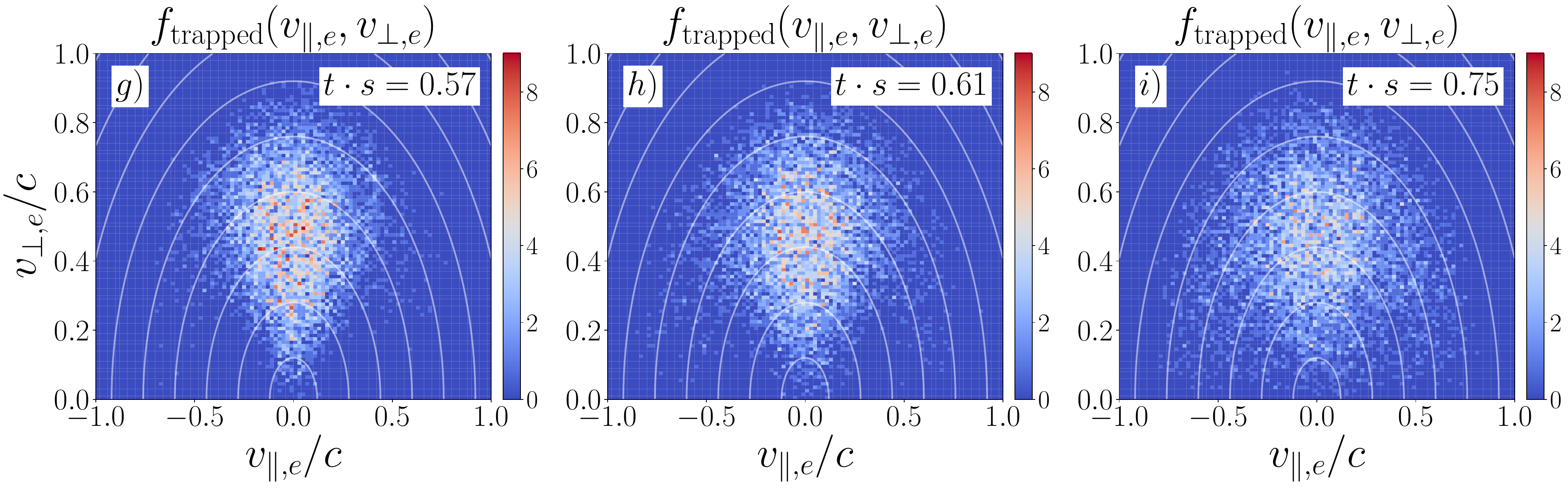}\\
         \includegraphics[width=0.95\linewidth]{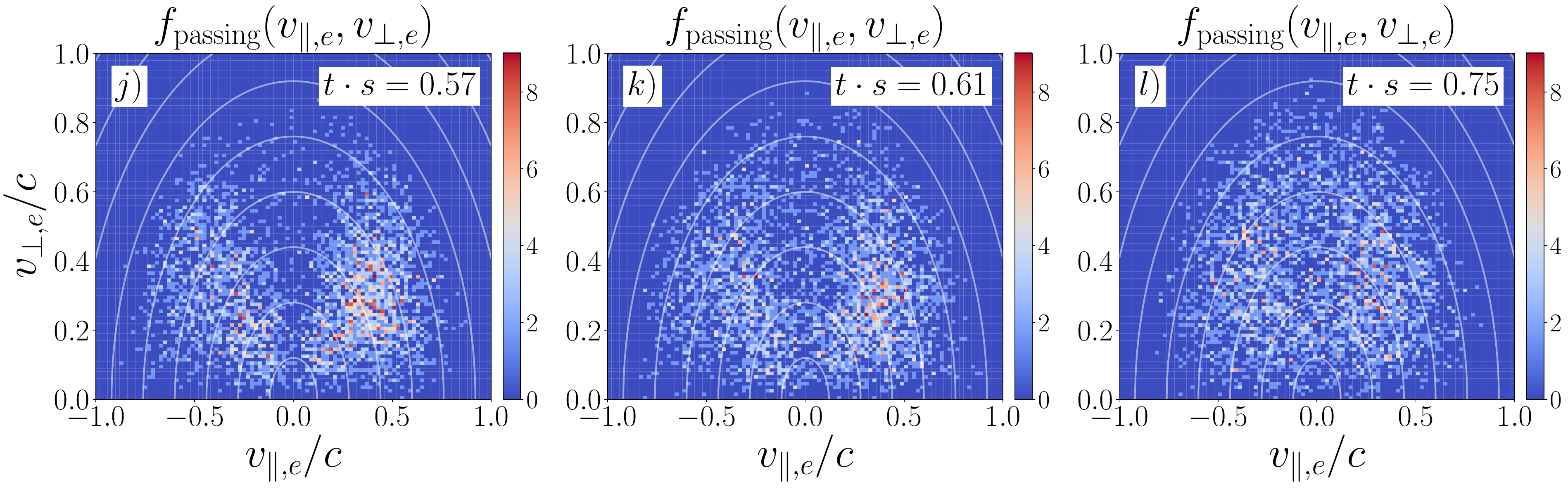}\\
    \end{tabular}
    \caption{The distribution function $f(v_{\parallel,j},v_{\perp,j})$ of trapped and passing ions and electrons at three different times: $t\cdot s = 0.57$ (first column), $t\cdot s = 0.61$ (second column), and $t\cdot s = 0.75$ (third column). The distribution function $f_{\text{trapped}}(v_{\parallel,i},v_{\perp,i})$ of the trapped ions is shown in the first row, $f_{\text{passing}}(v_{\parallel,i},v_{\perp,i})$ for the passing ions are shown in the second row, $f_{\text{trapped}}(v_{\parallel,e},v_{\perp,e})$ for the trapped electrons are shown in the third row, and $f_{\text{passing}}(v_{\parallel,e},v_{\perp,e})$ for the passing electrons are shown in the fourth row. In all the plots, the solid white curves denote contours of constant particle energy in the frame moving with the waves: $v_{\perp,j}^2 + (v_{\parallel,j} - \omega/k_{\parallel})^2 \approx v_{\perp,j}^2 + v_{\parallel,j}^2 = \text{const.}$ ($j=i,e$). An animation is available.}
    \label{fig:DistributionFunctionsIonsLecs}
\end{figure*}

Figure \ref{fig:DistributionFunctionsIonsLecs} shows the distribution functions of trapped and passing ions and electrons at three different times $t\cdot s = 0.57$, $t\cdot s = 0.61$, and $t\cdot s = 0.75$, spanning $\Delta \tau_{LR}$ and also part of mirror's saturated stage. In the following we describe the evolution of each population:

The distribution of trapped ions (figs. \ref{fig:DistributionFunctionsIonsLecs}$a$, \ref{fig:DistributionFunctionsIonsLecs}$b$, and \ref{fig:DistributionFunctionsIonsLecs}$c$) shows a clear loss-cone like form at $t\cdot s=0.57$ (all outside the loss-cone), meaning that all trapped ions are effectively trapped in mirror troughs. At this time, trapped ions have reached its maximum pressure anisotropy according to figure \ref{fig:AnisotropyTrapped}$a$. 

Once IC waves are excited, they interact with both trapped and passing ions via pitch-angle scattering in a quasilinear fashion (\cite{KennelEngelmann1966}). 
This diffusion process happens along paths of constant particle's energy in the frame moving with the waves (see e.g. \cite{Squire2022}): 
\begin{align}
    \label{eq:ConstantEnergyWaveFrame}
    v_{\perp,j}^2 + (v_{\parallel,j} - \omega/k_{\parallel})^2 = \text{const.}
\end{align}

We plot these contours in solid white lines in each plot of figure \ref{fig:DistributionFunctionsIonsLecs} as $v_{\perp,j}^2 + (v_{\parallel,j} - \omega/k_{\parallel})^2 \approx v_{\perp,j}^2 + v_{\parallel,j}^2 = \text{const.}$, as in a high-$\beta$ scenario, the phase velocity of an IC wave offers a small correction of order $v_A/v_{th,i} = \sqrt{1/\beta}$. Additionally, the IC waves in our simulations are destabilized in both parallel and anti-parallel directions to $\textbf{B}$. We can see that the relaxation of the distribution function of trapped ions by the quasi-linear interaction with IC waves agrees very well with these paths, by looking at $t\cdot s = 0.61$ and $t\cdot s = 0.75$.

The distribution of passing ions (figs. \ref{fig:DistributionFunctionsIonsLecs}$d$, \ref{fig:DistributionFunctionsIonsLecs}$e$, and \ref{fig:DistributionFunctionsIonsLecs}$f$) shows, on the one hand, a concentration of ions at low perpendicular velocities and relatively large parallel velocities, and it looks fairly symmetric in $v_{\parallel}$. This is consistent with having untrapped ions mainly streaming along the mean magnetic field in both directions. On the other hand, the population of large parallel velocity is also shown at $v_{\parallel}/c \approx 0.3$ (see section \ref{sec:LionRoarsTrappedPassing}). Interestingly, the passing ions also interact quasilinearly with IC waves, and this is particularly evident in the evolution of passing ions. Indeed, we can clearly see how the large parallel velocity population of passing ions evolves along the contours of of constant particle energy with excellent agreement at $t\cdot s = 0.61$ and $t\cdot s = 0.75$. We can understand the evolution of this population by looking at the gyroresonance condition

\begin{align}
    \omega - k_{\parallel}v_{\parallel,i} = \pm \omega_{c,i}.
    \label{eq:gyroresonance}
\end{align}

If we look at the peak power at positive frequencies in the power spectrum shown in fig. \ref{fig:PowerSpectrumOmega}$c$, we can estimate the frequency and wavenumber at which most of the power of IC waves resides: $\omega/\omega_{c,i}^{\text{init}} \approx 0.2$, and $ck_{\parallel}/\omega_{p,i}^{\text{init}} \approx \pm 0.15$. From eq. (\ref{eq:gyroresonance}) we can estimate then the parallel velocity of the ions interacting gyroresonantly with these IC waves:

\begin{align}
    \frac{v_{\parallel,i}}{c} = \frac{\omega/\omega_{c,i}^{\text{init}}\mp 1}{(ck_{\parallel}/\omega_{p,i}^{\text{init}})(m_ic^2/k_BT_i^{\text{init}})^{1/2}(\beta_i^{\text{init}}/2)^{1/2}},
\end{align}

which gives $v_{\parallel,i}/c \approx 0.36$ and $v_{\parallel}/c \approx -0.24$, which falls in the range of the large parallel velocity population. The quasilinear evolution also happens with the population with smaller parallel velocity.

The population of trapped electrons (figs. \ref{fig:DistributionFunctionsIonsLecs}$g$, \ref{fig:DistributionFunctionsIonsLecs}$h$, and \ref{fig:DistributionFunctionsIonsLecs}$i$) shows a very similar evolution to that of trapped ions; the loss-cone like distribution is also apparent. The evolution of this distribution is also consistent with a quasilinear interaction now between the electron and whistler waves, driving the distribution towards isotropy along paths of constant particle energy, as can be seen at later times in figure \ref{fig:DistributionFunctionsIonsLecs}.

Finally, the population of passing electrons (figs \ref{fig:DistributionFunctionsIonsLecs}$j$, \ref{fig:DistributionFunctionsIonsLecs}$k$, and \ref{fig:DistributionFunctionsIonsLecs}$l$) also shows a very similar evolution to that of the ions. The populated loss-cone shape of the distribution is also apparent, and we can see the quasilinear evolution of the distribution function along constant particle energy contours at later times.

This way, we have provided evidence for the source of both IC and whistler waves observed in our simulations. Once ions and electrons get trapped in regions of low magnetic field strength of mirror modes, they get significantly anisotropic with a loss-cone like distribution, which is able to destabilize parallel-propagating IC and whistler waves, respectively. These waves then interact with both population of trapped and passing particles in a quasilinear fashion, driving both populations of trapped and passing ions and electrons towards a more isotropic state. Consequently, this mechanism can contribute to regulate the global anisotropy of ions and electrons, and can thus be a pathway for particle escape and consequent saturation of mirror modes (\cite{Kunz2014}).

\section{Mass-Ratio Dependence}
\label{sec:MassRatioDependence}

In this section, we compare simulations with different mass ratios: $m_i/m_e=8$, $m_i/m_e=32$, but with the same initial conditions for ions, as shown for runs b20m8w800, b20m32w800,and b20m64w800 in Table \ref{table:SimulationList}, although with somewhat different temperatures. We see that IC and whistler waves' signatures do appear in all three simulations, and thus they do not seem to present a strong dependence on mass ratio.

\begin{figure}[t]
    \centering
    \includegraphics[width=\linewidth]{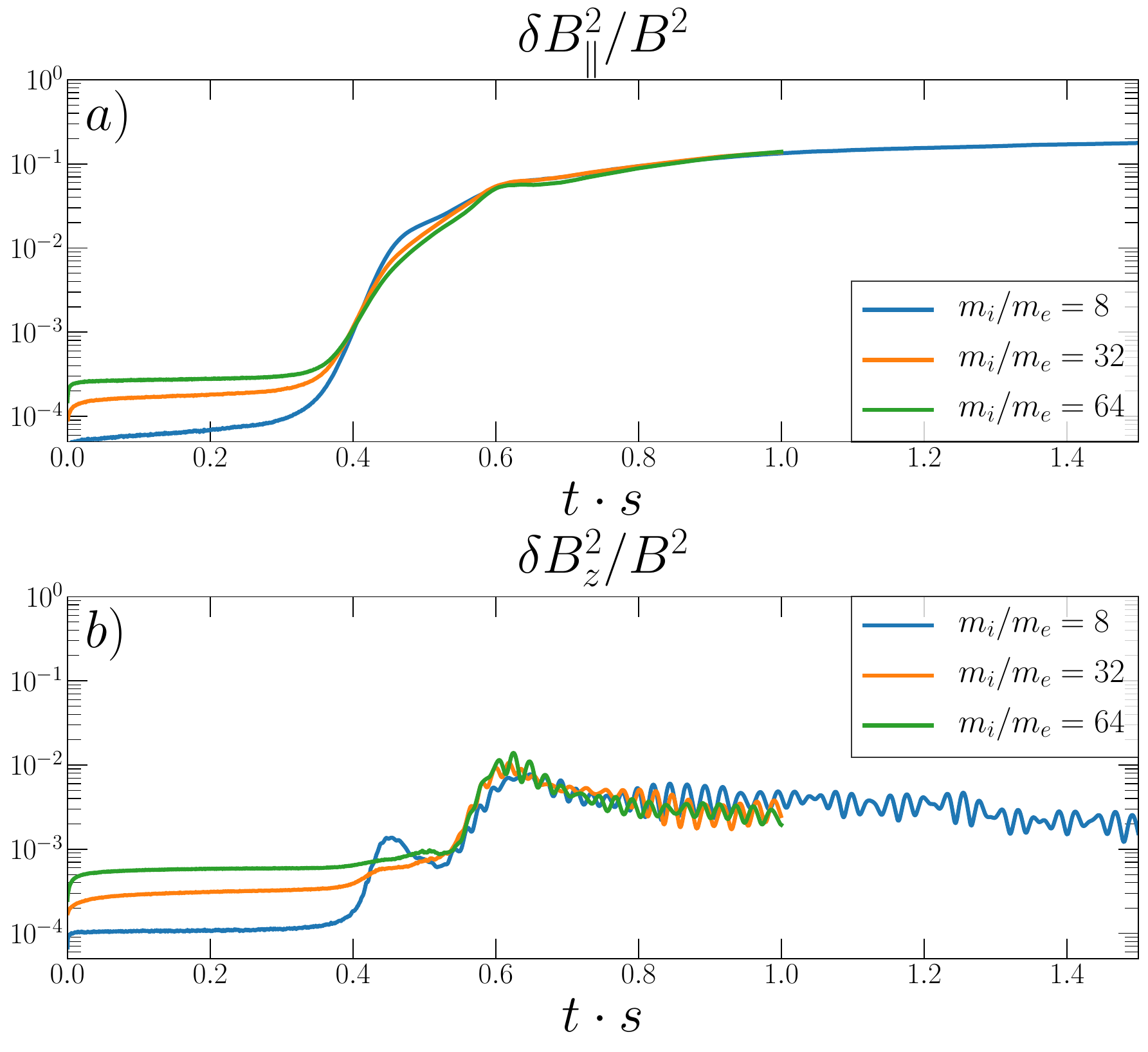}
    \caption{Panel a: The energy in the parallel component of the magnetic field fluctuations $\delta \textbf{B}$, for three simulations with different mass ratios: $m_i/m_e=8$ (run b20m8w8, blue line), $m_i/m_e=32$ (run b20m32w8, orange line), and $m_i/m_e=64$ (run b20m64w8, green line). Panel b: same as in panel a but for the perpendicular component of $\delta \textbf{B}$ out of the plane of the simulation in the same runs.}
    \label{fig:MassRatiodeltaB}
\end{figure}

Figure \ref{fig:MassRatiodeltaB} shows the evolution of $\delta B_{\parallel}^2$ (panel $a$) and $\delta B_{z}^2$ (panel $b$) for the three runs with mass ratios: $m_i/m_e=8,32,$ and $64$ (runs b20m8w800, b20m32w800, and b20m64w800 in table \ref{table:SimulationList}). We can see a very consistent evolution of $\delta B_{\parallel}^2$ in all three runs, meaning that $m_i/m_e$ does not play a significant role on the early evolution and saturation of the mirror instability. Similarly, $\delta B_z^2$ shows the same features in all three runs, especially during mirrors' secular growth and saturated stages ($t\cdot s \approx 0.5$ onwards). The early peak in $\delta B_{\parallel}^2$ at $t\cdot s \approx 0.4$ corresponding to the early whistler burst is also seen in the three runs, but more prominently in the simulation with $m_i/m_e=8$. This is possibly due to an enhancement of this wave activity by the ions, which are able to weakly feel the presence of whistlers, as the mass separation is not very large. This effect disappears as the mass ratio increases, and the early whistlers only affect the electrons. More importantly, for $t\cdot s > 0.5$, all three runs show a very similar evolution of $\delta B_{\parallel}^2$. 

\begin{figure}[hbtp]
    \centering
    \includegraphics[width=\linewidth]{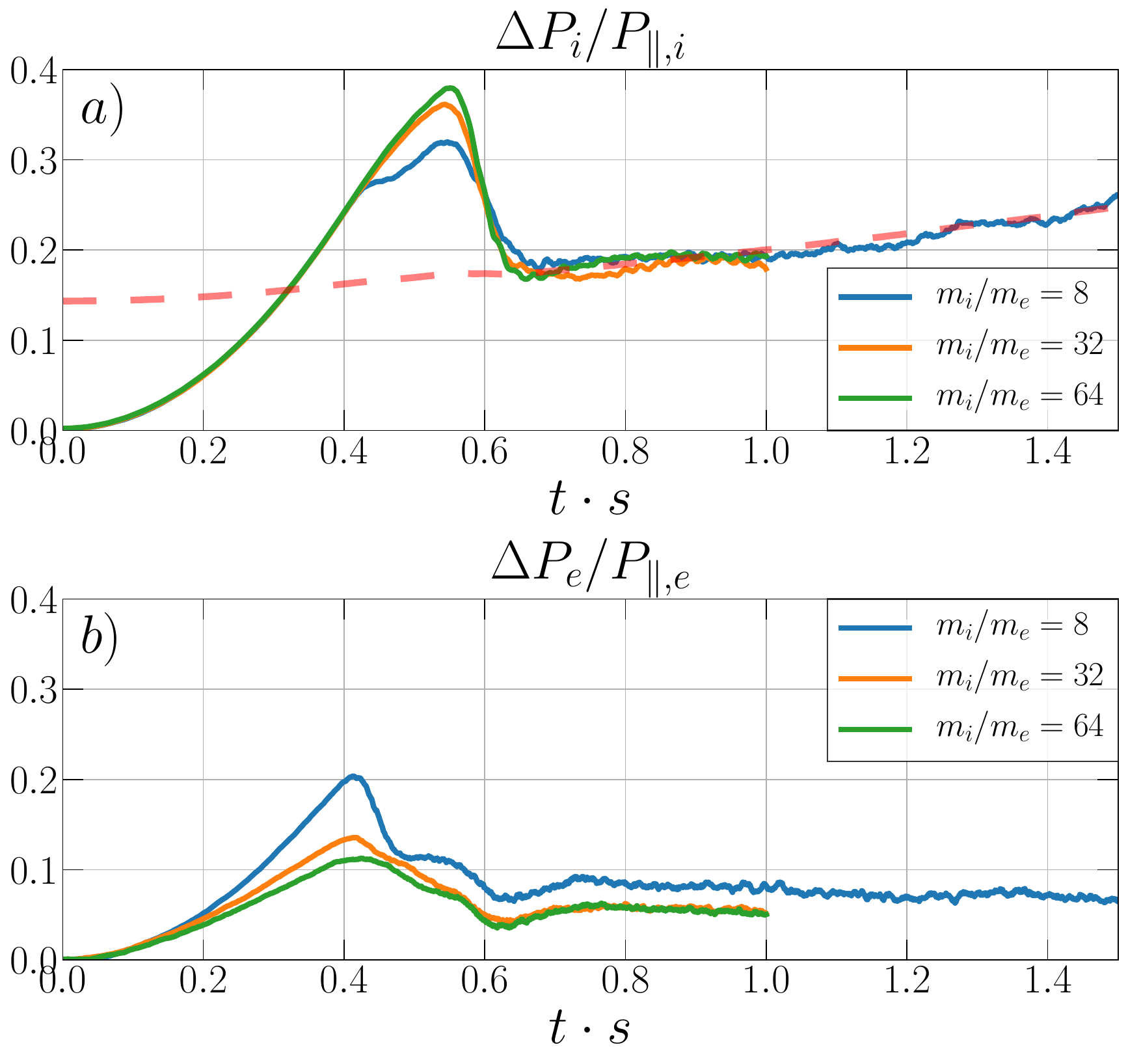}
    \caption{Panel a: Evolution of the ion pressure anisotropy for three simulations with different mass ratios: $m_i/m_e=8$ (run b20m8w8, blue line), $m_i/m_e=32$ (run b20m32w8, orange line), and $m_i/m_e=64$ (run b20m64w8, green line). The dashed red line indicates the best-fit the threshold shown figure \ref{fig:IonLecAnisotropy}$a$, $\Delta P_i/P_{\parallel,i}\propto \beta_{\parallel,i}^{-0.45}$. Panel b: same as in panel a but for the electron pressure anisotropy in the same runs.}
    \label{fig:MassRatioAnisotropy}
\end{figure}

Figure \ref{fig:MassRatioAnisotropy} shows the evolution of the pressure anisotropy of ions (panel $a$) and electrons (panel $b$) for the same three runs. In the case of the ions, we can see an overall evolution that is very consistent in all three runs, both in early and late stages. We can see a smaller anisotropy overshoot for the simulation with $m_i/m_e=8$ at $t\cdot s \approx 0.4$, coincident with the enhancement seen in $\delta B_z^2$, during the early whistler burst, suggesting that ions can weakly interact with the whistlers at this mass ratio, and consequently their anisotropy does not reach the same overshoot as the rest of the runs. Notwithstanding the foregoing, we can see how all three runs display a very similar pressure anisotropy evolution afterwards, which is also well described by the best-fit threshold $\Delta P_i \propto \beta_i^{-0.45}$ shown in fig. \ref{fig:IonLecAnisotropy}.

In the case of the electron pressure anisotropy $\Delta P_e$, we can also see a similar evolution overall in fig. \ref{fig:MassRatioAnisotropy}$b$. The overshoot at $t\cdot s \approx 0.4$ is larger for decreasing mass ratios, possibly due to the fact that the whistler amplitude required for efficient scattering decreases as $m_i/m_e$ increases, as explained above. This means that, after $\Delta P_e/P_{e,\parallel}$ has surpassed the threshold for efficient growth of the whistler modes, the simulations with larger $m_i/m_e$ take shorter times to reach the necessary whistler amplitude to efficiently scatter the electrons. This implies that the overshoot decreases for higher mass ratios. During late stages, we can see a very similar evolution of $\Delta P_e$ in all three runs, that is even more evident for $m_i/m_e = 32$ and $m_i/m_e = 64$ (orange and green curves in fig. \ref{fig:MassRatioAnisotropy}$a$),  which essentially lie on top of each other.

\begin{figure}
    \centering
    \begin{tabular}{cc}
    \includegraphics[width=0.5\linewidth]{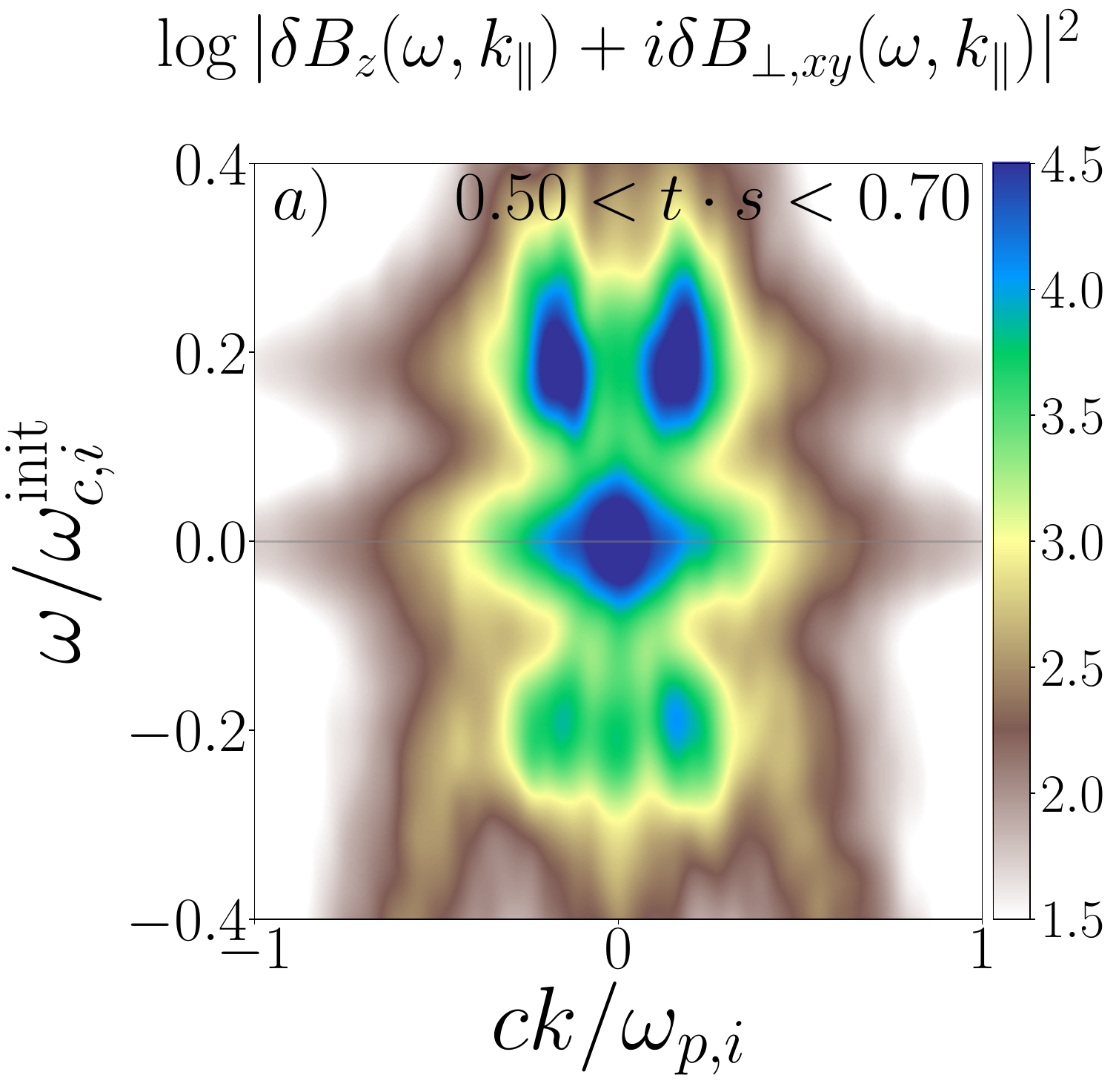}
         &  
    \includegraphics[width=0.45\linewidth]{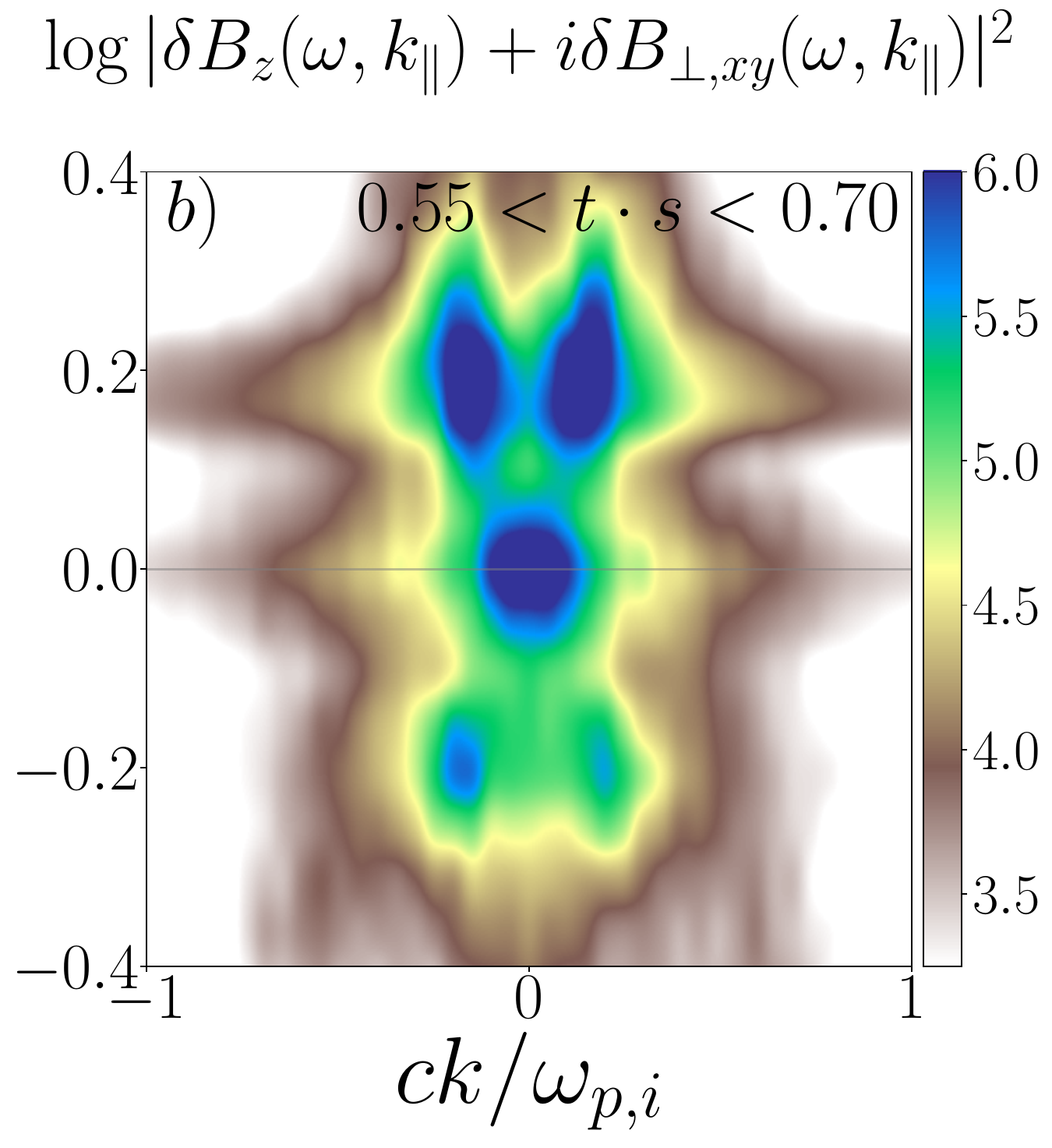}
    \end{tabular}
    \caption{The power spectrum of $\delta B_z(\omega,k_{\parallel}) + i\delta B_{\perp}(\omega,k_{\parallel})$ at $0.5 < t\cdot s < 0.7$ for $m_i/m_e = 32$ (run b20m32w800, left panel) and $m_i/m_e = 64$ (run b20m64w800, right panel). Positive and negatives frequencies show the power in left-hand and right-hand polarized waves, respectively. } 
    \label{fig:PowerSpectrumMassRatio}
\end{figure}

Finally, figure \ref{fig:PowerSpectrumMassRatio} shows the power spectrum of $\delta B_z(\omega,k_{\parallel}) + i\delta B_{\perp,xy}(\omega,k_{\parallel})$ for the simulation with $m_i/m_e=32$ (fig. \ref{fig:PowerSpectrumMassRatio}$a$) and with $m_i/m_e=64$ (fig. \ref{fig:PowerSpectrumMassRatio}$b$). Here we also see a very similar power distribution at both mass ratios, showing both left-hand and right-hand polarized waves (positive and negative frequencies, respectively). The peak power is also observed at the same frequencies and wavenumbers as in fig. \ref{fig:PowerSpectrumOmega} for both polarizations.

This way, we can see that the linear and nonlinear evolution of the mirror instability and the late IC and whistler evolution are well captured in our simulations, and it does not strongly depend on mass ratio.

\section{Dependence on initial plasma $\beta$}
\label{sec:betadependence}

We tested whether the IC and whistler waves' activity is present in simulations with $\beta_i^{\text{init}}=2$ (i.e, total $\beta^{\text{init}}=4$), and $\beta_i^{\text{init}}=40$ (i.e. total $\beta^{\text{init}}=80$), and compare them with our fiducial simulation at $\beta_i^{\text{init}}=20$. We confirm that the mirror instability can develop in all simulations, and both IC and whistler waves do appear at nonlinear stages. 

\begin{figure*}[t]
    \centering
    \begin{tabular}{ccc}
         \includegraphics[width=0.33\linewidth]{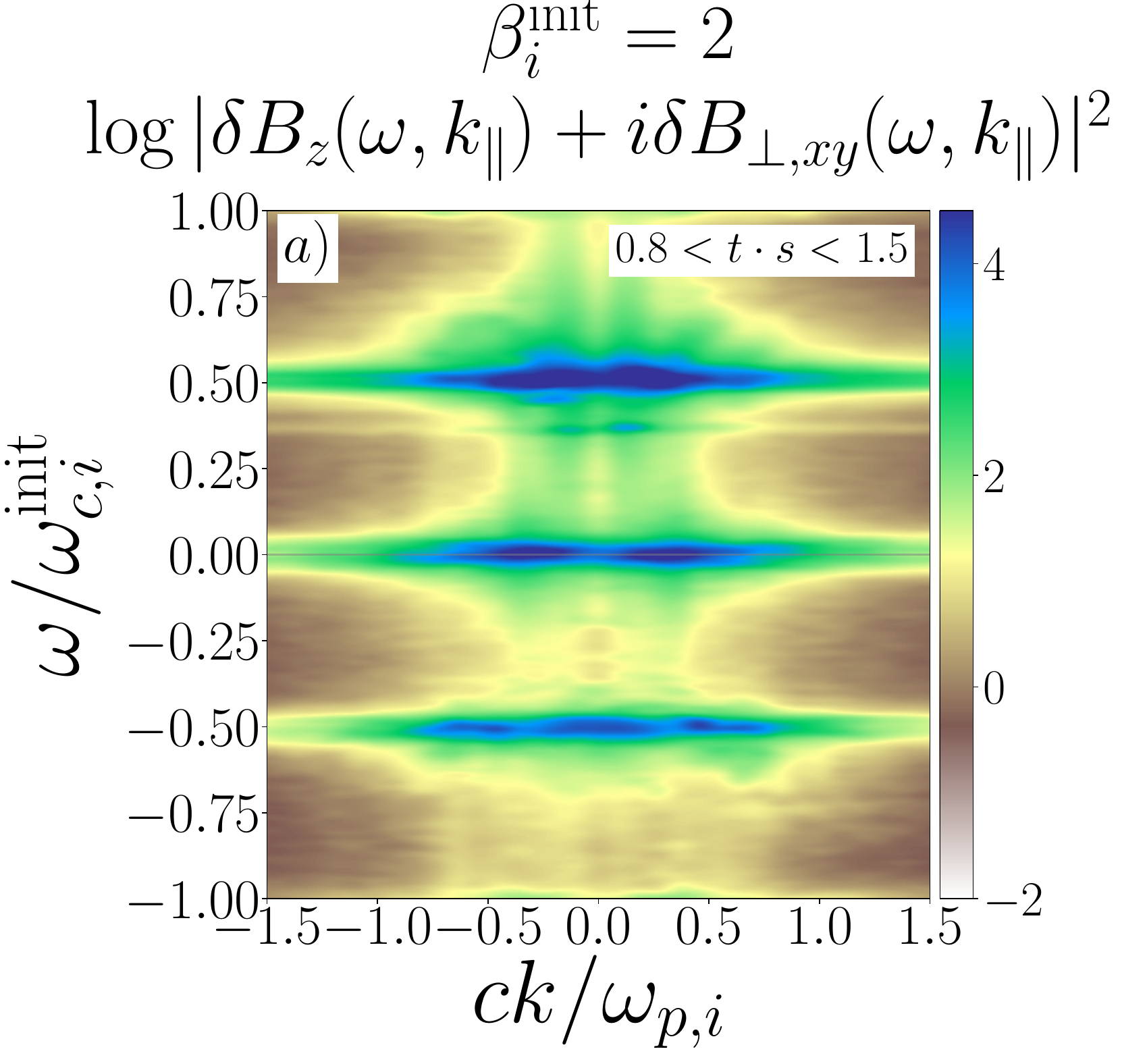} &  
         \includegraphics[width=0.33\linewidth]{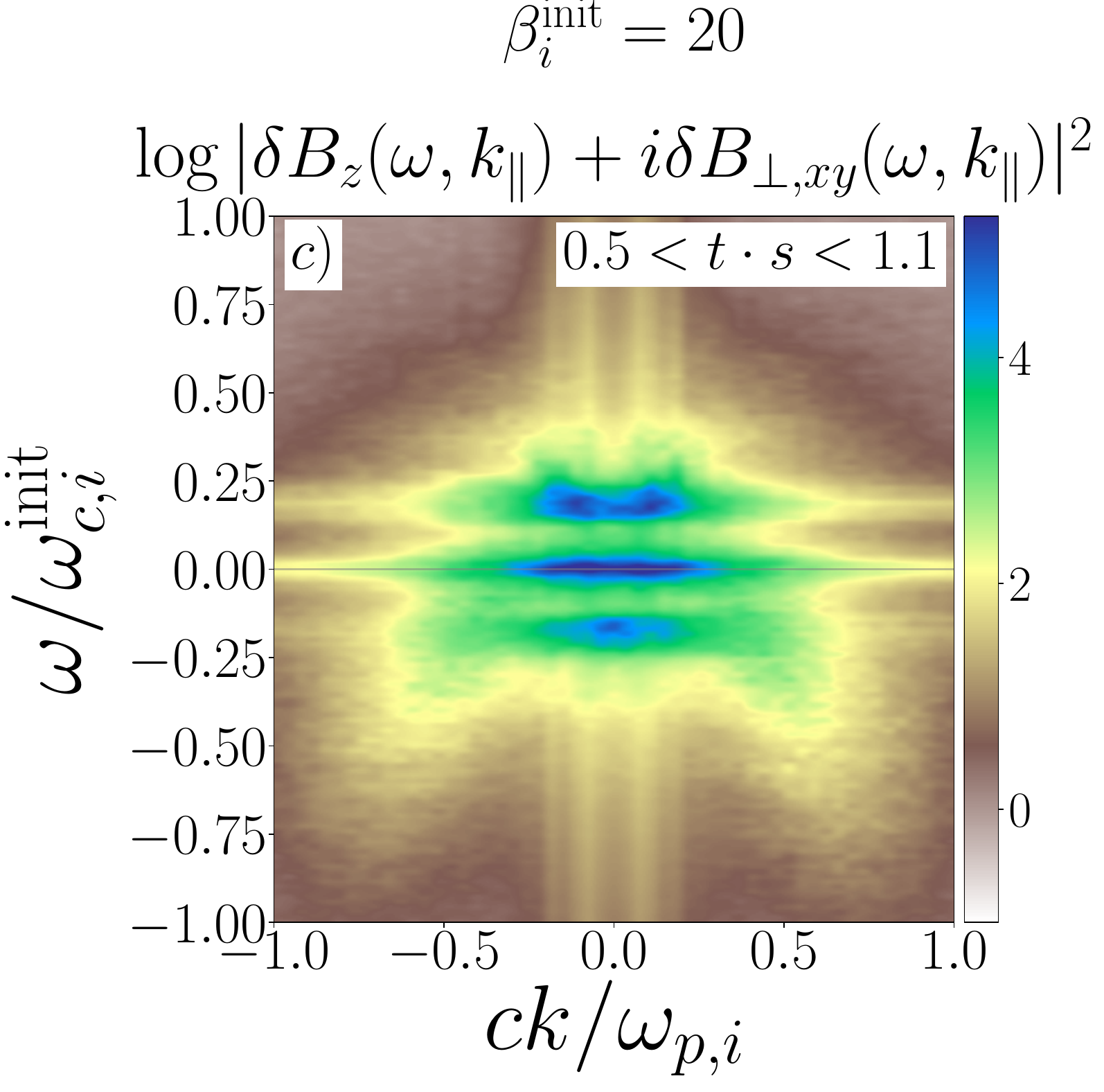} &  
         \includegraphics[width=0.33\linewidth]{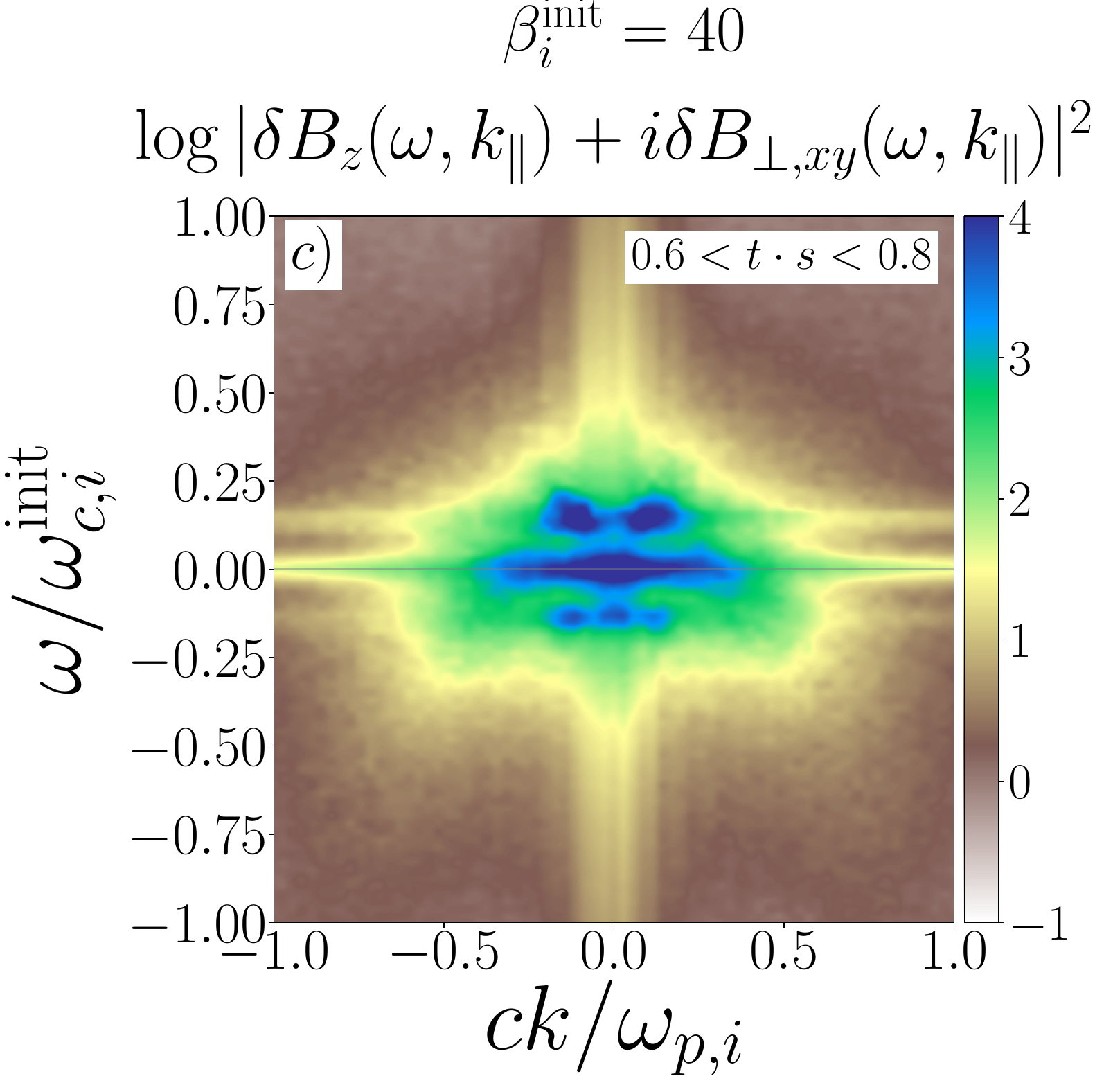}
    \end{tabular}
    \caption{The power spectrum of $\delta B_z(\omega,k_{\parallel}) + i \delta B_{\perp,xy}(\omega,k_{\parallel})$ for three runs with different initial ion beta: $\beta_{i}^{\text{init}}=2$ (panel $a$, run b2m8w800), $\beta_{i}^{\text{init}}=20$ (panel $b$, run b20m8w800), and $\beta_{i}^{\text{init}}=40$ (panel $c$, run b40m8w800). Positive and negatives frequencies show the power in left-hand and right-hand polarized waves, respectively.}
    \label{fig:PowerSpectrumBeta}
\end{figure*}

The power spectrum of $\delta B_z(\omega,k_{\parallel}) + i \delta B_{\perp,xy}(\omega,k_{\parallel})$ is shown in figure \ref{fig:PowerSpectrumBeta}, and we can see that it is similar among the three $\beta_{i}$ cases. In all three cases we see the power concentrated at $\omega \sim 0$ corresponding to mirror modes. In addition, we also see a concentration of power in right and left polarized waves, so both IC and whistler waves are also present, although their peak frequency changes. For the $\beta_i^{\text{init}}=2$ case we see that the peak frequency is at $\omega/\omega_{c,i}^{\text{init}} \approx 0.5$, whereas in the $\beta_i^{\text{init}}=40$ case it shifts to smaller values, $\omega/\omega_{c,i}^{\text{init}} \approx 0.1$. This shift in peak frequency can also be explained by the IC and whistler dispersion relations analogous to our discussion in section \ref{sec:LionRoars}.

\begin{figure}[hbtp]
    \centering
    \begin{tabular}{c}
        \includegraphics[width=\linewidth]{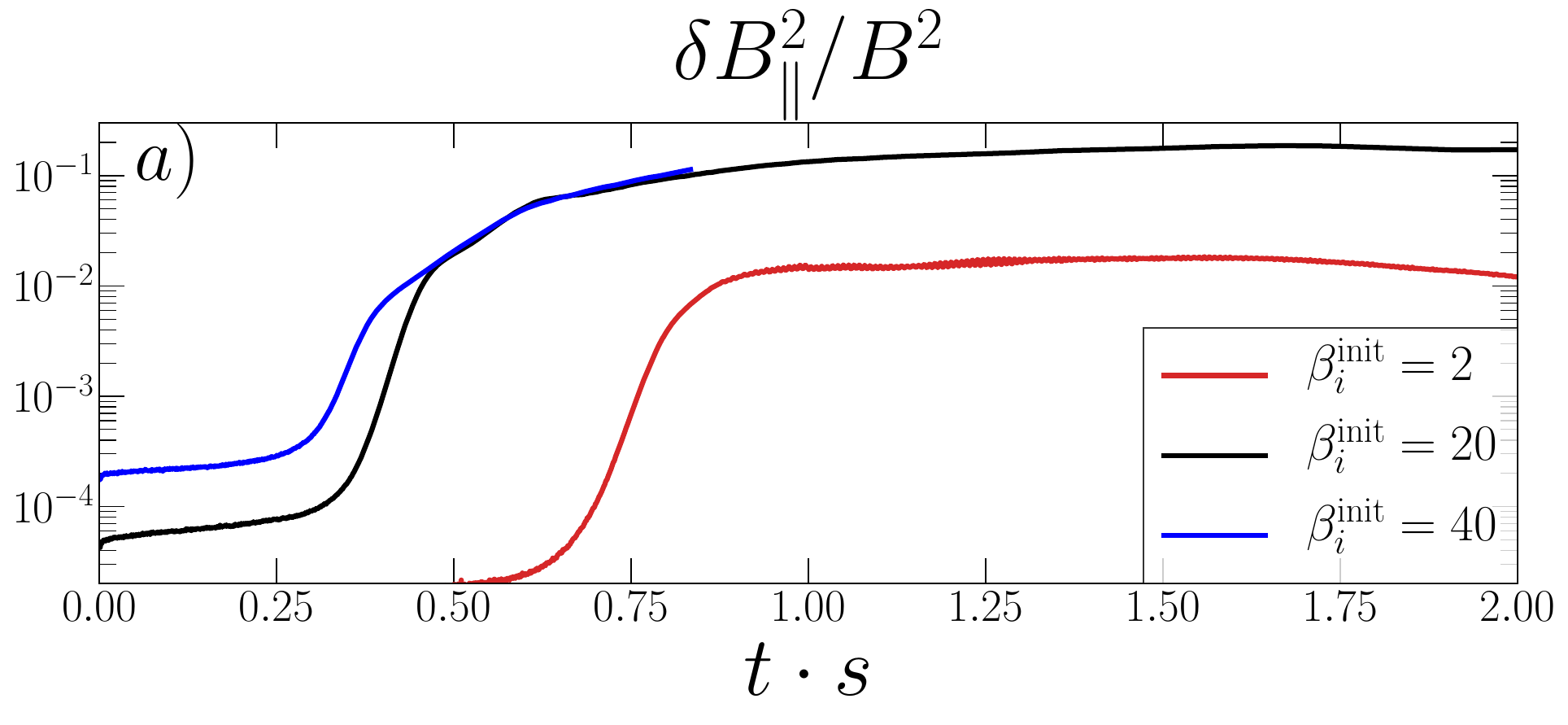} \\
         \includegraphics[width=\linewidth]{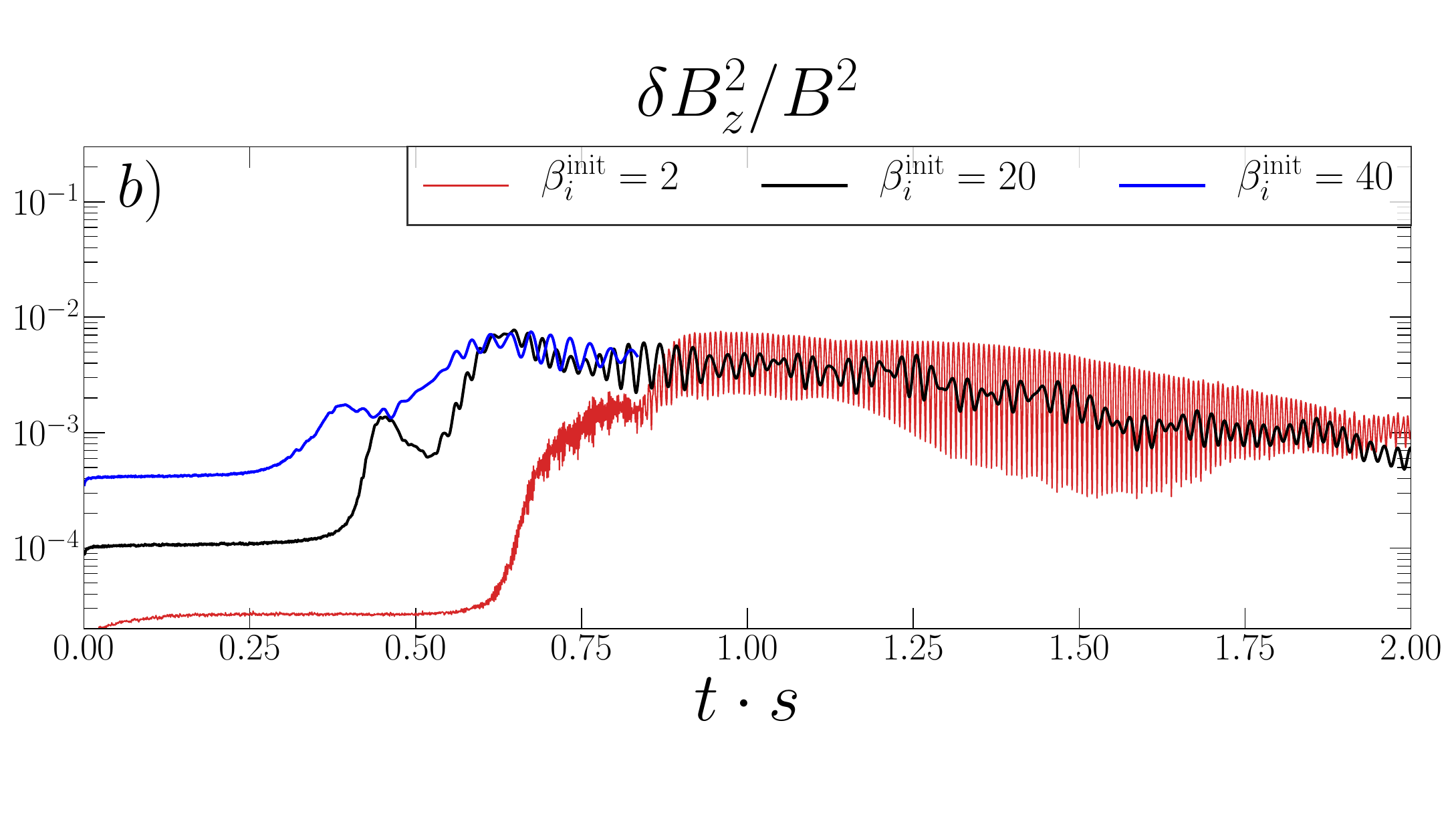}
    \end{tabular}
    \caption{Panel $a:$ Evolution of $\delta B_{\parallel}^2$ for three simulations with different initial ion beta: $\beta_{i}^{\text{init}}=2$ (solid red line, run b2m8w800), $\beta_{i}^{\text{init}}=20$ (solid black line, run b20m8w800), and $\beta_{i}^{\text{init}}=40$ (solid blue line, run b40m8w800). Panel $b$: Evolution of $\delta B_{z}^2$ for the same three simulations shown in panel $a$.}
    \label{fig:betadependence_deltaB}
\end{figure}

Figure \ref{fig:betadependence_deltaB} compares the evolution of $\delta B_{\parallel}^2$ (i.e., mainly the development of the mirror instability) for the three runs with different initial $\beta^{\text{init}}$ (the other phyiscal parameters are the same, see table \ref{table:SimulationList}). In all three cases we can see an exponential phase followed by the secular and saturated stages characteristic of the mirror instability, which develops earlier for higher initial $\beta^{\text{init}}$, consistent with the smaller anisotropy threshold for the growth of the mirror instability at larger beta. The amplitude of $\delta B_{\parallel}^2$ at the saturated stage is comparable for both $\beta^{\text{init}}=20$ and $\beta^{\text{init}}=40$ runs, and is smaller for the $\beta^{\text{init}}=2$ run, as also seen by previous works (e.g. \cite{Riquelme2015}).

Indeed, when we look at the evolution of $\delta B_z^2$, we can see that for both $\beta^{\text{init}}=20$ and $\beta^{\text{init}}=40$ runs, the evolution is similar: both display an early whistler burst at $t\cdot s \approx 0.4$, and a IC/whistler excitation stage ($t\cdot s \approx 0.5$ onwards) at almost the same amplitude. In the case of the $\beta^{\text{init}}=2$ run, we can see that the first exponential growth in $\delta B_z^2$ at $t\cdot s \approx 0.6$ is consistent with an IC burst (see e.g. \cite{Ley2019}), after which we see the typical oscillation pattern that the excitation of late IC and whistler waves produces, from $t\cdot s \approx 0.8$ onwards, saturating at a similar amplitude than the rest of the runs, and displaying a very high-frequency oscillation. 

\begin{figure}[hbtp]
    \centering
    \begin{tabular}{cc}
        \includegraphics[width=\linewidth]{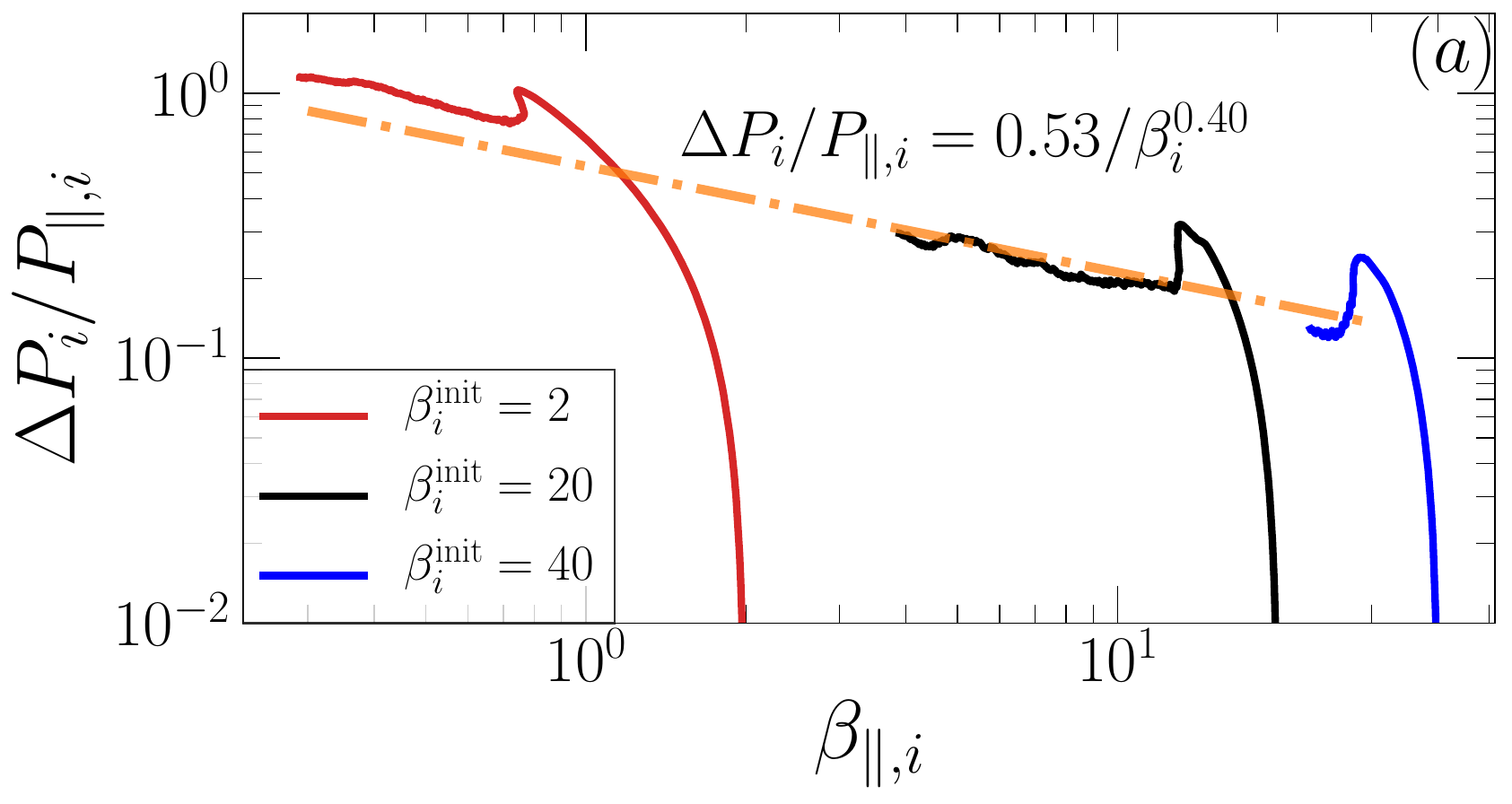}\\
        \includegraphics[width=\linewidth]{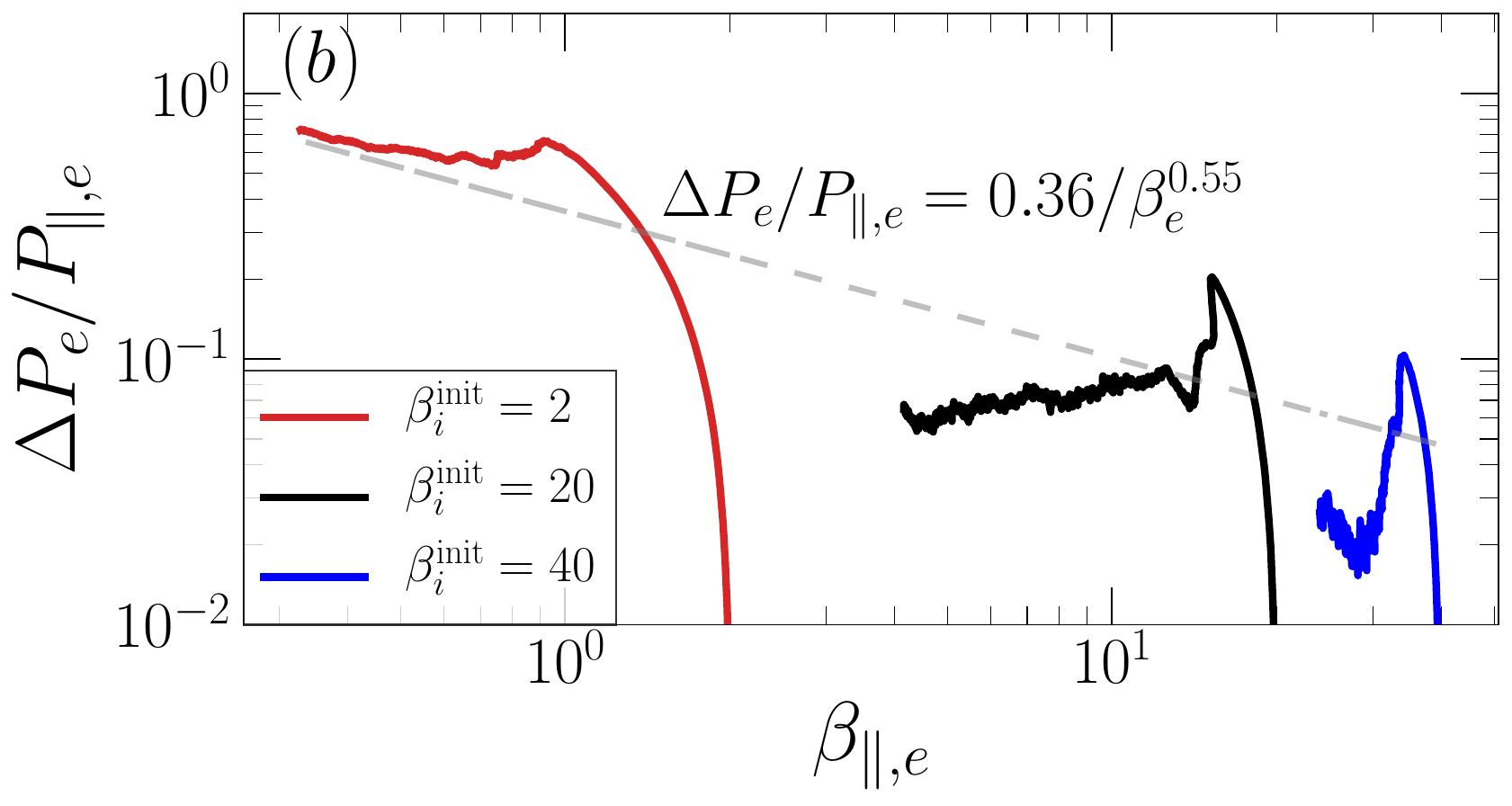}
    \end{tabular}
    \caption{Panel $a$: Ion Anisotropy, $\Delta P_i/P_{\parallel,i}$ as a function of parallel ion beta, $\beta_{\parallel,i}$ (with respect to the main magnetic field $\textbf{B}$) for three different simulations with different initial ion beta:  $\beta_{i}^{\text{init}}=2$ (solid red line, run b2m8w800), $\beta_{i}^{\text{init}}=20$ (solid black line, run b20m8w800), and $\beta_{i}^{\text{init}}=40$ (solid blue line, run b40m8w800). The dotted-dashed orange line shows the IC threshold $\Delta P_{i}/P_{\parallel,i} = 0.53/\beta_{\parallel,i}^{0.4}$ from \cite{GaryLee1993} for $\gamma_{IC}/\omega_{c,i}=10^{-2}$. Panel $b$: Electron anisotropy $\Delta P_e/P_{\parallel,e}$ as a function of parallel electron beta, $\beta_{\parallel,e}$ for the same three simulations shown in panel $a$. The dashed gray line in this case shows the threshold for the whistler instability, $\Delta P_e/P_{\parallel,e} = 0.36\beta_{\parallel,e}^{-0.55}$ for growth rate $\gamma = 0.01\omega_{c,e}$, from \cite{GaryWang1996}.}
    \label{fig:betadependence_anisotropy}
\end{figure}

In figure \ref{fig:betadependence_anisotropy}, we compare the evolution of the ion and electron pressure anisotropy plotted as a function of their parallel plasma $\beta_i$ for the three simulations with different initial $\beta_i$ (As in all our simulations the mean magnetic field strength is continuously increasing, so the particles' $\beta_i$ decreases over time, and therefore the simulation evolves towards the left in fig. \ref{fig:betadependence_anisotropy}.).

In the case of the ions (fig. \ref{fig:betadependence_anisotropy}$a$), we can see a 
similar overshoot and subsequent regulation, but the overshoot occurs at a lower anisotropy value for increasing $\beta_i$. This is consistent with the inverse $\beta_i$ dependence of the mirror instability threshold: mirror modes are excited earlier at higher $\beta_i$, and therefore have relatively more time to regulate the anisotropy before it reaches a higher overshoot. Interestingly, the saturated stage of the ion pressure anisotropy is consistent with the theoretical IC threshold from \cite{GaryLee1993}: $\Delta P_i/P_{\parallel,i} =  0.53\beta_{\parallel,i}^{-0.40}$ for $\gamma_{IC}/\omega_{c,i}=10^{-2}$ (see fig. \ref{fig:IonLecAnisotropy}$a$) in all three runs, suggesting a universality in the threshold that $\Delta P_i/P_{\parallel,i}$ follows, as a consequence of the excitation of IC waves during mirrors' saturated stage. (In the case of the $\beta_{i}^{\text{init}}=40$ run, however, it is more unclear whether it can follow the above mentioned threshold at late stages, given the short duration of this run.)

In the case of electrons (fig. \ref{fig:betadependence_anisotropy}$b$), we can also see that the overshoot is reached at lower values of the pressure anisotropy $\Delta P_e/P_{\parallel, e}$ for increasing initial beta, consistent with an inverse-$\beta_i$ dependence now of the whistler instability anisotropy threshold. It is interesting to note that after the anisotropy overshoot, and during these late stages, the electron pressure anisotropy tends to be significantly smaller than the expectation from the threshold for the whistler instability in the higher initial $\beta_i$ runs ($\beta_{i}^{\text{init}}=20$ and $\beta_{i}^{\text{init}}=40$), irrespective of the generation of pressure anisotropy that the continuous amplification of the magnetic field produces as a consequence of the shear motion in the simulation. Notice, however, that in low magnetic field regions the electron pressure anisotropy is larger than the whistler threshold and, therefore, enough to excite whistlers (fig \ref{fig:AnisotropyBeta}). This shows the key role played by mirror-generated magnetic troughs in creating the conditions to excite whistlers despite the fact that, globally, the pressure anisotropy may be not be enough to make these waves unstable. On the other hand, in the $\beta_{i}^{\text{init}}=2$ run, $\Delta P_e/P_{\parallel, e}$ continues to weakly grow because of the continuous $B$ amplification, and this is done following a marginal stability state well described by the threshold of the whistler instability $\Delta P_e/P_{\parallel,e} \propto \beta^{-0.55}$ (\cite{GaryWang1996}), consistent with previous works at lower $\beta_{\parallel,e}$ (\cite{Ahmadi2018}).

The persistence of the late IC and whistler activity at different initial plasma $\beta_i$ suggests that this phenomenon is a natural consequence of the excitation of the mirror instability. In other words, in a weakly collisional plasma with an initial plasma $\beta_i$ sufficiently high to effectively excite the mirror instability, the excitation of IC and whistler waves at its late, saturated stages seems to be ubiquitous. 

\section{Summary and Discussion}
\label{sec:discussion}

In summary, we have performed fully kinetic PIC simulations of a collisionless plasma subject to a continuous amplification of the background magnetic field to study the nonlinear stages of the mirror instability and the ensuing excitation of secondary ion-cyclotron (IC) and whistler instabilities, in conditions where plasma pressure dominates over magnetic pressure (high-$\beta$). After mirror modes reach high-amplitudes and are able to trap ions and electrons within regions of low-$\textbf{B}$, we observe the excitation of sub-dominant left-hand polarized IC and right-hand polarized whistler waves that persist throughout the rest of the simulation, well into the nonlinear stages of the mirror instability (see section \ref{sec:LionRoars}). The whistler waves in our simulations seem to be consistent with the observations of whistler lion roars in the Earth's magnetosheath. 

By tracking ions and electrons through the simulation, we studied the excitation mechanism of both IC and whistler waves. We characterized the population of tracked particles as trapped and passing (i.e. untrapped) within mirror modes, and followed the evolution of their distribution functions. We observed that the trapped population of both ions and electrons become highly anisotropic while trapped inside mirror modes, contributing most of the anisotropy that allows the plasma to become unstable to IC and whistler waves, respectively. On the other hand, the passing ions and electrons developed features concentrated at small perpendicular and large parallel velocities, and fairly symmetric with respect to $v_{\parallel}$, with a clear absence at small parallel velocities (see section \ref{sec:LionRoarDistributionFunctions}). 

Once IC and whistlers are excited, they interact with both trapped and passing population of ions and electrons, respectively, via gyroresonant pitch-angle scattering. As a result of this interaction, both trapped ions and electrons reduce their anisotropy and escape from magnetic troughs of mirror modes, following the prediction of quasilinear theory. The passing ion and electron populations evolve in a similar manner (see fig. \ref{fig:DistributionFunctionsIonsLecs}). Interestingly, this process is observed to regulate the global anisotropy of ions and electrons in the simulation, driving the ion pressure anisotropy towards the IC instability threshold (\cite{GaryLee1993}), and the electron pressure anisotropy towards a global anisotropy much smaller than expected from theoretical whistler threshold. Given this low electron pressure anisotropy, the whistler excitation can be explained by the fact that, within mirror-generated magnetic troughs, the pressure anisotropy is locally larger than the whistler threshold (fig. \ref{fig:AnisotropyBeta}$i$). Thus, we interpret the whistler-driven regulation of electron pressure anisotropy as a local phenomenon, mainly produced by trapped electrons within non-linear mirror structures.

The excitation of the secondary IC and whistler waves is maintained as long as mirror modes are present and growing, and this also was observed in simulations of lower and higher initial plasma $\beta$. This way, IC and whistler waves could be a concomitant feature of the nonlinear evolution of the mirror instability, and provide an interesting physical connection between ion-scale instabilities and electron-scale physics.

In this work, we did not vary the scale-separation ratio $\omega_{c,i}/s$. In an environment like the ICM, turbulent eddies could drive the plasma locally through shear motions at kinetic scales with a wide range of frequencies $s$, and we typically expect larger kinetic energy at low frequencies (i.e., higher $\omega_{c,i}/s$). For larger values of $\omega_{c,i}/s$, previous works have shown that mirror modes can develop comparatively earlier in the simulations, therefore having relatively more time to saturate, and reaching similar amplitudes (\cite{Kunz2014,Melville2016,Riquelme2016,Ley_2023}). In this sense, we would expect a similar late excitation of IC and whistler waves once mirror modes have reached a saturated stage.

The excitation of IC and whistler waves at saturated stages of the mirror instability modulates its nonlinear evolution, and therefore could affect transport processes in the ICM in which mirror modes come into play. 

Particularly important is the pressure anisotropy regulation in the context of collisionless heating and dissipation via magnetic pumping in the ICM (\cite{Kunz2011,Ley_2023}). The marginal stability level that the ion pressure anisotropy reaches at the saturated stage, $\Delta P_i \propto \beta_{\parallel,i}^{0.45}$ (see fig. \ref{fig:IonLecAnisotropy}$a$, also correctly pointed out by \cite{SironiNarayan2015}) is larger than the usual mirror threshold $1/\beta_{\parallel,i}$ by a factor $\sim \beta_{\parallel,i}^{0.55}$. which directly translates into an excess heating of the same order.  Indeed, given that $\beta$ is estimated to be $\beta \sim 10-100$, and that the heating rate is directly proportional to the pressure anisotropy, this could imply a heating rate several times larger than predicted from the mirror threshold, enhancing the efficiency of the mechanism by draining more energy from the turbulent motions that drive the pumping. 

The structures of high and low magnetic field that mirror modes produce in the saturated stage seem to be persistent in time, and its energy $\delta B_{\parallel}^2$ does not decrease as long as the amplification of the mean magnetic field $B$ is maintained (see fig. \ref{fig:MagneticFluctuations}$g$). Even when this amplification is halted or reversed, the decaying timescales of mirror modes are large compared to the typical ion gyroperiod (\cite{Melville2016,Ley_2023}). This implies that the trapping process of ions and electrons also persists, along with the excitation of secondary IC and whistlers. This source of whistler waves can have interesting implications in the context of ICM thermal conduction models like whistler-regulated MHD (\cite{Drake_2021}), as they can dominate the electron scattering in the presence of mirror modes. 

This source of whistler waves associated to mirror modes can also influence the suppression of the effective heat conductivity in the plasma even in the absence of heat-fluxes (\cite{Komarov2016,Riquelme2016,Roberg-Clark2016,Roberg-Clark2018}), and this can have consequences in larger-scale instabilities such as the Magneto-thermal instability (MTI, \cite{Balbus2000,Berlok2021,Perrone2022a,Perrone2022b}). 

Future work aimed towards 3D fully kinetic PIC simulations would be required to have a full understanding of the consequences of the mirror instability and secondary IC/whistler excitation in these high-$\beta$ plasmas.

\begin{acknowledgements}
We thank Aaron Tran for providing the dispersion solver used in this work, and we thank Lorenzo Sironi, Jonathan Squire and Alexander Schekochihin for useful comments and discussion. F.L. acknowledges support from NSF Grant PHY-2010189. M.R. thanks support from ANID Fondecyt Regular grant No. 119167. This work used the Extreme Science and Engineering Discovery Environment (XSEDE), which is
supported by National Science Foundation grant No. ACI-1548562. This work used the XSEDE supercomputer Stampede2 at the Texas Advanced Computer Center (TACC)
through allocation TG-AST190019 (\cite{Towns+2014}). This
research was performed using the compute resources and
assistance of the UW-Madison Center For High Throughput
Computing (CHTC) in the Department of Computer Sciences.
This research was partially supported by the supercomputing
infrastructure of the NLHPC (ECM-02).
\end{acknowledgements}

\bibliography{refs}{}
\bibliographystyle{aasjournal}



\end{document}